\documentclass[apj]{emulateapj}

\newcommand {\Lya}    {Ly$\alpha$}   
\newcommand {\Lyb}    {Ly$\beta$}    
\newcommand {\Lyg}    {Ly$\gamma$}
\newcommand {\Lyd}    {Ly$\delta$}

\newcommand {\Lyeta}  {Ly$\eta$}
\newcommand {\HH}     {H$_2$}        
\newcommand {\HI}     {\ion{H}{1}}   
\newcommand {\OVI}    {\ion{O}{6}}   
\newcommand {\OVII}   {\ion{O}{7}}
\newcommand {\OVIII}  {\ion{O}{8}}
\newcommand {\CIII}   {\ion{C}{3}}   
\newcommand {\NV}     {\ion{N}{5}}
\newcommand {\CIV}    {\ion{C}{4}}
\newcommand {\SiIV}   {\ion{Si}{4}}
\newcommand {\SiIII}  {\ion{Si}{3}}
\newcommand {\SiII}   {\ion{Si}{2}}
\newcommand {\FeIV}   {\ion{Fe}{4}}
\newcommand {\FeIII}  {\ion{Fe}{3}}
\newcommand {\FeII}   {\ion{Fe}{2}}
\newcommand {\CII}    {\ion{C}{2}}

\newcommand {\kms}    {km~s$^{-1}$}
\newcommand {\NHI}    {$N_{\rm HI}$}
\newcommand {\NOVI}   {$N_{\rm OVI}$}
\newcommand {\NCIII}  {$N_{\rm CIII}$}
\newcommand {\NCIV}   {$N_{\rm CIV}$}
\newcommand {\NSiIII} {$N_{\rm SiIII}$}
\newcommand {\NSiIV}  {$N_{\rm SiIV}$}
\newcommand {\NFeIII} {$N_{\rm FeIII}$}
\newcommand {\NNV}    {$N_{\rm NV}$}
\newcommand {\bHI}    {$b_{\rm HI}$}
\newcommand {\tnma}{\tablenotemark{a}}
\newcommand {\tnmb}{\tablenotemark{b}}
\newcommand {\tnmc}{\tablenotemark{c}}
\newcommand {\tnmd}{\tablenotemark{d}}
\newcommand {\tnme}{\tablenotemark{e}}

\newcommand {\nd}  {\nodata}
\newcommand {\lam}    {$\lambda$}
\newcommand {\FUSE}  {{\it FUSE}} 
\newcommand {\HST}  {{\it HST}}
\newcommand {\Chandra} {{\it Chandra}}
\newcommand {\dndz}  {$d{\cal N}/dz$}
\newcommand {\dndx}  {$d{\cal N}/dX$}
\newcommand {\etal}  {et~al.} 
\newcommand {\cd}    {cm$^{-2}$}  

\begin{document}

\title{The Low-$z$ Intergalactic Medium. III. H\,I and Metal Absorbers at $z<0.4$}
\author{Charles W. Danforth and J. Michael Shull}
\affil{CASA, Department of Astrophysical and Planetary Sciences, University of Colorado, 389-UCB, Boulder, CO 80309; danforth@casa.colorado.edu, mshull@casa.colorado.edu}


\begin{abstract} 
We conduct an ultraviolet (\HST\ and \FUSE) spectroscopic survey of
\HI\ (Lyman lines) and seven metal ions (\OVI, \NV, \CIV, \CIII,
\SiIV, \SiIII, \FeIII) in the low-redshift intergalactic medium (IGM)
at $z<0.4$.  We analyzed 650 \Lya\ absorbers over redshift pathlength
$\Delta z=5.27$, detecting numerous absorbers: 83 \OVI\ systems, 39
\CIII, 53 \SiIII, 24 \CIV, 24 \NV, and so on.  Our survey yields
distributions in column density and estimates of the IGM baryon
content and metallicities of C, N, O in the IGM.  In the low-$z$ IGM,
we have accounted for $\sim40$\% of the baryons: 30\% in the
photoionized \Lya\ forest and 10\% in the ($T=10^{5-6}$~K) warm-hot
intergalactic medium (WHIM) traced by \OVI.  Statistical metallicities
of C, N, O ions are consistent with the canonical ($z=0$) value of
10\% solar, with considerable scatter.  Improved statistics for weak
\OVI\ absorbers allows us to estimate $\Omega_{\rm WHIM}/\Omega_b=
0.073\pm0.008$ down to log\,$N_{\rm OVI}=13.4$ and $0.086\pm0.008$
down to log\,$N_{\rm OVI}=13.0$.  The \OVI\ absorber line frequency,
$d{\cal N}/dz=40^{+14}_{-8}$, down to 10 m\AA\ equivalent width
suggests a 250-300 kpc extent of metals around dwarf galaxies
(M$_V=-17$ to $-18$).  Many absorbers appear to contain multiphase
gas, with both collisional ionization and photoionization determining
the ionization state.  \NV\ absorption is well-correlated with \OVI\
and both ions show similarly steep power-law indices $d{\cal
N}/dz\propto N^{-\beta}$ with $\beta_{\rm OVI} \approx\ \beta_{\rm
NV}\approx 2$ while $\beta_{\rm HI}=1.7$.  We conclude that \OVI\ and
\NV\ are reliable tracers of the portion of the WHIM at
$T\approx10^{5-6}$~K.  \CIV\ may be present in both collisional and
photoionized phases; \NCIV\ correlates poorly with both \NHI\ and
\NOVI\ and $\beta_{\rm HI}<\beta_{\rm CIV}<\beta_{\rm OVI}$.  The ions
\CIII, \SiIII, and \SiIV\ are well correlated with \HI\ and show 
patterns typical of photoionization.  We detect \FeIII\ in 14
absorbers with column densities higher than expected, [\FeIII/\SiIII]
~$\approx0.5\pm0.3$.  Adjacent ion stages of the same element
(C\,III/IV and Si\,III/IV) provide useful constraints on the
photoionization parameter, log\,$U\approx-1.5\pm0.5$.  Comparison of
\SiIV\ and \CIV\ with high-$z$ surveys shows a modest increase in line
density, consistent with increasing IGM metallicity at recent epochs.
\end{abstract}



\section{Introduction}


We continue our far-ultraviolet (FUV) surveys of the baryon content and metallicity of the low-redshift intergalactic medium (IGM) by analyzing 650 \Lya\ absorbers, the largest database to date from the {\it Space Telescope Imaging Spectrograph} (STIS) aboard the {\it Hubble Space Telescope} (\HST) and the {\it Far Ultraviolet Spectroscopic Explorer} (\FUSE).  These spectrographs offer sensitive probes of \HI\ (Lyman lines) together with associated metal-line absorbers.  Of particular importance are UV absorption lines from high ion stages (\OVI, \NV) produced primarily in hot, collisionally ionized gas, as well as lower ion stages (\CIV, \CIII, \SiIV, \SiIII, \FeIII) that can arise in both photoionized gas and hot gas.  The advent of these sensitive FUV spectrographs has stimulated an increase in our understanding of the low-redshift IGM.

In this survey, we address several astronomical questions facilitated by a more accurate census of baryons and heavy elements in the IGM.  From observations with \HST, \FUSE, and \Chandra, it seems likely \citep{Shull03,Stocke06a,Nicastro05} that most of the baryons still reside in the IGM, even at low redshift.  But in what thermal phase and ionization state do they exist, and with what spatial distribution?  What are the physical properties (temperature, density) of the low-$z$ absorbers?  From their metallicities and nucleosynthetic patterns, can we uncover their stellar sources?  By comparing the spatial proximity of metal-line absorbers to neighboring galaxies, can we estimate the extent of metal transport into the IGM?  In this paper, we attempt to make progress on all of these issues.  

Over the past decade, the Colorado group and others \citep[e.g.,][]{Tripp08} have focused on UV studies of \HI\ and \OVI\ absorbers in the low-$z$ IGM \citep{Penton1,Penton4,Tripp00,Savage02}.  Danforth \& Shull (2005) and Danforth \etal\ (2006), hereafter denoted Papers~1 and~2, respectively, analyzed IGM absorption at $z\leq0.3$ in \FUSE\ sight lines toward 31 AGN with high-quality data from STIS and FUSE.  Those studies began with a set of 171 strong \Lya\ absorbers (equivalent widths $W_{\lambda}\geq 80$~m\AA) taken from the literature \citep[and other sources]{Penton1,Penton4}.  Because \Lya\ lines are subject to instrumental broadening and line saturation, we endeavored to measure higher-order Lyman transitions (\Lyb--\Lyeta) in the FUSE data at $z_{\rm abs}<0.3$ in order to obtain accurate curve-of-growth (CoG) solutions to the doppler parameter, $b_{\rm HI}$, and neutral hydrogen column density, \NHI.  As first noted in \FUSE\ studies of \Lyb/\Lya\ \citep{Shull00} and confirmed in Paper~2, the \Lya\ measurements alone will systematically underestimate \NHI, a discrepancy that increases in stronger lines, and overestimate $b_{\rm HI}$.  Paper~2 also confirmed the result of \citet{Penton4} that the column-density distribution of \HI\ absorbers follows a power law, $d{\cal N}/dN_{\rm HI}\propto N_{\rm HI}^{-\beta}$, with $\beta_{\rm HI}=1.68\pm0.11$.  The CoG-derived $b$-values show that most of the \HI\ absorbers arise in gas with $T<10^5$~K, incompatible with their existence in a hot (shock-heated) IGM phase.  A small population of broad, shallow \Lya\ absorbers may probe trace amounts of \HI\ in hot, highly ionized gas \citep{Richter04,Lehner07}.
 
Along with the \HI\ lines, Papers 1 and 2 measured corresponding absorption lines of the metal ions \OVI\ \lam\lam1031.926, 1037.617 and \CIII\ \lam977.020.  We found \OVI\ to be the most accessible tracer of the warm-hot ionized medium (WHIM), thought to arise in shock-heated gas at $T=10^{5-6}$~K \citep{Dave99,CenOstriker99}.  The FUV lines of \OVI\ are currently the best tracer of the ``cooler'' portions of these shocks, reaching a peak ionization fraction, $f_{\rm OVI}\approx 0.22$ at temperature $T_{\rm max}=10^{5.45}$~K in collisional ionization equilibrium or CIE \citep{SutherlandDopita93} and similar values in non-equilibrium cooling \citep{GnatSternberg07}.  Owing to the relatively high solar oxygen abundance, $\rm (O/H)_{\odot}\approx 5\times10^{-4}$, and the substantial oscillator strength of the stronger line of the \OVI\ doublet (1031.9261~\AA, $f=0.1325$), the \OVI\ ultraviolet transitions are detectable down to column densities $N_{\rm OVI}\approx10^{13}$~\cd, more than two orders of magnitude below the current sensitivity of the X-ray absorption lines of \OVII\ (21.602~\AA, $f=0.696$) and \OVIII\ (18.969~\AA, $f=0.416$) studied with \Chandra\ and {\it XMM-Newton}.  Although the ({\it 1s--2p}) X-ray transitions are crucial for detecting the hotter WHIM at $T>10^6$~K, their intrinsic line strengths, $f\lambda$, are much weaker than the ({\it 2s--2p}) \OVI\ ultraviolet transitions and the detection limits and velocity resolution of current X-ray instruments make them much less sensitive.  

Papers 1 and 2 represented a significant statistical improvement over previous studies of small numbers of sight lines \citep[e.g.][]{Tripp00,Savage02} with 40 \OVI\ detections and $\Delta z\approx 2.2$ vs.\ approximately 10 \OVI\ absorbers over $\Delta z\sim0.5$.  The detection statistics were sufficient for us to draw a number of intriguing conclusions.  We found that \NOVI\ was poorly correlated with \NHI, and the power-law distribution of \OVI\ column density, $\beta_{\rm OVI}=2.2\pm0.1$, was significantly steeper than for \HI.  These results implied a multiphase IGM with hot, shocked WHIM (traced by \OVI) kinematically associated with cooler, photoionized \HI.  We inferred a total cosmological mass fraction in the WHIM of at least $\Omega_{\rm WHIM}=(0.0022\pm0.0003)\,[h_{70}\,(Z_{\rm O}/0.1 Z_{\odot})\,(f_{\rm OVI}/0.2)]^{-1}$ or $4.8\pm0.9\%$ of the total baryonic mass at $z<0.15$.  Unlike \OVI, the 30 \CIII\ detections showed reasonable column density correlation with \HI, and the \dndz\ distributions showed similar slopes, suggesting that \CIII\ is predominantly photoionized.

In this survey (Paper~3), we expand the parameter space of the low-$z$ IGM survey, using a much larger data set of 650 \Lya\ absorbers, observed over a total redshift pathlength $\Delta z=5.27$ (compared to $\Delta z=2.2$ in Papers~1 and~2).  There were several limitations inherent in our initial survey.  Our combined STIS/E140M$+$\FUSE\ spectral survey provides an expansion and improvement in three important ways:

(1) In Paper 1, we limited our search to ``strong'' \Lya\ systems ($W_{\rm Ly\alpha}>80$ m\AA).  However, \citet{Penton4} showed that the median $W_{\rm Ly\alpha}=68$~m\AA, so this equivalent-width threshold eliminated more than half of the possible IGM systems from study.  Furthermore, Paper~1 found little if any correlation between \NHI\ and \NOVI.  Indeed, at low column densities, log\,$N_{\rm HI}\leq13.4$, we found many absorbers with $N_{\rm OVI}\approx N_{\rm HI}$.  Thus, measurable \OVI\ absorption is possible even in weak \HI\ absorbers.  We relax this ``\Lya\ bias'' here by measuring metal ion absorption in all detected \Lya\ systems. 

(2) The combined STIS/E140M$+$\FUSE\ dataset covers a large wavelength range: 905--1710~\AA\ in all cases and out to 1729~\AA\ in most of the datasets.  We use the longer wavelengths available in the E140M data to measure \HI, \OVI, and \CIII\ at higher redshifts than were possible in the \FUSE\ data alone ($z_{\rm OVI}\leq0.15$).  We now cover \Lya\ out to $z\leq0.4$ and measure higher-order Lyman lines in strong systems to even higher redshifts in some sight lines.  This expanded redshift space more than doubles the \OVI\ absorber pathlength.

(3) We use the broad spectral coverage of STIS to measure metal ions not found in \FUSE\ data.  Lines of \NV\ and \CIV\ can exist at temperatures nearly as high as \OVI, and they may provide additional tracers of the cooler WHIM.  We also measure lines of \SiIV, \SiIII, and \FeIII, which trace predominantly photoionized material.  By observing adjacent ionization states of the same elements (C\,III/IV, Si\,III/IV), we can make direct ionization measurements without the complication of relative abundance uncertainties.

\noindent In \S~2, we discuss our selection of AGN sight lines and the rationale behind our choice of spectral lines.  In \S~2.2 and \S~2.3, we discuss our technique for \Lya\ line identification and the measurement and verification of metal ion lines.  In \S~2.4 and \S~2.5, we detail our analytic procedure and methods for deriving quantities of cosmological interest such as \dndz, $\beta_{\rm ion}$, metallicity, and $\Omega_{\rm ion}$.  Our basic detection results are presented in \S~3.1.  We compare new \HI, \OVI, and \CIII\ results to those of previous work and discuss the new ions (\CIV, \NV, \SiIII, \SiIV, \FeIII) in this context.  In \S~3.2, we compare observed line ratios with a set of CLOUDY models to derive typical IGM ionization parameters and temperatures.  We discuss the implications of these results to cosmology in \S~3.3. In \S~4 we summarize our conclusions.

\section{Observations and Data Analysis}

\subsection{Sight Line Selection and Data Reduction}

We have performed detailed \FUSE+STIS analyses on sight lines toward 35 bright AGN.  The primary objective of this census is to remove as much ``rich sight line'' bias as possible, while accumulating the longest available pathlength through the low-redshift IGM.  Therefore, we included {\it all} available AGN with both \FUSE\ and STIS/E140M data of reasonable quality.  The combined dataset covers 905--1710~\AA\ at a resolution of 20~\kms\ or better, at a signal to noise per resolution element $(S/N)_{\rm res}\ga5$ in most cases.  This gives us a potential redshift pathlength out to $z\ga0.4$ in \Lya, \Lyb, \OVI, \CIII, and most of the other species, but only out to $z\approx 0.1$ in \CIV.  Five of our sight lines (3C\,273, Akn\,654, Mrk\,509, Mrk\,205, and PG\,1116+215) are lacking the longest wavelength order of the STIS echelle data and truncate at $\lambda=1709$~\AA.  The other 30 sight lines include the zeroth-order data at $1711<\lambda<1729$~\AA, which we include where possible.

\begin{deluxetable}{lcclrrc}
\tabletypesize{\footnotesize}
\tablecolumns{7} 
\tablewidth{0pt} 
\tablecaption{{\it FUSE}+STIS IGM Sight Lines}
\tablehead{\colhead{Sight Line}   &
           \colhead{R.A.}         &
	   \colhead{Decl.}        & 
	   \colhead{$z_{\rm AGN}$}&
	   \colhead{STIS}         &
           \colhead{FUSE}         &
	   \colhead{Papers}       \\
	   \colhead{}   &        
	   \colhead{(J2000.0)}    &
	   \colhead{(J2000.0)}    &
	   \colhead{}& 
	   \colhead{(ksec)}       &
	   \colhead{(ksec)}       &
	   \colhead{I \& II\tablenotemark{a}} }
\startdata 
Mrk\,335         & 00 06 19.5  & $+$20 12 10 &  0.025785 & 17.1 &  97.0  & Y\\ 
HE\,0226$-$4410  & 02 28 15.2  & $-$40 57 16 &  0.495    & 43.8 &  33.2  & Y\\ 
PKS\,0312$-$770  & 03 11 55.2  & $-$76 51 51 &  0.223000 &  8.4 &   5.5  & N\\ 
PKS\,0405$-$123  & 04 07 48.4  & $-$12 11 37 &  0.572590 & 27.2 &  71.1  & Y\\ 
HS\,0624$+$6907  & 06 30 02.5  & $+$69 05 04 &  0.370000 & 62.0 & 112.3  & N\\ 
PG\,0953$+$414   & 09 56 52.4  & $+$41 15 22 &  0.234100 &  8.0 &  72.1  & Y\\ 
Ton\,28          & 10 04 02.5  & $+$28 55 35 &  0.329700 & 33.0 &  11.2  & N\\ 
3C\,249.1        & 11 04 13.7  & $+$76 58 58 &  0.311500 & 68.8 & 216.8  & N\\ 
PG\,1116$+$215   & 11 19 08.6  & $+$21 19 18 &  0.176500 & 26.5 &  77.0  & Y\\ 
PG\,1211$+$143   & 12 14 17.7  & $+$14 03 13 &  0.080900 & 42.5 &  52.3  & Y\\ 
PG\,1216$+$069   & 12 19 20.9  & $+$06 38 38 &  0.331300 &  5.8 &  12.0  & N\\ 
Mrk\,205         & 12 21 44.0  & $+$75 18 38 &  0.070846 & 62.1 & 203.6  & N\\ 
3C\,273          & 12 29 06.7  & $+$02 03 09 &  0.158339 & 18.7 &  42.3  & Y\\ 
Q\,1230$+$0115   & 12 30 50.0  & $+$01 15 23 &  0.117000 &  9.8 &   4.0  & N\\ 
PG\,1259$+$593   & 13 01 12.9  & $+$59 02 07 &  0.477800 & 95.8 & 668.3  & Y\\ 
PKS\,1302$-$102  & 13 05 33.0  & $-$10 33 19 &  0.278400 &  4.8 & 142.7  & Y\\ 
Mrk\,279         & 13 53 03.4  & $+$69 18 30 &  0.030451 & 54.6 & 228.5  & Y\\ 
NGC\,5548        & 14 17 59.5  & $+$25 08 12 &  0.017175 & 69.8 &  55.0  & N\\ 
Mrk\,1383        & 14 29 06.6  & $+$01 17 06 &  0.086470 & 10.5 &  63.5  & Y\\ 
PG\,1444$+$407   & 14 46 45.9  & $+$40 35 06 &  0.267300 & 48.6 &  10.0  & N\\ 
Mrk\,876         & 16 13 57.2  & $+$65 43 10 &  0.129000 & 29.2 &  46.0  & Y\\ 
3C\,351          & 17 04 41.4  & $+$60 44 31 &  0.371940 & 77.0 & 141.9  & N\\ 
H\,1821$+$643    & 18 21 57.3  & $+$64 20 36 &  0.297000 & 50.9 & 132.3  & Y\\ 
Mrk\,509         & 20 44 09.7  & $-$10 43 25 &  0.034397 &  7.6 &  62.3  & Y\\ 
PHL\,1811        & 21 55 01.5  & $-$09 22 25 &  0.190000 & 33.9 &  75.0  & Y\\ 
PKS\,2155$-$304  & 21 58 52.0  & $-$30 13 32 &  0.116000 & 10.8 & 123.2  & Y\\ 
Akn\,564         & 22 42 39.3  & $+$29 43 31 &  0.024684 & 10.3 &  60.9  & N\\ 
NGC\,7469        & 23 03 15.6  & $+$08 52 26 &  0.016317 & 22.8 &  44.3  & N
\enddata      
\tablenotetext{a}{Sight line included in Papers I \& II?}
\end{deluxetable}

Of the 35 sight lines, we excluded six targets (NGC\,3516, NGC\,3783, NGC\,4051, NGC\,4151, NGC\,4395, and NGC\,4593) at $z_{\rm AGN}\leq0.01$ because the available redshift pathlength is minimal and any absorbers could be intrinsic to either our Galaxy or the AGN.  We also excluded the high-redshift object HS\,1700$+$6416 ($z=2.72$) since its interpretation was greatly complicated by high-$z$ EUV lines from several partial Lyman-limit systems \citep{Fechner06}.  Our final sample of 28 AGN sight lines is presented in Table~1.  They cover a total \Lya\ redshift pathlength of $\Delta z=5.27$ with $\langle z_{\rm Ly\alpha}\rangle\approx0.14$ and $\langle z_{\rm AGN}\rangle=0.21$.

We retrieved the \FUSE\ data from the Multimission Archive at the Space Telescope Science Institute (MAST) archive and reduced them as in Paper~2.  We binned the data by three instrumental pixels.  The \FUSE\ instrumental resolution of $\sim20$~\kms\ typically represents 8--10 pixels or $\sim3$ bins.  We used primarily \FUSE\ data from the highest-throughput LiF1 channel \citep{Moos00}, but went to LiF2 or SiC channels when LiF1 did not cover the region of interest or to provide a check on marginal LiF1 detections.  The archival STIS/E140M data were reduced at the University of Colorado by S. V. Penton.  We smoothed the data over three pixels to better match the nominal instrumental resolution of $\sim7$~\kms.  Data from both instruments were normalized in overlapping 10~\AA\ segments.  We selected line-free continuum regions interactively and fitted the spectral continuum using low-order Legendre polynomials.  For each 10~\AA\ segment, we calculated the local S/N from the standard deviation of the normalized continuum data about the fitted continuum.

\begin{figure*}
  \epsscale{1} \plotone{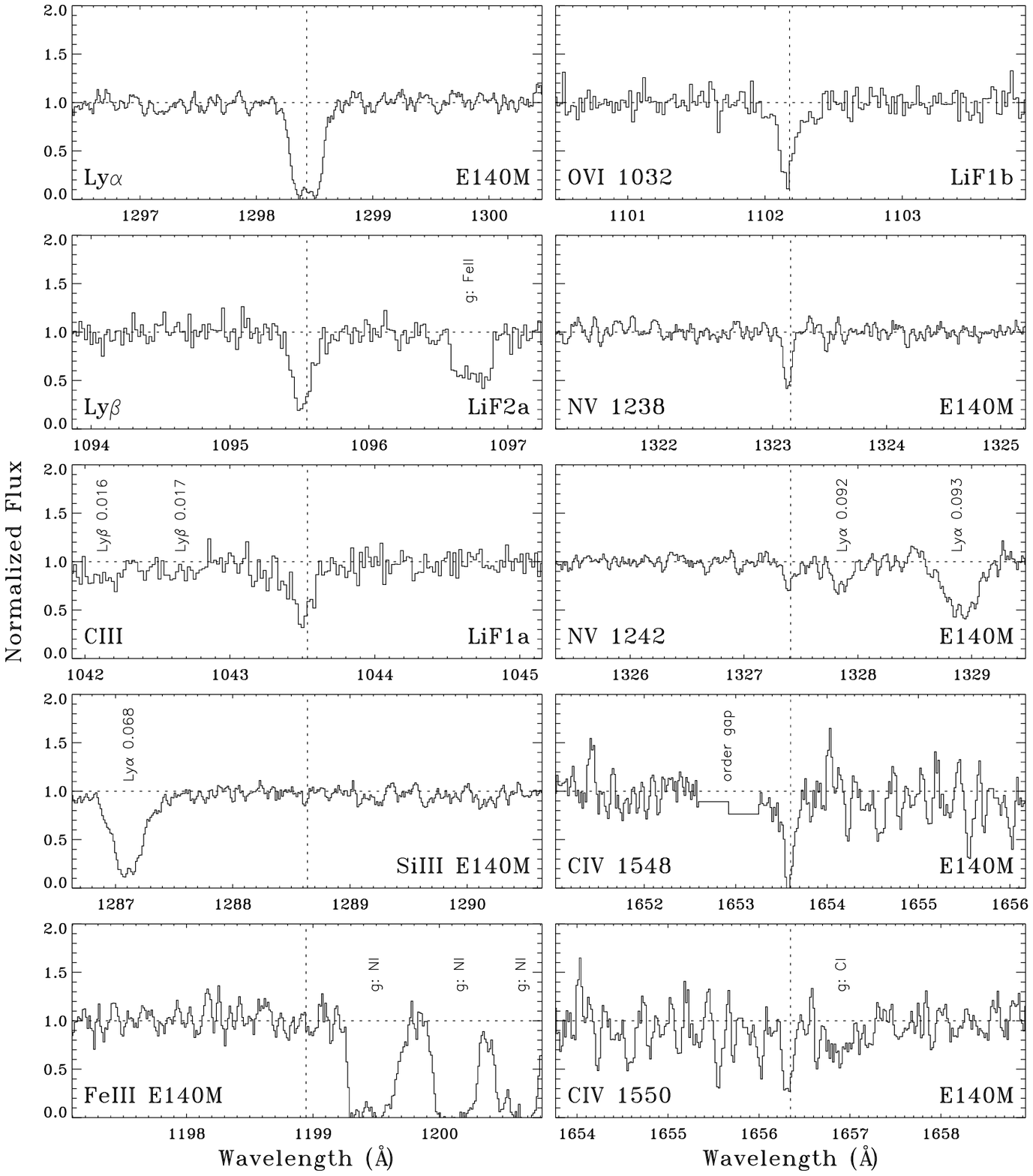} 
  \caption{\small IGM absorption in the $z=0.06808$ absorber toward
  PG\,0953$+$414 showing typical, normalized \FUSE\ and STIS/E140M
  data.  \Lya\ and \Lyb\ (top left) show strong, consistent profiles
  ($W_{\rm Ly\alpha}=284\pm13$ m\AA\ and $W_{\rm Ly\beta}=128\pm16$
  m\AA).  C\,III is detected but Si\,III and Fe\,III are
  non-detections.  High ions are depicted in the right hand panels:
  O\,VI \lam1032 and both N\,V lines (center left) are detected with
  consistent profiles.  The O\,VI \lam1038 line is blended with a
  strong \HH\ transition and is not shown.  C\,IV shows noisy but
  consistent detections in both bands of the doublet as well.  The two
  Si\,IV transitions are not shown.  Each panel is centered at the
  redshifted wavelength of the transition and covers $\pm500$~\kms\ in
  either direction.  Other detected features in the data are
  identified.  The ``g:'' prefix denotes a Galactic absorption line,
  while a numerical suffix denotes the redshift of an IGM line.  The
  source channel (e.g. STIS/E140M or \FUSE\ LiF2a) is indicated in the
  lower right.}  \label{fig:stack1}
\end{figure*}

\begin{figure*}
  \epsscale{1}\plotone{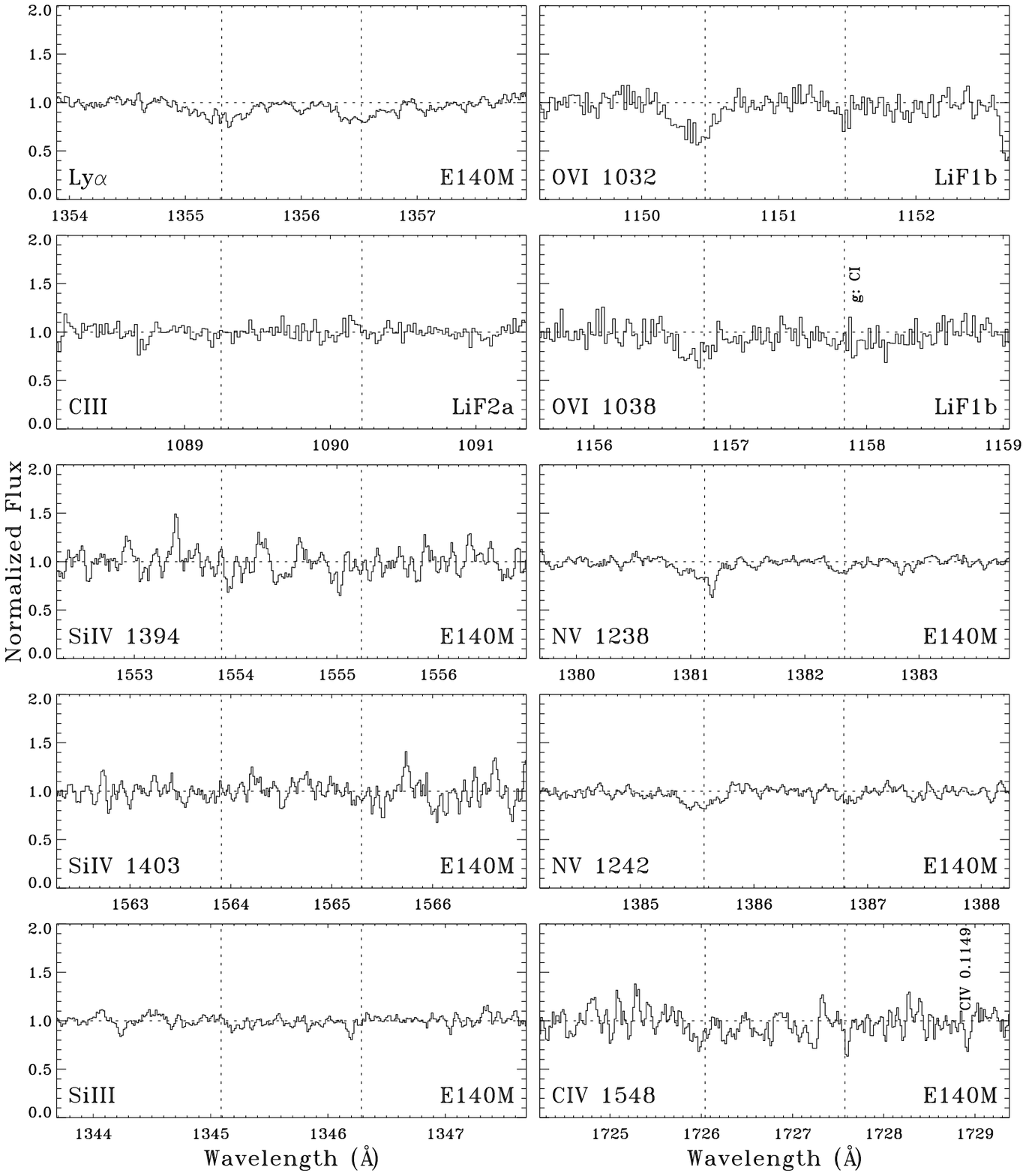} 
  \caption{\small Same as Fig. \ref{fig:stack1} for two absorbers in
  the Mrk\,876 sight line.  The vertical dashed lines in the panels
  show the redshifted locations of the $z=0.11487$ (left) and
  $z=0.11586$ (right) absorbers.  Both absorbers show weak, broad
  \Lya\ profiles ($W_{\rm Ly\alpha}=87\pm13$ and $74\pm5$~m\AA,
  respectively), and the expected \Lyb\ equivalent widths are below
  the level of the noise and are not shown.  C\,III shows a clean
  non-detection at both redshifts.  Data from the stronger Si\,IV
  transition ($\lambda$1394) is consistent with noise and a better
  upper limit is given by the weaker Si\,IV $\lambda$1403 line.
  Si\,III shows a clean non-detection at $z=0.11487$, but a marginal
  detection at $z=0.11586$.  The Li-like ions are a different matter:
  O\,VI and N\,V absorption is detected at both redshifts in both
  lines of both ions (O\,VI $\lambda$1038 line at $z=0.11586$ is
  blended with a weak Galactic C\,I line and is taken as an upper
  limit).  Both systems show weak C\,IV 1548\AA\ detections.  C\,IV
  1550\AA\ is detected for $z=0.11487$, but is redshifted out of the
  data range for $z=0.11586$.}  \label{fig:stack2}
\end{figure*}

\subsection{Identification of \Lya\ Absorbers}

We identified the associated higher Lyman lines and metal absorption systems using the \Lya\ lines as ``signposts''.  In the software, we displayed the normalized STIS/E140M spectrum in 10~\AA\ segments, marking the ISM lines and IGM line positions corresponding to previously-identified \Lya\ systems (e.g., Figures~\ref{fig:stack1} and~\ref{fig:stack2}).  Any unidentified line was centroided, measured, and initially assumed to be \Lya.  For weak lines, including broad shallow absorbers, we expect the optical depths to be in the ratio, $\tau({\rm Ly}\alpha)/\tau({\rm Ly}\beta)\approx 6.2$.  If the wavelength and equivalent width at the position of the corresponding \Lyb\ and higher-order Lyman lines were consistent with the \Lya\ identification, we accepted the absorber and entered it into a database.  For strong absorbers ($W_{\rm Ly\alpha}\ga150$~m\AA), we would expect to detect \Lyb\ or higher-order Lyman lines.  Quite a few potential strong lines, initially assumed to be \Lya, were ruled out by this criterion.  Line identification was an iterative process:  we assumed weak, unidentified lines to be \Lya\ until proven otherwise.  More careful analysis occasionally showed that lines identified as weak \Lya\ were better explained as metal lines of stronger IGM systems, high-order Lyman lines of a strong higher-redshift system, intrinsic AGN features, high velocity components of Galactic lines, or instrumental artifacts.

Since the rest-frame line density of FUV absorbers increases dramatically at shorter wavelengths, we began our \Lya\ search at long wavelengths and worked our way to the blue.  The identification process was carried out iteratively until all absorption features were either identified or ruled out as not being \Lya\ systems.  In several cases, \Lya\ lines were blended or redshifted beyond the long wavelength end of the data (1710 or 1729~\AA, $z_{\rm abs}>0.406$ or $0.422$, depending on sight line), and strong \HI\ absorbers were then identified through their \Lyb\ absorption.  Our \Lya\ identifications were informed by those of previous work, in particular \citet{Penton1,Penton4}, but we searched each sight line and verified each absorber independently. 

Although some fraction of strong IGM systems are composed of multiple velocity components, this is not always apparent in the often-saturated \Lya\ line profiles.  In cases where higher Lyman lines showed multiple, unambiguously separable components in \Lya\ or \Lyb, we treated each component as a separate absorber.  However, most profiles were consistent with single components or showed some asymmetry which we could not split unambiguously into multiple features.  These cases (the majority) are treated as single absorbers.  Our counting statistics are intermediate between the ``component'' and ``system'' nomenclature of \citet{Tripp08}, but generally closer to the latter classification.

\begin{deluxetable*}{llccccccc}
\tabletypesize{\footnotesize}
\tablecolumns{9} 
\tablewidth{0pt} 
\tablecaption{Sample H\,I Absorbers}
\tablehead{\colhead{Sight Line}    &
   \colhead{$z_{\rm abs}$}         &
   \colhead{W$\rm_r(Ly\alpha)$}    &
   \colhead{W$\rm_r(Ly\beta)$}     &
   \colhead{W$\rm_r(Ly\gamma)$}    &
   \colhead{W$\rm_r(Ly\delta)$}    &
   \colhead{$b_{\rm HI}$}          &
   \colhead{log\,\NHI}             &
   \colhead{method\tnma}          \\
\colhead{}                         &
   \colhead{}                      &
   \colhead{(m\AA)}                &
   \colhead{(m\AA)}                &
   \colhead{(m\AA)}                &
   \colhead{(m\AA)}                &
   \colhead{(\kms)}                &
   \colhead{($N$ in cm$^{-2}$)}           &
   \colhead{}                      }
\startdata
Mrk\,335       & 0.00653& $  210\pm  19$& $   59\pm   8$& \nodata       & \nodata       & $ 20^{+ 7}_{ -4}$& $ 13.99^{+0.10}_{-0.11}$& COG \\
Mrk\,335       & 0.00759& $   78\pm   3$& $<  13       $& \nodata       & \nodata       & $ 33\pm 4      $& $ 13.24^{+0.06}_{-0.05}$& Ly$\alpha$ AOD\\
Mrk\,335       & 0.02086& $  149\pm   4$& $   42\pm   7$& $< 205       $& $<  29       $& $ 60\pm 2      $& $ 13.51\pm0.02         $& Ly$\alpha$ AOD\\
&&&&&&&&\\
HE\,0226$-$4410& 0.01216& $   50\pm   3$& $<  18       $& \nodata       & \nodata       & $ 41\pm 7      $& $ 13.06^{+0.08}_{-0.07}$& Ly$\alpha$ AOD\\
HE\,0226$-$4410& 0.02682& $   66\pm  10$& $   23\pm   4$& \nodata       & \nodata       & $ 30\pm 4      $& $ 13.17^{+0.12}_{-0.05}$& Ly$\alpha$ AOD\\
HE\,0226$-$4410& 0.02729& $   30\pm   5$& $<  18       $& \nodata       & \nodata       & $ 31\pm 5      $& $ 12.79^{+0.13}_{-0.09}$& Ly$\alpha$ AOD\\
HE\,0226$-$4410& 0.04129& $   40\pm   4$& $<  19       $& \nodata       & \nodata       & $ 32\pm 7      $& $ 12.93^{+0.09}_{-0.07}$& Ly$\alpha$ AOD\\
HE\,0226$-$4410& 0.04609& $  157\pm  12$& $   41\pm   3$& \nodata       & \nodata       & $ 16^{+ 4}_{ -3}$& $ 13.84^{+0.06}_{-0.09}$& COG \\
HE\,0226$-$4410& 0.05177& $   26\pm   3$& $<  26       $& \nodata       & \nodata       & $\sim 17         $& $ 12.76^{+0.27}_{-0.12}$& Ly$\alpha$ AOD\\
HE\,0226$-$4410& 0.05814& $   30\pm  10$& \nodata      & $<  14       $& $<  29       $& $ 12\pm 3      $& $ 12.77\pm0.14         $& Ly$\alpha$ fit
  \enddata
  \tablenotetext{a}{H\,I solution method: AOD=integrated apparent optical depth, fit=Voigt fit, COG=multi-line curve of growth solution, COG$^\dag$=poorly-constrained curve of growth.}
  \label{tab:h1}
\end{deluxetable*}

In total, we identified 660 \HI\ absorbers via \Lya\ and/or \Lyb\ absorption and found several discrepancies with published line lists.  We truncated the line list at $z=0.4$ since data quality falls rapidly beyond 1700~\AA.  This left us with a total of 650 IGM systems, presented\footnote{Truncated sample tables are presented here.  Full versions can be seen in the electronic version.} in Table~\ref{tab:h1}. 

\subsection{Line Measurements and Consistency Checks}

\begin{deluxetable*}{lrlcccc}
\tabletypesize{\footnotesize}
\tablecolumns{7} 
\tablewidth{0pt} 
\tablecaption{IGM Diagnostic Lines}
\tablehead{\colhead{ion}                         &
           \colhead{$\lambda_{\rm rest}$ (\AA)}  &
           \colhead{$f$~\tnma}                    &
           \colhead{$[X/H]_\sun$\tnmb}           &
           \colhead{I.P.\tnmc (eV)}              &	
	   \colhead{$f_{\rm ion, CIE}$~\tnmc}          &
           \colhead{log\,$T_{\rm max}$}          }	
	\startdata	
	H\,I    &  1215.67 &  0.4164 &0.00    & $<$13.6 &\nodata&$<$4.19\tnmd    \\   
	        &  1025.72 &  0.07914&        &         &     &                \\
	Fe\,III &  1122.52 &  0.0544 &$-4.55$ & 16-31   &0.89 &4.27-4.58\tnmd  \\
	Si\,III &  1206.50 &  1.63   &$-4.49$ & 16-35   &0.90 &4.30-4.85\tnmd  \\
	C\,III  &  977.02  &  0.757  &$-3.61$ & 24-48   &0.83 &4.66-5.01\tnmd  \\	
	Si\,IV  &  1393.76 &  0.514  &$-4.49$ & 35-48   &0.35 &      4.8\tnme  \\
	        &  1402.70 &  0.255  &        &         &     &                \\
	C\,IV   &  1548.20 &  0.191  &$-3.61$ & 48-65   &0.29 &      5.0\tnme  \\
	        &  1550.77 &  0.095  &        &         &     &                \\
	N\,V    &  1238.82 &  0.157  &$-4.22$ & 78-98   &0.24 &      5.25\tnme \\
	        &  1242.80 &  0.078  &        &         &     &                \\
	O\,VI   &  1031.93 &  0.1325 &$-3.34$ & 114-138 &0.22 &      5.45\tnme \\
	        &  1037.62 &  0.0658 &        &         &     &                
	\enddata
	\tablenotetext{a}{~Line oscillator strength \citep{Morton03}}
	\tablenotetext{b}{~Solar photospheric element abundance log\,(X/H); \citet{Asplund05}}
	\tablenotetext{c}{~Peak ion fraction under CIE via \citet{SutherlandDopita93}; value used in $\Omega_{\rm ion}$ and $Z_{\rm ion}$ calculations over the listed range in ionization potential.}
	\tablenotetext{d}{~Temperature range over which ion is dominant under collisional ionization equilibrium, via \citet{SutherlandDopita93}}
	\tablenotetext{e}{~Ion is never dominant under CIE.  Listed temperature is peak CIE abundance.}
	\label{tab:atomicdata}
\end{deluxetable*}

The next step was to perform detailed measurements of 11 UV resonance absorption transitions of seven ion species (Table~\ref{tab:atomicdata}), spanning a range of ionization levels and strengths.  In particular, \OVI\ and \CIV\ are normally interpreted as tracers of hot gas, with peak ionization fractions ($f_{\rm ion}$) at temperature $T_{\rm max}$ in collisional ionization equilibrium (CIE).  As computed by \citet{SutherlandDopita93}, \OVI\ has maximum CIE abundance of $f_{\rm OVI}=0.22$ at log\,$T_{\rm max}=5.45$, \NV\ reaches $f_{\rm NV}=0.24$ at log\,$T_{\rm max}=5.25$, and $f_{\rm CIV}=0.29$ at log\,$T_{\rm max}=5.00$.  Non-equilibrium ionization effects may alter the post-shock ionization fractions \citep{RajanShull08}.  Because \NV\ lies intermediate in ionization potential between the two other Li-like ions, it should be a good WHIM tracer as well.  However, the solar nitrogen abundance is more than five times lower than oxygen, and nitrogen is often observed to be underabundant in Galactic HVCs \citep{Gibson01,Collins03}.  Until now, \NV\ absorption is largely an unknown quantity in the low-$z$ IGM.

The UV lines of \SiIV\ $\lambda\lambda$1393.755, 1402.770, \SiIII\ $\lambda$1206.500, \CIII\ $\lambda$ 977.020, and \FeIII\ $\lambda$1122.526 are all strong transitions of abundant metals, with production ionization potentials below 3~ryd (40.8~eV) and expected to probe photoionized, metal-enriched material.  In these lines, we also have two sets of adjacent ionization levels (C\,III/IV and Si\,III/IV), which may help to define the ionization state of the absorbers, independent of elemental abundance ratios.  Highly ionized metals were the primary focus of this survey and singly-ionized species were not explicitly studied.  However, low-ion absorption exists in some \Lya\ IGM systems \citep{Tripp08} and in many ionized HVCs \citep{Sembach99,Collins03,Collins07}.  In a future paper, we will examine the behavior of \SiII, \CII, \FeII, and similar species in the IGM and give further analysis options in adjacent ionization stages (Si\,II/III/IV, C\,II/III/IV, Fe\,II/III).

We examined all 13 transitions for each IGM absorber;  Figures~1 and~2 show examples.  In each case, we measured the line using the apparent optical depth (AOD) method \citep{SembachSavage92} and/or fitted the line with a Voigt profile (see discussion in Paper~2 and sources therein).  While the Voigt and AOD methods generate equivalent line measurements and uncertainties in most cases, each has its advantages; AOD is preferable for weak, asymmetric, or noisy lines, while Voigt fits are more reliable for saturated or blended lines.  In cases where no line was apparent, we measured a $4\sigma$ upper limit on equivalent width, $W_{\lambda}$, according to 
\begin{equation}
  W_{\rm min}=\frac{4~\lambda}{R~(S/N)_{\lambda}} \; , \label{wmineq} 
\end{equation} 
where we adopted instrumental resolutions, $R=\lambda/\Delta\lambda$, for the two spectrographs of $R_{\rm FUSE}=15,000$ and $R_{\rm STIS/E140M}=42,000$.  In cases where the measured $W_{\lambda}\le W_{\rm min}$, we declared the line to be an upper limit.  Heavily blended lines, in which no information was available, and lines that were redshifted out of the data range were noted but ignored.  Measurements of less severely blended lines and lines with ambiguous identification were measured and taken to be upper limits.

In the second stage of processing, we checked our detections for consistency.  We required the \Lyb\ detections or upper limits to be consistent with \Lya\ measurements, and the metal doublets (\CIV, \NV, \OVI, \SiIV) to be consistent within the error bars in log\,$N$.  The singlet lines (\CIII, \SiIII, \FeIII) were harder to verify this way, but we were able to confirm or rule out many based on detections or non-detections in similar ions in the same absorption system. 

We carried out a third-stage vetting process of all metal-line detections to eliminate likely fixed pattern detector noise \citep{Sahnow00} masquerading as weak detections, instrumental artifacts, and other inconsistencies.  In some cases a ``detection'' was indistinguishable from adjacent data fluctuations and there was no other supporting evidence such as detections in other ions in the same absorber.  Such cases were deemed suspect and demoted to upper limits even though they were technically $>4\sigma$.  The effects of this process on the detection statistics varied.  Unfortunately, an unknown number of weak but bona-fide detections may have been eliminated in this check, but the remaining detections are much more robust than in the earlier stages.  The elimination of some marginal detections is apparent in the low-$N$ turn-overs seen in some \dndz\ plots (see below for discussion).  If we assume that absorbers follow a power-law distribution with $N$, we can assess the effect of this final detection verification step.  However, the low-$N$ distribution of metal-line absorbers may also be truncated owing to the limited volumes over which galaxies can disperse metals \citep[][Paper~1]{TumlinsonFang05}.

The final processing stage involved deriving column densities and $b$-values for the different species.  We determined most (69\%) of the \HI\ values directly from \Lya\ measurements.  Paper~2 showed that \Lya-only measurements underestimate \NHI, owing to saturation and the lack of CoG information from higher Lyman lines.  The effect is strongest in saturated lines and is probably negligible for $W_{\rm Ly\alpha}\la100$~m\AA.  The median equivalent width in our sample is $W_{\rm Ly\alpha}=94$~m\AA.  Eleven \HI\ absorbers were measured from \Lyb\ only, when \Lya\ was strongly blended or redshifted out of the E140M spectral range ($z\ga0.4$).  Values of column density and doppler width for strong \HI\ absorbers were usually (30\%) determined from concordance CoG plots (see Paper~2 for discussion).  In some cases, we measured \Lyg\ and \Lyd\ to further constrain the \HI\ solution.  Lyman-line equivalent widths and the derived \bHI\ and \NHI\ measurements are listed in Table~\ref{tab:h1}.

We took column densities and $b$-values of the doublet ions (\OVI, \NV, \CIV, \SiIV) as the error-weighted mean of the two lines, when possible, or as measured if only one line was available.  Values for the single lines (\CIII, \SiIII, \FeIII) were taken as measured by the AOD or Voigt profile methods.  We present truncated tables of the measured equivalent widths, $b$-values, and column densities for all seven ions in Tables~4--10.  Note that AOD-measured $b$-values are equivalent to Voigt fit measurements for strong lines, but become increasingly uncertain for weak and noisy lines.  Some of the poorly-determined $b$-values are listed as approximate and AOD $b$-values in the tables should generally be treated with caution.  Full versions of the Tables, including both detections and upper limits, are available in the electronic edition.

\begin{deluxetable*}{llccccccl}
\tabletypesize{\footnotesize}
\tablecolumns{9} 
\tablewidth{0pt} 
\tablecaption{Sample O\,VI Detections}
\tablehead{\colhead{Sight Line}        &
   \colhead{$z_{\rm abs}$}             &
   \colhead{$b$ (1032)}  &
   \colhead{W$\rm_r$(1032)}&
   \colhead{$b$ (1038)}  &
   \colhead{W$\rm_r$(1038)}&
   \colhead{$b$(O\,VI)}&
   \colhead{log\,$N$(O\,VI)}&
   \colhead{notes}                    \\
  \colhead{}                           &
   \colhead{}                          &
   \colhead{(\kms)}                    &
   \colhead{(m\AA)}                    &
   \colhead{(\kms)}                    &
   \colhead{(m\AA)}                    &
   \colhead{(\kms)}                    &
   \colhead{($N$ in cm$^{-2}$)}               &
   \colhead{}                          }
\startdata
HE\,0226$-$4410& 0.20700& $  34\pm  1 $ & $  184\pm 10$& $  34\pm  8 $ & $  111\pm  4$& $ 34\pm  8$& $  14.33\pm0.06$ & fit, AOD \\
HE\,0226$-$4410& 0.22004& $ \sim 16  $ & $  12\pm  1$& \nodata & $<  31       $& $ \sim 16$& $  13.05\pm0.15$ & AOD \\
HE\,0226$-$4410& 0.24515& $ \sim 29  $ & $  23\pm  1$& $ \sim 18  $ & $  10\pm  1$& $ \sim 25$& $  13.29\pm0.15$ & AOD, AOD \\
HE\,0226$-$4410& 0.31833& $   6\pm  3 $ & $    9\pm  2$& \nodata & $<   7       $& $  6\pm  3$& $  12.97\pm0.17$ & AOD \\
HE\,0226$-$4410& 0.34031& $  21\pm  2 $ & $   74\pm 11$& \nodata & $<  82       $& $ 21\pm  2$& $  13.75\pm0.05$ & fit \\
HE\,0226$-$4410& 0.35493& $  13\pm  9 $ & $    9\pm  3$& \nodata & $<   8       $& $ 13\pm  9$& $  12.94\pm0.26$ & AOD \\
HE\,0226$-$4410& 0.35524& $  44\pm  6 $ & $   63\pm  2$& $  35\pm  5 $ & $   22\pm  1$& $ 41\pm  6$& $  13.69\pm0.08$ & AOD, AOD \\
PKS\,0312$-$770& 0.20192& \nodata & \nodata      & $ \sim 40  $ & $  31\pm  1$& $ \sim 40$& $  13.76\pm0.15$ & AOD \\
PKS\,0312$-$770& 0.20275& $  62\pm  2 $ & $  527\pm 28$& $  66\pm  2 $ & $  323\pm 16$& $ 63\pm  2$& $  14.85\pm0.07$ & fit, fit \\
PKS\,0405$-$123& 0.09657& $  40\pm 20 $ & $   59\pm 31$& $  14\pm 13 $ & $   32\pm 27$& $ 31\pm 20$& $  13.71\pm0.21$ & fit, fit \\
PKS\,0405$-$123& 0.16714& $  82\pm  4 $ & $  470\pm 40$& $  61\pm  5 $ & $  198\pm 18$& $ 75\pm 10$& $  14.68\pm0.05$ & fit, AOD \\
  \enddata
\end{deluxetable*}
\begin{deluxetable*}{llccccccl}
\tabletypesize{\footnotesize}
\tablecolumns{9} 
\tablewidth{0pt} 
\tablecaption{Sample C\,IV Detections}
\tablehead{\colhead{Sight Line}        &
   \colhead{$z_{\rm abs}$}             &
   \colhead{$b$ (1548)}  &
   \colhead{W$\rm_r$(1548)}&
   \colhead{$b$ (1550)}  &
   \colhead{W$\rm_r$(1550)}&
   \colhead{$b$(C\,IV)}&
   \colhead{log\,$N$(C\,IV)}&
   \colhead{notes}                    \\
  \colhead{}                           &
   \colhead{}                          &
   \colhead{(\kms)}                    &
   \colhead{(m\AA)}                    &
   \colhead{(\kms)}                    &
   \colhead{(m\AA)}                    &
   \colhead{(\kms)}                    &
   \colhead{($N$ in cm$^{-2}$)}               &
   \colhead{}                          }
\startdata
PKS\,0312$-$770& 0.10877& $  22\pm  1 $ & $   78\pm 11$& \nodata & $<  55       $& $ 22\pm  1$& $  13.52\pm0.25$ & AOD \\
HS\,0624$+$6907& 0.06348& $  25\pm  1 $ & $  113\pm  3$& $ \sim 27  $ & $  44\pm  3$& $ 25\pm  1$& $  13.49\pm0.09$ & AOD, AOD \\
HS\,0624$+$6907& 0.07573& $  10\pm  1 $ & $   45\pm  4$& $  12\pm  3 $ & $   21\pm  6$& $ 10\pm  3$& $  13.06\pm0.07$ & fit, fit \\
PG\,0953$+$414 & 0.04960& $  17\pm  5 $ & $   33\pm  6$& \nodata & $<  22       $& $ 17\pm  5$& $  12.98\pm0.16$ & AOD \\
PG\,0953$+$414 & 0.06808& $  11\pm  1 $ & $  135\pm 24$& $  10\pm  1 $ & $   93\pm 18$& $ 10\pm  1$& $  13.83\pm0.09$ & fit, fit \\
PG\,1211$+$143 & 0.05104& $  46\pm  1 $ & $  260\pm  4$& $  33\pm  2 $ & $  143\pm  2$& $ 41\pm  6$& $  14.04\pm0.04$ & AOD, AOD \\
PG\,1211$+$143 & 0.06447& $  46\pm  5 $ & $   73\pm 12$& $  39\pm  6 $ & $   49\pm 10$& $ 43\pm  6$& $  13.32\pm0.07$ & fit, fit \\
PG\,1211$+$143 & 0.06493& $  24\pm  5 $ & $   31\pm  5$& $ \sim 46  $ & $  38\pm  5$& $ 24\pm  5$& $  13.10\pm0.21$ & fit, AOD \\
Mrk\,205       & 0.00443& $  50\pm 10 $ & $  362\pm 14$& $  55\pm  2 $ & $  235\pm  9$& $ 51\pm 10$& $  14.05\pm0.18$ & AOD, AOD \\
Mrk\,205       & 0.02746& \nodata & $< 131       $& $ \sim 17  $ & $  53\pm  1$& $ \sim 17$& $  13.55\pm0.10$ & AOD \\
Q\,1230$+$0115 & 0.04886& $  21\pm  9 $ & $   39\pm  1$& \nodata & $<  21       $& $ 21\pm  9$& $  13.00\pm0.16$ & fit \\
  \enddata
\end{deluxetable*}
\begin{deluxetable*}{llccccccl}
\tabletypesize{\footnotesize}
\tablecolumns{9} 
\tablewidth{0pt} 
\tablecaption{Sample N\,V Detections}
\tablehead{\colhead{Sight Line}        &
   \colhead{$z_{\rm abs}$}             &
   \colhead{$b$ (1238)}  &
   \colhead{W$\rm_r$(1238)}&
   \colhead{$b$ (1242)}  &
   \colhead{W$\rm_r$(1242)}&
   \colhead{$b$(N\,V)}&
   \colhead{log\,$N$(N\,V)}&
   \colhead{notes}                    \\
  \colhead{}                           &
   \colhead{}                          &
   \colhead{(\kms)}                    &
   \colhead{(m\AA)}                    &
   \colhead{(\kms)}                    &
   \colhead{(m\AA)}                    &
   \colhead{(\kms)}                    &
   \colhead{($N$ in cm$^{-2}$)}               &
   \colhead{}                          }
\startdata
PKS\,0312$-$770& 0.20192& $   8\pm  1 $ & $   25\pm  4$& $   9\pm  2 $ & $   17\pm  3$& $  8\pm  2$& $  13.12\pm0.07$ & fit, fit \\
PKS\,0312$-$770& 0.20275& $  67\pm  4 $ & $  133\pm 10$& $  56\pm  6 $ & $   63\pm  7$& $ 63\pm  6$& $  13.74\pm0.04$ & fit, fit \\
PKS\,0405$-$123& 0.16714& $  51\pm  5 $ & $  143\pm 16$& \nodata & $< 155       $& $ 51\pm  5$& $  13.83\pm0.04$ & fit \\
PKS\,0405$-$123& 0.33400& \nodata & \nodata      & $ \sim 12  $ & $  28\pm  1$& $ \sim 12$& $  13.52\pm0.12$ & AOD \\
PKS\,0405$-$123& 0.36079& $  24\pm  8 $ & $   46\pm 11$& \nodata & \nodata      & $ 24\pm  8$& $  13.47\pm0.12$ & AOD \\
HS\,0624$+$6907& 0.29530& $  18\pm  5 $ & $   30\pm  2$& $   7\pm  3 $ & $   11\pm  1$& $ 14\pm  5$& $  13.17\pm0.13$ & AOD, AOD \\
PG\,0953$+$414 & 0.06808& $   8\pm  1 $ & $   45\pm  4$& $  15\pm  2 $ & $   30\pm  5$& $ 10\pm  3$& $  13.46\pm0.05$ & fit, fit \\
PG\,0953$+$414 & 0.23350& $   9\pm  3 $ & $   11\pm  5$& \nodata & $<   8       $& $  9\pm  3$& $  12.68\pm0.14$ & fit \\
3C\,249.1      & 0.11925& $ \sim 15  $ & $  11\pm  2$& \nodata & $<  10       $& $ \sim 15$& $  12.78\pm0.20$ & AOD \\
3C\,249.1      & 0.17371& $  24\pm  7 $ & $   29\pm  4$& \nodata & $<  17       $& $ 24\pm  7$& $  13.19\pm0.13$ & AOD \\
PG\,1116$+$215 & 0.05904& $ \sim 27  $ & $  20\pm  2$& $ \sim 19  $ & $  18\pm  2$& $ \sim 24$& $  13.16\pm0.12$ & AOD, AOD \\
  \enddata
\end{deluxetable*}
\begin{deluxetable*}{llcccl}
\tabletypesize{\footnotesize}
\tablecolumns{7} 
\tablewidth{0pt} 
\tablecaption{Sample C\,III Detections}
\tablehead{\colhead{Sight Line}        &
   \colhead{$z_{\rm abs}$}             &
   \colhead{W$\rm_r$(977)}&
   \colhead{$b$(C\,III)}&
   \colhead{log\,$N$(C\,III)} &
   \colhead{notes}                    \\
  \colhead{}                           &
   \colhead{}                          &
   \colhead{(m\AA)}                    &
   \colhead{(\kms)}                    &
   \colhead{($N$ in cm$^{-2}$)}               &
   \colhead{}                          }
\startdata
HE\,0226$-$4410& 0.20700& $  219\pm 66$& $ 23\pm  5$& $  13.98\pm0.24$ & fit \\
HE\,0226$-$4410& 0.34031& $   23\pm  3$& $ 32\pm  4$& $  12.63\pm0.10$ & AOD \\
PKS\,0312$-$770& 0.20192& $  270\pm 17$& $ 57\pm  4$& $  13.63\pm0.20$ & AOD \\
PKS\,0312$-$770& 0.20275& $  600\pm 35$& $ 94\pm  5$& $  13.97\pm0.35$ & AOD \\
PKS\,0405$-$123& 0.16714& $  369\pm 73$& $ 16\pm  2$& $  14.78\pm0.71$ & fit \\
PKS\,0405$-$123& 0.33400& $   36\pm  4$& $ 38\pm  2$& $  12.82\pm0.07$ & AOD \\
PKS\,0405$-$123& 0.36079& $  141\pm 10$& $ 17\pm  1$& $  13.61\pm0.05$ & fit \\
HS\,0624$+$6907& 0.32088& $   10\pm  3$& $ 12\pm  4$& $  12.10\pm0.13$ & fit \\
PG\,0953$+$414 & 0.06808& $  112\pm 20$& $ 28\pm  4$& $  13.35\pm0.06$ & fit \\
PG\,0953$+$414 & 0.11827& $   36\pm  3$& $ 26\pm  8$& $  12.83\pm0.09$ & AOD \\
PG\,0953$+$414 & 0.14232& $   30\pm 16$& $ 20\pm 12$& $  12.65\pm0.19$ & fit \\
  \enddata
\end{deluxetable*}
\begin{deluxetable*}{llccccccl}
\tabletypesize{\footnotesize}
\tablecolumns{9} 
\tablewidth{0pt} 
\tablecaption{Sample Si\,IV Detections}
\tablehead{\colhead{Sight Line}        &
   \colhead{$z_{\rm abs}$}             &
   \colhead{$b$ (1394)}  &
   \colhead{W$\rm_r$(1394)}&
   \colhead{$b$ (1403)}  &
   \colhead{W$\rm_r$(1403)}&
   \colhead{$b$(Si\,IV)}&
   \colhead{log\,$N$(Si\,IV)}&
   \colhead{notes}                    \\
  \colhead{}                           &
   \colhead{}                          &
   \colhead{(\kms)}                    &
   \colhead{(m\AA)}                    &
   \colhead{(\kms)}                    &
   \colhead{(m\AA)}                    &
   \colhead{(\kms)}                    &
   \colhead{($N$ in cm$^{-2}$)}               &
   \colhead{}                          }
\startdata
PKS\,0312$-$770& 0.20192& $  35\pm  8 $ & $  177\pm 54$& $  40\pm  2 $ & $   80\pm  7$& $ 36\pm  8$& $  13.39\pm0.11$ & fit, AOD \\
PKS\,0312$-$770& 0.20275& $  76\pm  3 $ & $  595\pm 99$& $  75\pm  1 $ & $  470\pm 12$& $ 75\pm  3$& $  14.17\pm0.19$ & fit, AOD \\
PKS\,0405$-$123& 0.16714& $  33\pm  2 $ & $  164\pm 12$& $  25\pm  3 $ & $  117\pm 21$& $ 30\pm  4$& $  13.38\pm0.09$ & fit, fit \\
HS\,0624$+$6907& 0.06348& $  24\pm  2 $ & $   47\pm  4$& \nodata & $<  54       $& $ 24\pm  2$& $  12.80\pm0.10$ & AOD \\
PG\,1116$+$215 & 0.13850& $  17\pm  1 $ & $   34\pm  2$& \nodata & $<  30       $& $ 17\pm  1$& $  12.69\pm0.08$ & AOD \\
PG\,1211$+$143 & 0.05104& $  15\pm  1 $ & $   73\pm  5$& $  17\pm  1 $ & $   45\pm  4$& $ 15\pm  1$& $  13.02\pm0.03$ & fit, fit \\
Mrk\,205       & 0.00443& $  30\pm  1 $ & $  115\pm  9$& $  24\pm  9 $ & $   74\pm  7$& $ 28\pm  9$& $  13.30\pm0.06$ & AOD, AOD \\
Q\,1230$+$0115 & 0.07805& $ \sim 10  $ & $  29\pm  4$& $ \sim 20  $ & $  24\pm  4$& $ \sim 13$& $  12.70\pm0.16$ & AOD, AOD \\
PG\,1259$+$593 & 0.23951& $  10\pm  8 $ & $   37\pm 30$& \nodata & \nodata      & $ 10\pm  8$& $  12.61\pm0.29$ & fit \\
PKS\,1302$-$102& 0.04225& $ \sim 18  $ & $  18\pm  2$& \nodata & $<  45       $& $ \sim 18$& $  12.39\pm0.23$ & AOD \\
PKS\,1302$-$102& 0.09485& $ \sim 18  $ & $  18\pm  7$& \nodata & $<  12       $& $ \sim 18$& $  12.36\pm0.27$ & AOD \\
  \enddata
\end{deluxetable*}
\begin{deluxetable*}{llcccl}
\tabletypesize{\footnotesize}
\tablecolumns{7} 
\tablewidth{0pt} 
\tablecaption{Sample Si\,III Detections}
\tablehead{\colhead{Sight Line}        &
   \colhead{$z_{\rm abs}$}             &
   \colhead{W$\rm_r$(1207)}&
   \colhead{$b$(Si\,III)}&
   \colhead{log\,$N$(Si\,III)} &
   \colhead{notes}                    \\
  \colhead{}                           &
   \colhead{}                          &
   \colhead{(m\AA)}                    &
   \colhead{(\kms)}                    &
   \colhead{($N$ in cm$^{-2}$)}               &
   \colhead{}                          }
\startdata
HE\,0226$-$4410& 0.16340& $   19\pm  1$& $ \sim  9$& $  12.07\pm0.09$ & AOD \\
HE\,0226$-$4410& 0.19859& $   20\pm  3$& $ 26\pm  3$& $  12.07\pm0.12$ & AOD \\
HE\,0226$-$4410& 0.20700& $   49\pm  3$& $ 24\pm  4$& $  12.47\pm0.05$ & AOD \\
HE\,0226$-$4410& 0.30565& $   30\pm  7$& $ 12\pm  7$& $  12.29\pm0.11$ & AOD \\
PKS\,0312$-$770& 0.05946& $   21\pm  5$& $ 21\pm  6$& $  12.07\pm0.15$ & AOD \\
PKS\,0312$-$770& 0.17976& $   13\pm  8$& $  8\pm  5$& $  11.77\pm0.21$ & fit \\
PKS\,0312$-$770& 0.20192& $  123\pm  5$& $ 30\pm  1$& $  12.97\pm0.05$ & AOD \\
PKS\,0312$-$770& 0.20275& $  647\pm151$& $ 36\pm  2$& $  14.92\pm0.27$ & fit \\
PKS\,0405$-$123& 0.16714& $  266\pm 12$& $ 30\pm  1$& $  13.34\pm0.02$ & fit \\
PKS\,0405$-$123& 0.21070& $  104\pm 12$& $ 40\pm  9$& $  12.86\pm0.05$ & AOD \\
PKS\,0405$-$123& 0.35210& $  128\pm 17$& $ 36\pm  4$& $  12.74\pm0.05$ & fit \\
  \enddata
\end{deluxetable*}
\begin{deluxetable*}{llcccl}
\tabletypesize{\footnotesize}
\tablecolumns{7} 
\tablewidth{0pt} 
\tablecaption{Sample Fe\,III Detections}
\tablehead{\colhead{Sight Line}        &
   \colhead{$z_{\rm abs}$}             &
   \colhead{W$\rm_r$(1123)}&
   \colhead{$b$(Fe\,III)}&
   \colhead{log\,$N$(Fe\,III)} &
   \colhead{notes}                    \\
  \colhead{}                           &
   \colhead{}                          &
   \colhead{(m\AA)}                    &
   \colhead{(\kms)}                    &
   \colhead{($N$ in cm$^{-2}$)}               &
   \colhead{}                          }
\startdata
HE\,0226$-$4410& 0.16236& $   23\pm  4$& $ \sim 20$& $  13.64\pm0.14$ & AOD \\
PKS\,0312$-$770& 0.20192& $   20\pm  7$& $  6\pm  2$& $  13.52\pm0.11$ & fit \\
PKS\,0312$-$770& 0.20275& $  127\pm  5$& $ 64\pm  1$& $  14.46\pm0.04$ & AOD \\
PKS\,0405$-$123& 0.16714& $   22\pm  5$& $ 10\pm  2$& $  13.56\pm0.09$ & fit \\
PKS\,0405$-$123& 0.35210& $   28\pm 15$& $ 16\pm  6$& $  13.58\pm0.17$ & fit \\
3C\,249.1      & 0.10834& $   15\pm  1$& $ \sim 17$& $  13.47\pm0.22$ & AOD \\
3C\,249.1      & 0.14183& $   36\pm  4$& $ 19\pm  5$& $  13.85\pm0.08$ & AOD \\
3C\,249.1      & 0.24396& $   16\pm  3$& $ \sim  4$& $  13.63\pm0.12$ & AOD \\
PG\,1116$+$215 & 0.13850& $   14\pm  1$& $ \sim  9$& $  13.44\pm0.10$ & AOD \\
PKS\,1302$-$102& 0.07695& $   24\pm  3$& $ \sim 18$& $  13.69\pm0.13$ & AOD \\
Mrk\,876       & 0.11487& $   25\pm  7$& $ 16\pm  5$& $  13.60\pm0.11$ & fit \\
  \enddata
\end{deluxetable*}

\subsection{Absorber Statistics}

With such a large database of detections, we are able to analyze the absorber statistics with considerably more confidence.  We constructed histograms showing the number of absorbers, ${\cal N}_i$, in logarithmic bins of width 0.2 dex in column density.  In general, the median column density uncertainty, $\delta\log\,N$, for the metal ion detections is larger than half a bin size.  We analyzed the distribution ${\cal N}(\log\,N)$ using a ``distributed-power'' histogram technique to spread the power of each detection across several log\,$N_i$ bins.  The total distribution is then,
\begin{eqnarray} 
{\cal N}({\rm log}\,N)&=&\sum_i {\cal N}_i ({\rm log}\,N_i) \\    
                      &=& \sum_{i} \frac{0.2}{\delta {\rm log} \, \nonumber 
        N_i\sqrt{2\pi}}~{\rm exp}\biggl(\frac{-[{\rm log}\,N-{\rm log}\,N_i]^2} {2 [\delta {\rm log}\,N_i]^2}\biggr). 
\end{eqnarray} 
The difference between this distributed-power method and simple histogram binning is subtle;  sharp peaks and valleys in the simple histogram are smoothed to some degree, but the general shape remains unchanged.  We believe this method takes measurement uncertainty into better account.

At low column densities we must correct the survey for incompleteness (Papers~1 and~2).  To do so, we divide the number of absorbers of a given equivalent width $W$ by an ``effective redshift pathlength'', $\Delta z_{\rm eff}(W)$, which accounts for absorption-line sensitivity.  For each species, we calculate the number of absorbers per unit redshift, \dndz, as a function of column density, $N$, to account for incompleteness in the weaker absorbers,
\begin{equation} 
  \frac {d{\cal N}(\log\,N)}{dz}=\sum_{i} {\frac{{\cal N}_i(\log\,N_i)}{\Delta z_i}}\; . 
\end{equation} 

We measured the sensitivity by the signal-to-noise ratio, $S/N$, determined for each sight line as the ratio of mean flux to standard deviation in the flux in the line-free continuum of each 10~\AA\ segment of \FUSE\ or STIS data.  In regions of overlapping data (e.g., the overlap of \FUSE\ LiF1 and LiF2 channels or the overlap between \FUSE\ and STIS/E140M coverage), we used the higher $S/N$ value.  The vector, $(S/N)_{\lambda}$, for a given wavelength $\lambda$ was rebinned to a resolution comparable to that of the data.  

We then modified $(S/N)_{\lambda}$ to take into account regions in which IGM absorption lines would not be detected.  We used two methods to assess the sensitivity of the effective pathlength, $\Delta z$, to different $(S/N)_{\lambda}$ algorithms.  In the first method, we set $(S/N)_{\lambda}=0$ in $\sim50$ regions of strong interstellar lines and instrumental artifacts that appear in most or all of the data sets.  In the second method, we let the smoothed, normalized flux data mask the $(S/N)_{\lambda}$ profile.  Regions of strong absorption ($\tau_{\lambda}\gg 1$) were completely masked, $(S/N)_{\lambda}=0$, while areas of unsaturated absorption were masked proportional to the optical depth, $(S/N)_{\lambda} \propto (\tau_{\lambda}+1)^{-1}$.  This effectively removed fixed ISM lines and instrumental artifacts, as well as strong IGM lines and intrinsic AGN features.  It also accounted for degraded sensitivity in regions with moderately strong absorption lines of any type (AGN, IGM, or ISM).

Next, we converted each $(S/N)_{\lambda}$ vector to a $4\sigma$ minimum equivalent-width vector using equation~\ref{wmineq}.  For single lines, we determined the total redshift pathlength, $\Delta z(W)$, at a given equivalent width sensitivity by summing the available pathlength over which $W_\lambda>W_{\rm min}$ in all 28 sight lines.  For multiplet lines, including Ly$\alpha$ and Ly$\beta$, we scaled a second vector, $W_{\rm min}(z)$, by the line-strength ratio, $f\lambda$, and shifted to the rest frame of the weaker line.  In this case, $\Delta z(W)$ is the sum of all available pathlength for which either $W_\lambda>W_{\rm min}(z_1)$ or $W_\lambda>[W_{\rm min}(z_2)] (f_1\lambda_1/f_2\lambda_2)$.  In all cases, we excluded pathlength within $\Delta z<0.005$ (1500~\kms) of the intrinsic AGN redshift and within $\Delta z<0.0017$ (500~\kms) of the Local Standard of Rest (LSR).  Thus, systems intrinsic to the AGN (outflows) or to our Galaxy (high velocity clouds) were not included.

\citet{Tripp08} study both intrinsic and intervening systems and adopt a more restrictive velocity limit ($\Delta z>0.017$, $v>5000$~\kms) than we use here.  As they note, there are numerous examples of high-$z$ quasars where intrinsic absorption components are seen at $v>5000$~\kms.  They find the line density for intrinsic absorbers is roughly a factor of two higher than for intervening systems.  Nevertheless, Tripp \etal\ note that the excess of systems is largely confined to the velocity range $-1000\la v_{\rm displ}\la 2500$~\kms\ and that intrinsic systems are relatively rare at $v_{\rm displ}>2500$~\kms.  

Both methods of $(S/N)_{\lambda}$ masking reduced the total $\Delta z$ in the data by 3-4\% relative to the unmasked $S/N$ profile.  However, the difference in total $\Delta z$ between these two methods was less than 1\%.  The second, self-masking method allocates relatively less pathlength for weak lines and relatively more for stronger lines.  We deemed this method to be more realistic and used it throughout our analysis.

In our previous work, the absorber redshifts were generally small and the redshift is linearly proportional to comoving distance, $d\ell=(c/H_0)\,dz$.  In the new survey, we included absorbers out to redshifts $z=0.4$, where a given a linear distance $d\ell$ covers a larger redshift interval $dz$ than at $z=0$.  We therefore use the relations,
\begin{eqnarray}    
  \frac{d\ell}{dz}&=&\frac{1}{(1+z)}~\frac{c}{H(z)} \nonumber \\
  &=&\frac{c}{H_0}~\frac{1}{(1+z)}~[\Omega_m(1+z)^3+\Omega_\Lambda]^{-1/2} \; ,  
\end{eqnarray}
where we assumed a flat ($\Omega_m$, $\Lambda$) cosmology with $H(z)=H_0~[\Omega_m(1+z)^3+\Omega_\Lambda]^{1/2}$.  By including the $(1+z)^3$ density scaling and normalizing to $H_0$, it is conventional \citep{BahcallPeebles69} to define the dimensionless ``absorption pathlength function'', $dX=[H_0/H(z)](1+z)^2\,dz$, so that
\begin{equation}
    dX \equiv (1+z)^2~[\Omega_m(1+z)^3+\Omega_\Lambda]^{-1/2}~dz \; .
\end{equation}

We calculated $\Delta X(N_{\rm min})$ from the individual spectra in the same manner as $\Delta z(N_{\rm min})$ above using numerical integration and adopting $\Omega_m=(0.261\pm0.016)h^{-2}_{70}$ and $\Omega_\Lambda=0.716\pm0.055$ \citep{Spergel07}.  For low-redshift absorbers, such as all \CIV\ detections ($z_{\rm abs}\le0.11$), the difference between $\Delta z$ and $\Delta X$ is negligible.  We list both \dndz\ and \dndx\ in the discussion, to provide both observational and theoretical perspectives. 

Uncertainties in \dndz\ arise as uncertainties in both ${\cal N}$ and $\Delta z$.  In the numerator, one-sided Poisson statistics \citep{Gehrels86} for $d{\cal N}_i$ are assumed.  While the uncertainty in $\Delta z(W)$ is negligible as discussed above,, we must still take into account the changing $\Delta z$ over the width of a column density bin.  For weak lines where the $\Delta z_{\rm eff}(W)$ is falling sharply, there is a significant difference in $\Delta z$ between one side of an equivalent width or column density bin and the other.  We approximate this uncertainty as $d\Delta z_i=\frac{1}{8}[(\Delta z)_{i+1}-(\Delta z)_{i-1}]$ or roughly the change in $\Delta z$ over a quarter of a bin, the typical $1\sigma$ dispersion of column densities from bin center.  The \dndz\ uncertainties from numerator and denominator are added in quadrature.  Uncertainty from Poisson statistics dominates in almost all cases, but $d\Delta z_i$ becomes significant for numerous absorbers (\HI, \OVI, \SiIII) in the weakest column density bins.  Uncertainties in \dndx\ are completely analogous.

\subsection{Baryon Fraction}

Through measurements of D/H and acoustic peaks in the CMB, we now have an accurate estimate of the cosmological density of baryons, $\Omega_b\,h_{70}^2=0.0455\pm0.0015$ \citep{Spergel07}.  However, most of these baryons do not exist in collapsed form, and probably are distributed throughout various phases of the IGM \citep{CenOstriker99,Shull03}.  This is certainly the case at high redshift, where the \Lya\ forest is the repository for large quantities of gas that has not yet collapsed into galaxies.  But even at low redshifts, the IGM likely contains between 50\% and 80\% of the baryonic matter.  Because low-density gas at a variety of temperatures can be difficult to detect, a complete cosmological baryon census requires a combination of ultraviolet and X-ray spectrographs.

Here, we use UV spectroscopic column densities and absorber frequencies of \HI\ and various metal ions to derive the baryon density contained in IGM thermal phases at $T\leq10^6$~K.  To compare to similar estimates in the literature, it is convenient to define two such measures.  The first, directly measured quantity provides the contribution to the critical mass density, of the IGM absorber mass in a given ion,
\begin{eqnarray}    
  \Omega_{\rm ion} &=& \left( \frac {H_0 \, m_{\rm ion}} {c\,\rho_{\rm cr,0}}\right) \times \\ 
 & & \int_{N_{\rm min}}^{N_{\rm max}} \left[ \frac{d{\cal N}(\log\,N_{\rm ion})}{dX} \right]  \, 
    \langle N_{\rm ion} \rangle \, d \log\,N_{\rm ion} \; .  \nonumber  
\end{eqnarray}
The integration is over the column density distribution and the corrected redshift pathlength parameter $dX$.  Here, $m_{\rm ion}$ is the ion mass, and the critical density is $\rho_{\rm cr,0}=(3H_0^2/8\pi G)=(1.879\times 10^{-29}~{\rm g~cm}^{-3}) h^2$.  The integration is over the column density distribution and the corrected redshift pathlength parameter $dX$.

Uncertainty in $\Omega_{\rm ion}$ is taken as the asymmetric Poisson uncertainty on $d{\cal N}(>N)$ evaluated at a $d{\cal N}$-weighted mean $dX$, $\langle z\rangle$, and $\langle N\rangle$.  The value of $\Omega_{\rm ion} \propto [Z\,(M/H)_{\odot}\,f_{\rm ion} h_{70}]^{-1}$, so we do not include systematic uncertainties in these quantities.  Non-equilibrium ionization may produce higher ion fractions for \SiIV\ and the Li-like ions while photoionization may lower the assumed near-unity CIE ion fractions of \CIII, \SiIII, and \FeIII.  Similarly non-solar abundance ratios or deviations from the canonical 10\% solar metallicity often assumed for IGM clouds will introduce systematic differences.

The second parameter, $\Omega^{\rm (ion)}_{\rm IGM}$, estimates the density of {\it all} baryons traced by the particular ion after correcting for abundances and ionization stages.  In Paper~1, we calculated the fraction of the universe composed of baryons at $T=10^{5-6}$~K, using \OVI\ as a tracer for baryonic WHIM gas in general.  This parameter includes corrections for the metallicity, $Z$, of the element relative to solar abundances, and for the ionization fraction, $f_{\rm OVI}$, of the particular ion.  Our adopted formula was
\begin{eqnarray}
  \Omega_{\rm IGM}^{\rm (OVI)} &=& \left( \frac {H_0} {c\,\rho_{\rm cr,0}} \right)\frac {\mu m_H} {Z\,(O/H)_{\sun}\,f_{\rm OVI}}\times  \\
& & \int_{N_{\rm min}}^{N_{\rm max}} \left[ \frac {d{\cal N}(\log\,N_{\rm OVI})}{dz}\right]\,\langle N_{\rm OVI} \rangle \, d \log\,N_{\rm OVI} \; ,\nonumber  
\end{eqnarray} 
where $\mu=1.32$ for low-metallicity gas and primordial helium, $Y_p\approx0.2477\pm0.0029$ \citep{Peimbert07}.  

We can easily generalize this formula for any ion dividing $\Omega_{\rm ion}$ by an assumed metallicity, $Z=0.1\,Z_\sun$, abundance of element $M$ relative to hydrogen, $(M/H)_\sun$, and ionization fraction (assumed to be the peak CIE ion abundance $f_{\rm ion}$ listed in Table~\ref{tab:atomicdata}).  Because the redshift range in our current work is larger than that covered in Paper~1, we adopt pathlength coordinate $X$ instead of $z$, as discussed above, and we perform the sum over bins of equal $\Delta X$.
\[   \Omega^{\rm (ion)}_{\rm IGM} = \left( \frac {H_0} {c\,\rho_{\rm cr,0}} \right)\frac {\mu m_H} {Z\,(M/H)_{\sun}\,f_{\rm ion}}\times \]
\[ \int_{0}^{X_{\rm max}}\int_{N_{\rm min}}^{N_{\rm max}} \left[\frac {d{\cal N}_{\rm ion}(\log\,N_{\rm ion})} {dX} \right]\,\langle N_{\rm ion} \rangle \, d\log\,N_{\rm ion} \, dX  \]
\begin{equation}
  \Omega^{\rm (ion)}_{\rm IGM} = \frac {1.83\times10^{-23}~h^{-1}_{70}~{\rm cm}^2}{Z\,(M/H)_\sun\,f_{\rm ion}}\times \end{equation}
\[ ~\sum_{j=0}^{X_{\rm max}}\sum_{i=N_{\rm min}}^{N_{\rm max}} \left[ \frac {d{\cal N}(\log\,N_{\rm ion})}{dX} \right]_{i,j} \frac{\langle N_{\rm ion} \rangle_{i,j} \,\Delta \log\,N_{\rm ion} \Delta X_j} {X_{\rm max}} \; . \]
As in our previous work, we adopt a fiducial IGM metallicity of $Z=0.1\,Z_{\odot}$ and a Hubble constant $H_0=(70\,{\rm km~s}^{-1}~{\rm Mpc}^{-1})h_{70}$.  We sum \dndx\ for each ion over bins in column density $N_{\rm ion}$ and normalized pathlength $dX$.  The final term, $\Delta X_j/X_{\rm max}$, averages the \dndx\ contributions from each $dX$ bin.

\subsubsection{Baryons in the \Lya\ Forest}

While the main focus of this paper is metal ions, our database of \HI\ absorbers is the largest to date in the low-$z$ IGM and can provide good constraints on $\Omega_{\rm IGM}^{{\rm (HI)}}$, the fraction of IGM baryons in the \Lya\ forest.  For this we use the methods of \citet{Penton2} and \citet{Schaye01}, which are slightly different from the methods employed for $\Omega_{\rm IGM}^{\rm (ion)}$ above.  In the first method, the IGM absorbers are assumed to be isothermal spheres, and the total cloud mass is inferred from the \HI\ column density and an assumed impact parameter.  The neutral fraction is derived from photoionization conditions.  Following \citet{Penton2}, we define
\[  \Omega^{(\rm HI)}_{\rm IGM} = \frac {(1.59\times10^9 \, M_\sun) \,H_0\,\bigl [\ell_{100}^5\,J_{-23}\,\bigl( \frac{4.8}{\alpha_s+3}\bigr)\bigr]^{1/2}} {c\,\pi\,\ell^2\,\rho_{\rm cr}} \times \]
\[ \int_{N_{\rm min}}^{N_{\rm max}}\frac{d{\cal N}(N_{\rm HI})}{dz}\,N_{14}^{1/2}\,dN_{\rm HI} \]
\[ \Omega^{(\rm HI)}_{\rm IGM} = 8.73\times10^{-5}\,h_{70}^{-1}\, \left[ J_{-23}\,\ell_{100}\, \frac{4.8}{\alpha_s+3} \right]^{1/2} \times \]
\begin{equation} \sum_{X=0}^{X_{\rm max}}\sum_{\log\,N_{\rm min}}^{\log\,N_{\rm max}} \frac{d{\cal N}(\log\,N_{\rm HI})}{dz}\,N_{14}^{1/2}\,\frac {\Delta X} {X_{\rm max}}  \; , 
  \label{eq:omega_penton}
\end{equation}
where $\ell_{100}$ is the typical scale of \Lya\ forest clouds scaled to 100~kpc, $J_{-23}$ is the ionizing radiation field intensity at low redshift \citep{Shull99} in units of $10^{-23}\rm~ergs~cm^{-2}~s^{-1}~Hz^{-1}~sr^{-1}$, $\alpha_s$ is the spectral index of the radiation field above 1 ryd scaled to $\alpha_s=1.8$, and $N_{14}$ is the \HI\ column density in units of $10^{14}\rm~cm^{-2}$.  Secondly, we use the formalism of \citet{Schaye01}, which assumes that IGM clouds are gravitationally bound and that their observed column densities are typical over a Jeans length:
\[  \Omega^{(\rm HI)}_{\rm IGM} = (1.46\times10^{-4})\,h_{70}^{-1}\, \Gamma_{-12}^{1/3}\,T_4^{0.59}\times \]
\[ \int_{N_{\rm min}}^{N_{\rm max}}\frac {d{\cal N}(N_{\rm HI})}{dz} \, N_{14}^{1/3} \, dN_{\rm HI} \]
\[  \Omega^{(\rm HI)}_{\rm IGM} = (1.46\times10^{-4})\,h_{70}^{-1}\,\Gamma_{-12}^{1/3}\,T_4^{0.59}\times \]
\begin{equation}~\sum_{X=0}^{X_{\rm max}}\sum_{\log\,N_{\rm min}}^{\log\,N_{\rm max}}\frac{d{\cal N}(\log\,N_{\rm HI})}{dz}\,N_{14}^{1/3}\,\frac{\Delta X}{X_{\rm max}} \; .
  \label{eq:omega_schaye}
\end{equation}
As with $\Omega_{\rm ion}$, we perform a two-dimensional sum over both column density and redshift bins assuming $T=2\times10^4$~K in the photoionized gas and a hydrogen photionization rate $\Gamma_{\rm HI}=0.03\times10^{-12}~\rm s^{-1}$ \citep{Shull99,Weymann01}.  Uncertainties in $\Omega_{\rm IGM}^{(\rm HI)}(\log\,N)$ are the quadratic sum of the one-sided Poisson error for total redshift pathlength $d{\cal N}(d \log N,dX)$.  \citet{Penton4} showed that uncertainty due to cosmic variance is small relative to Poisson error for $\Delta z\ga1$.  Since we are summing over $\Delta z>5$ for all but the lowest column density bins, we discount cosmic variance as a significant source of error.
 
\section{Results and Discussion}

%
%
\begin{deluxetable}{lrccccccccc}
\tabletypesize{\footnotesize}
\tablecolumns{11} 
\tablewidth{0pt} 
\tablecaption{IGM Detections and Results Summary}
\tablehead{\colhead{ion}                         &
           \colhead{${\cal N}$}                  &
           \colhead{$z_{\rm abs}$}               &
           \colhead{$\Delta z_{\rm max}$}        &
           \colhead{$\alpha_{14}$}               &
           \colhead{$C_{14}$}                    &
           \colhead{$Z/Z\rm_\sun$\,\tnma}         &
           \colhead{$d{\cal N}/dz$\,\tnmb}         &
           \colhead{$d{\cal N}/dX$\,\tnmb}         &
           \colhead{$\beta$}                     &
	     \colhead{Source\tnmc} }
\startdata
 O\,VI   & 83& $<0.40 $& 5.22 & $0.71\pm0.03 $&$2.0\pm0.1 $&$0.15\pm0.01 $&$15^{+3}_{-2} $&$18^{+3}_{-2} $&$1.98\pm0.11 $& 1 \\
 N\,V    & 24& $<0.396$& 5.30 & $0.98\pm0.07 $&$5.3\pm0.7 $&$0.39\pm0.05 $&$2\pm1        $&$3\pm1        $&$1.87\pm0.17 $& 1 \\
 C\,IV   & 24& $<0.116$& 2.42 & $0.87\pm0.06 $&$5.5\pm0.7 $&$0.055\pm0.007 $&$10^{+3}_{-2} $&$10^{+4}_{-2} $&$1.79\pm0.17 $& 1 \\
 C\,III  & 39& $<0.40 $& 4.84 & $0.56\pm0.05 $&$9.5\pm0.9 $&$0.015\pm0.001 $&$10^{+3}_{-2} $&$12^{+4}_{-2} $&$1.79\pm0.10 $& 1 \\
 Si\,IV  & 20& $<0.24 $& 4.21 & $0.74\pm0.06 $&$12\pm2 $&$0.19\pm0.03 $&$4^{+2}_{-1}  $&$5^{+2}_{-1}  $&$1.9\pm0.2 $& 1 \\
 Si\,III & 53& $<0.40 $& 5.14 & $0.81\pm0.03 $&$41\pm3 $&$0.023\pm0.002 $&$6^{+2}_{-1}  $&$7^{+2}_{-1}  $&$1.80\pm0.09 $& 1 \\
 Fe\,III & 14& $<0.40 $& 5.11 & $0.71\pm0.05 $&$1.6\pm0.2 $&$0.7\pm0.1  $&$\sim1        $&$\sim1        $&$2.2\pm0.4$& 1 \\
 H\,I    &650& $<0.40 $& 5.27 & \nodata&\nodata&\nodata&$129\pm6      $&$151^{+7}_{-6}$&$1.73\pm0.04$& 1 \\
\cutinhead{Previous Work/Literature Values}
 O\,VI   & 40 & $<0.15$& 2.21 &0.9$\pm$0.1  &2.5$\pm$0.2  &0.09     & 17$\pm$3      &\nd            & 2.2$\pm$0.1   & 2\\
 O\,VI   & 44 &0.12$-$0.57& $\sim1.9$&\nd          & \nd         &\nd      & 23$\pm$4      &\nd            & \nd           & 3\\
 O\,VI   & 53 &$<0.5$  &3.18  &$\sim0.9$    &$\sim2$      &\nd     & $18.3^{+3.0}_{-2.6}$&\nd       & \nd           & 4\\
 C\,III  & 30 & $<0.21$&2.41&0.73$\pm$0.08&11.5$\pm$1.0\tnmd&0.01\tnmd&12$^{+3}_{-2}$&\nd           & 1.68$\pm$0.04 & 5\\
 H\,I    &187 &$<0.069$& 0.770&\nd          & \nd         &\nd      & 165$\pm$15    &\nd            & 1.65$\pm$0.07 & 6\\
\cutinhead{This Work - Equivalent Sample}
 O\,VI   & 35&$<0.15$& 2.96 & 0.81$\pm$0.06 & 1.96$\pm$0.19 & 0.11$\pm$0.01  & 19$^{+6}_{-4}$ & 20$^{+7}_{-4}$ & 2.26$\pm$0.20 & 1\\
 C\,III  & 28&$<0.21$& 3.485& 0.65$\pm$0.05 & 7.69$\pm$0.91 & 0.016$\pm$0.002& 14$^{+5}_{-3}$ & 14$^{+7}_{-4}$ & 2.06$\pm$0.15 & 1\\
\enddata
\tablenotetext{a}{Scaled by $f_{\rm ion}$, the maximum CIE abundance (Sutherland \& Dopita 1993).}
\tablenotetext{b}{Number of absorbers per unit pathlength $(dX)$, integrated down to equivalent width $W>30$ m\AA}
\tablenotetext{c}{1-this work; 2-Danforth \& Shull 2005; 3-Tripp et al. 2005; 4-Tripp et al. 2007; 5-Danforth et al. 2006; 6-Penton et al. 2004}
\tablenotetext{d}{Published value of parameter $C_{14}$ (Paper~2) and hence $Z_C=12\%$ were incorrect.  Corrected values are listed here.}
\label{tab:results1}
\end{deluxetable}

In Table~\ref{tab:results1}, we present a compilation of our results.  The first three columns list ion name, total number of detections (at {\it any} equivalent width), and redshift coverage of the survey.  Column~4 lists the maximum total redshift pathlength sampled to {\it any} sensitivity;  because of (S/N) dependence, a smaller total pathlength is sensitive to weaker absorbers.

For each ion, we produce a plot of the ``multiphase ratio'', log\,$(N_{\rm HI}/N_{\rm ion})$ vs.\ log\,\NHI.  This ratio was initially defined for \OVI\ (Paper~1) to characterize the potential connections between different tracers of baryonic mass (\HI\ and \OVI) observed to be kinematically associated in the low-$z$ IGM.  We assume that absorbing clouds in the IGM have both a photoionized component (traced by sharp \Lya\ lines and low-ionization metal ions) and shock-heated gas (WHIM, traced by \OVI\ or other high-ions).  Collisionally ionized \CIII, \SiIII, etc. are possible, but we consider this less common due to the fast cooling rates at $T<10^5$~K, where these ions reach their peak CIE abundance (see Paper~2 for discussion).  In columns~5 and~6, we list the slope, $\alpha_{14}$, and the intercept, $C_{14}$, at column density \NHI$=10^{14}$~\cd\ of an error-weighted linear fit to the multiphase correlation.  These two parameters are used, together with various assumptions about ionization corrections, to infer a statistical metallicity of the IGM (see the derivations and discussion in Papers~1 and~2).  We quote each metallicity as a fraction of the solar abundance \citep{Asplund05} and scale them by $f_{\rm ion}$, the peak ion abundance in collisional ionization equilibrium \citep{SutherlandDopita93} or in photoionization equilibrium (see Table~\ref{tab:atomicdata}).  Our inferred metallicities are an average for the entire sample; we make no claims as to the metallicity of individual absorbers.  The observed doubly-ionized species (\CIII, \SiIII, \FeIII) have peak CIE abundances close to unity, so these metalicity values are probably lower limits.  The ion \SiIV\ and the Li-like ions (\CIV, \NV, \OVI) have peak CIE ionization fractions $f_{\rm peak}\sim0.2-0.3$.

In column~8, we provide the total absorber frequency per unit redshift, \dndz, integrated down to fiducial equivalent width 30 m\AA\ (the survey goes deeper than this limit in many cases).  Column~9 gives values of the proper-length-corrected \dndx\ for $W\ge30$~m\AA.  The cumulative bivariate distribution of absorbers per redshift interval as a function of column density was fitted with a power law, $[d^2{\cal N}(>N)/dz\,d\log\,N]\propto N^{-\beta}$, and the index $\beta$ is listed in column~10.  The power-law slope becomes systematically steeper by 0.05--0.1 dex when using traditional histogram techniques, as some of the power is scattered to the high and low ends of the distribution using our ``distributed power'' technique.  We determined $\beta$ by an error-weighted fit to the cumulative \dndz.  However, the index fitted to cumulative \dndx\ is comparable, within the uncertainties in all cases.

Where possible, Table~\ref{tab:results1} lists the equivalent values from Papers~1 and~2, as well as several other large surveys of the low-$z$ IGM.  For direct comparison of the current survey to our previous work, we calculate \OVI\ statistics for $z<0.15$ and \CIII\ statistics for $z<0.21$, consistent with the limits in Papers~1 and~2, respectively.  We have not adopted the $W_{\rm Ly\alpha}>80$ m\AA\ criterion used in those papers, nor have we accounted for the somewhat less stringent detection requirements adopted in our earlier work (detections in Papers~1 and~2 were $3\sigma$, while the current work requires a minimum of $4\sigma$).

We continue our results in Table~\ref{tab:results2}, listing $\Omega_{\rm ion}$ and $\Omega^{\rm (ion)}_{\rm IGM}$ for absorbers with equivalent width limits $W_\lambda>10$~m\AA\ and $W_\lambda>30$~m\AA.  The $\Omega^{\rm ion}_{\rm IGM}$ values are relative to the fiducial baryon fraction $\Omega_{\rm baryon}=0.0455 h_{70}^{-2}$ \citep{Spergel07} and are scaled by $(f_{\rm peak}/f_{\rm ion})\,(Z/0.1)^{-1}\,h_{70}^{-1}$ as described above.  As in Table~\ref{tab:results1} above, we list related values from previous surveys and compare to equivalent values from previous surveys. 

\subsection{Discussion of Individual Species}

\begin{figure*}
  \epsscale{1.1}\plotone{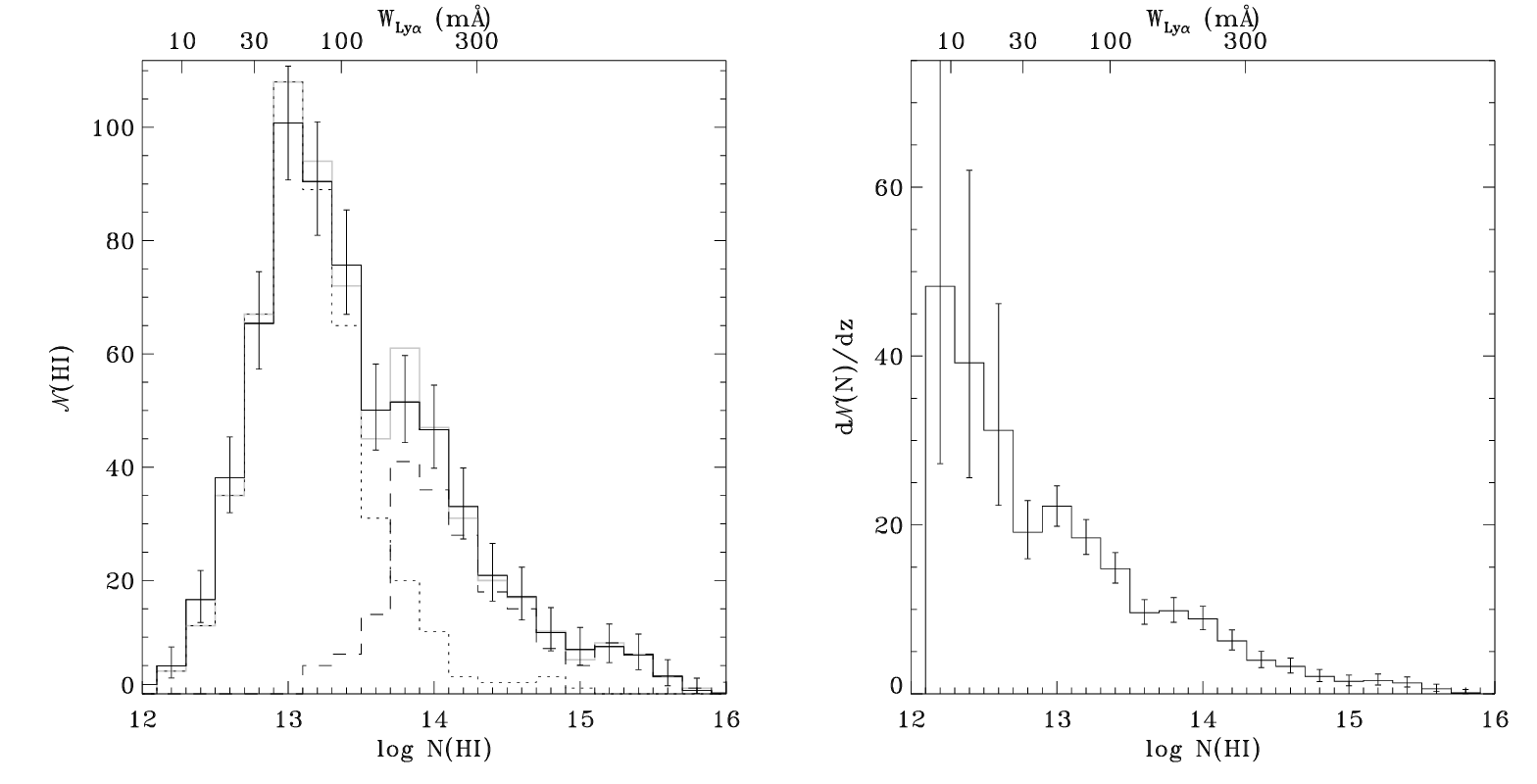} 
  \caption{H\,I detection statistics and \dndz.  Left panel shows the
  number of H\,I absorbers per logarithmic bin, $\Delta\log\,N_{\rm
  HI}=0.2$, uncorrected for completeness.  Note the turnover at
  log\,$N_{\rm HI}<13$.  Dashed curve shows absorbers for which
  higher-order Lyman lines were used to constrain the (\NHI, \bHI)
  solution.  Dotted curve shows \Lya-only solutions.  Right panel
  shows completeness-corrected differential \dndz\ vs.\ \NHI.  Error
  bars on the left panel are based on one-sided Poisson counting
  errors \citep{Gehrels86}, while the error in \dndz\ includes a
  component of $\Delta z$ uncertainty as well.  See text for
  discussion.}  \label{fig:h1}
\end{figure*}

\subsubsection{\HI}

%
%
\begin{deluxetable*}{lccccc}
\tabletypesize{\footnotesize}
\tablecolumns{6} 
\tablewidth{0pt} 
\tablecaption{$\Omega_{\rm ion}$ and $\Omega^{(\rm ion)}_{\rm IGM}$ Summary}
\tablehead{\colhead{ion}                                                   &
           \colhead{$\Omega_{\rm ion}~[10^{-8}]$}                           &
           \colhead{$\Omega_{\rm ion}~[10^{-8}]$}                          &
           \colhead{$\Omega^{(\rm ion)}_{\rm IGM}/\Omega\rm_b$\tnma} &
           \colhead{$\Omega^{(\rm ion)}_{\rm IGM}/\Omega\rm_b$\tnma} &
	     \colhead{Source\tnmb}\\
           \colhead{}                                                   &
           \colhead{($>10$~m\AA)}                           &
           \colhead{($>30$~m\AA)}                           &
           \colhead{($>10$~m\AA)}                           &
           \colhead{($>30$~m\AA)}                           &
	     \colhead{} }
\startdata
 O\,VI            & $  49\pm4   $&$  41\pm5   $&$0.086\pm0.008 $&$0.073\pm0.008 $ & 1 \\
 N\,V             & $  3.6\pm0.8$&$  3.0\pm0.9$&$0.050\pm0.011 $&$0.042\pm0.012 $ & 1 \\
 C\,IV            & $  8.2\pm1.4$&$  7.7\pm1.5$&$0.027\pm0.005 $&$0.026\pm0.005 $ & 1 \\
 C\,III           & $  8.9\pm1.3$&$  7.8\pm1.4$&$0.010\pm0.002 $&$0.009\pm0.002 $ & 1 \\
 Si\,IV           & $  8.4\pm1.1$&$  8.3\pm1.1$&$0.075\pm0.010 $&$0.075\pm0.010 $ & 1 \\
 Si\,III          & $  11\pm2   $&$ 10\pm2    $&$0.037\pm0.006 $&$0.037\pm0.007 $ & 1 \\
 Fe\,III          & $  20\pm5   $&$ 11\pm7    $&$0.040\pm0.011 $&$0.022\pm0.015 $ & 1 \\
\cutinhead{Previous Work/Literature Values}
 O\,VI ($z<0.15$) &\nd           & $26\pm4$    &\nd             & 0.048$\pm$0.007 & 2\\
 O\,VI ($z>0.12$) &\nd           & \nd         &\nd             & 0.059           & 3\\
\cutinhead{This Work - Equivalent Sample}
 O\,VI ($z<0.15$) &$  49\pm4    $&$  33\pm5$   &$0.087\pm0.008$ &$0.058\pm0.008$& 1 \\
\enddata
\tablenotetext{a}{Scaled by $f_{\rm ion}$, the maximum CIE abundance (Sutherland \& Dopita 1993), $Z=0.1\,Z_\sun$, and $H_0=70\rm~km~s^{-1}~Mpc^{-1}$.  $\Omega_{\rm b}=0.0455\pm0.0015\,h_{70}^{-2}$ \citep{Spergel07}.  See text.}
\tablenotetext{b}{1-this work; 2-Danforth \& Shull 2005; 3-Tripp et al. 2005}
\label{tab:results2}
\end{deluxetable*}

Our new \HI\ results (Figure~\ref{fig:h1}) are compatible with previous studies.  \citet{Penton4} found \dndz\ $\approx 165\pm15$ for \Lya\ absorbers down to $W_\lambda=30$~m\AA\ equivalent width, with a median $W_{\rm Ly\alpha}=68$ m\AA\ and a power-law distribution in column density $d{\cal N}(N)/dz\propto N^{-\beta}$ with $\beta_{\rm HI}=1.65\pm0.07$.  In Paper~2, we searched for \OVI\ and \CIII\ absorbers in \Lya\ systems with $W_{\rm Ly\alpha}\geq80$~m\AA, correcting for \Lya\ saturation with full CoG determinations of column density in most cases.  Despite this correction for artificially low values of \NHI, we found only a minor change in index, $\beta_{\rm HI}=1.68\pm0.11$.  In our new work, we do not use a minimum \Lya\ equivalent width, but rather are limited only by the quality of the data.  We find a survey median of $W_{\rm Ly\alpha}=94$~m\AA, with the uncorrected distribution peaked at $W_{\rm Ly\alpha}\sim45$~m\AA.  The power-law index is $\beta_{\rm HI}=1.73\pm0.04$, and the distribution has an integrated line frequency of $d{\cal N}/dz=129^{+6}_{-5}$ down to 30 m\AA, significantly different from the result of \citet{Penton4}.  However, all the absorbers in the Penton \etal\ surveys were at $z<0.07$, while those in the current survey are at $z<0.4$ (with $\langle z\rangle=0.14$).  The proper-length-corrected absorber frequency is $d{\cal N}/dX=151^{+7}_{-6}$, more in line with the \citet{Penton4} result.

The \NHI\ histogram and \dndz\ plot show a small dip at log\,$N_{\rm HI}=13.6$ in an otherwise smooth and well-behaved distribution.  This is likely a selection effect of how we chose and measured \Lya\ absorbers.  In Paper~2, we showed that \HI\ measurements using only the \Lya\ line tend to underestimate \NHI\ because of line saturation. The dip at log\,$N_{\rm HI}=13.6$ lies at the transition between \Lya-only and CoG-determined column densities and is almost certainly an artifact.  Since the total number of absorbers is conserved, any effect on $\beta_{\rm HI}$, \dndz, or $\Omega$ from this artifact is likely negligible and is unique to the \HI\ data.
 
\begin{figure}
  \epsscale{1.1}\plotone{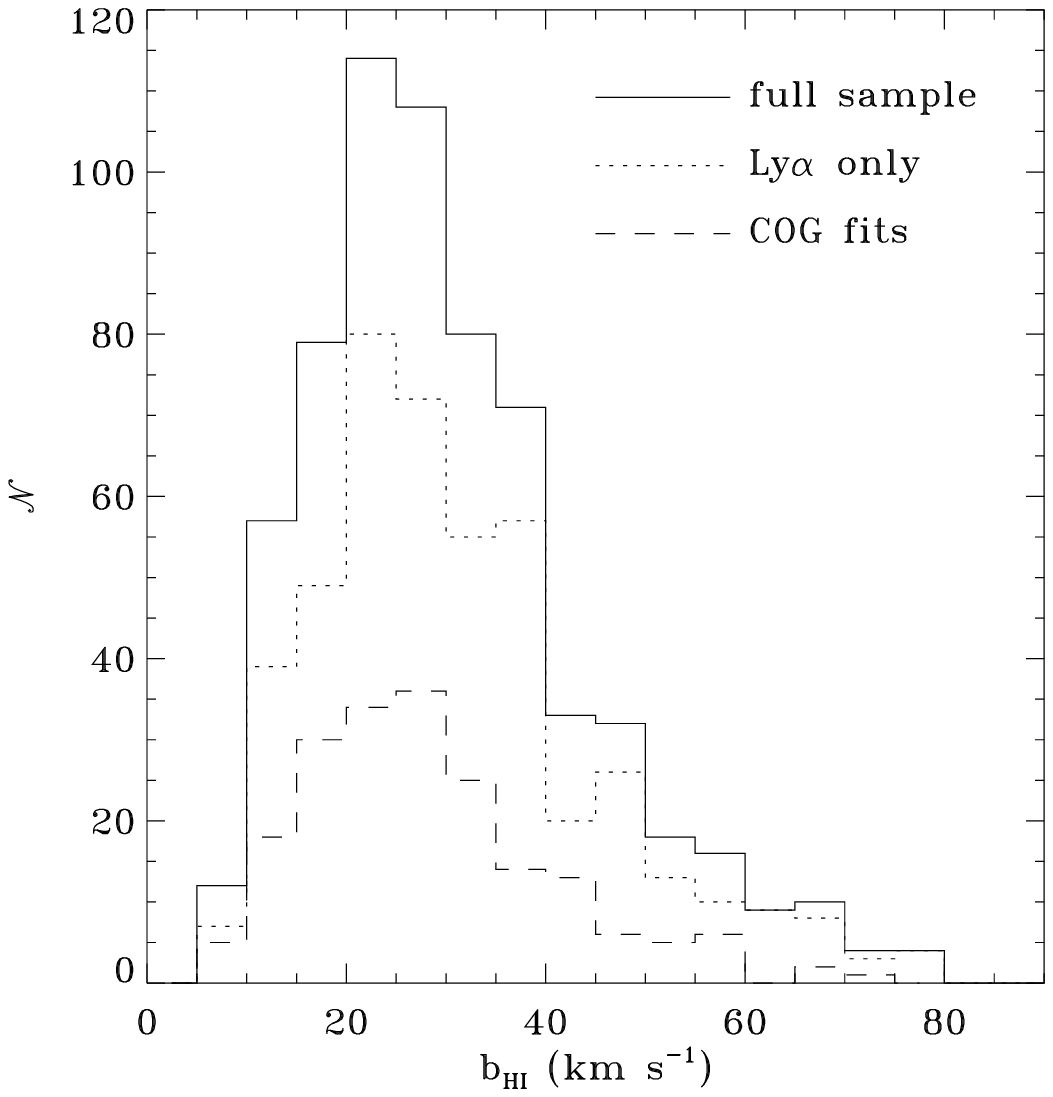} 
  \caption{Distribution of doppler $b$-values for 650 H\,I absorbers.
  Dashed line represents the subset of 195 systems for which a curve
  of growth $N$, $b$ solution is available.  Dotted line represents
  the remainder (455) for which only \Lya\ fits were used.  The median
  and standard deviations for each sample are $27\pm15$ \kms\ (total),
  $25\pm13$ \kms\ (CoG), and $28\pm16$ \kms\ (\Lya).  Broad \Lya\
  Absorbers (BLAs) $b_{\rm HI}>40$ \kms\ represent $\la20$\% of the
  absorbers.  These may indicate high gas temperatures ($T>10^5$~K),
  multiple components, or bulk flows within the absorber.  A number of
  the broadest systems may be multiple blended components (see Shull
  et al.(2000) and Paper~2 for discussion) so this BLA fraction is
  probably an upper limit.  Compare with Fig.~6 in Paper~2 and Fig.~12
  of \citet{Tripp08}.}  \label{fig:h1_b}
\end{figure}

Hydrogen is the lightest element and thus is particularly sensitive to thermal broadening.  The distribution of doppler paramters, \bHI, provides a valuable upper limit on gas temperature.  Figure~\ref{fig:h1_b} shows the distributions for the entire sample (650 lines) of \HI\ measurements, the corrected \Lya-only (455), and CoG-qualified (195) subsamples.  The distributions are similar for each sample, with a median and $1\sigma$ variation of $27\pm15$ \kms\ ($28\pm16$ \kms, $25\pm13$ \kms), consistent with the results of Paper~2.  Assuming that the entire line width is due to thermal broadening, $b_{\rm HI}=27$ \kms\ corresponds to a gas temperature $T=60.5\,b_{\rm HI}^2\approx44,000$~K.  However, some of the doppler width is undoubtedly due to velocity components and bulk flows.

\citet{Lehner07} find a large number of broad \Lya\ absorbers (BLAs) in many of the same sight lines with $b_{\rm HI}>40$~\kms\ as measured from \Lya\ lines.  In some cases, they report $b_{\rm HI}>100$~\kms.  If the line width is entirely due to gas temperature, these measurements imply $T\geq10^5$~K.  However, very broad lines are difficult to detect against all but extremely well-characterized AGN continua and subject to a number of interpretations.  Only extremely strong \Lya\ systems can be both broad and deep at the same time, and these are often not unambiguously a single component.  Shallow, broad features are far more common, but may arise from continuum variations or instrumental artifacts such as ``echelle ripples''.  We find that 15-20\% of our \HI\ systems show $b_{\rm HI}>40$ \kms, and the fraction falls off rapidly at higher \bHI.  The details of these individual detections are beyond the scope of this paper, but we will address these BLA systems in detail in an upcoming paper (Danforth \etal, in preparation).

\subsubsection{\OVI} 

\begin{figure*}
  \epsscale{1.1}\plotone{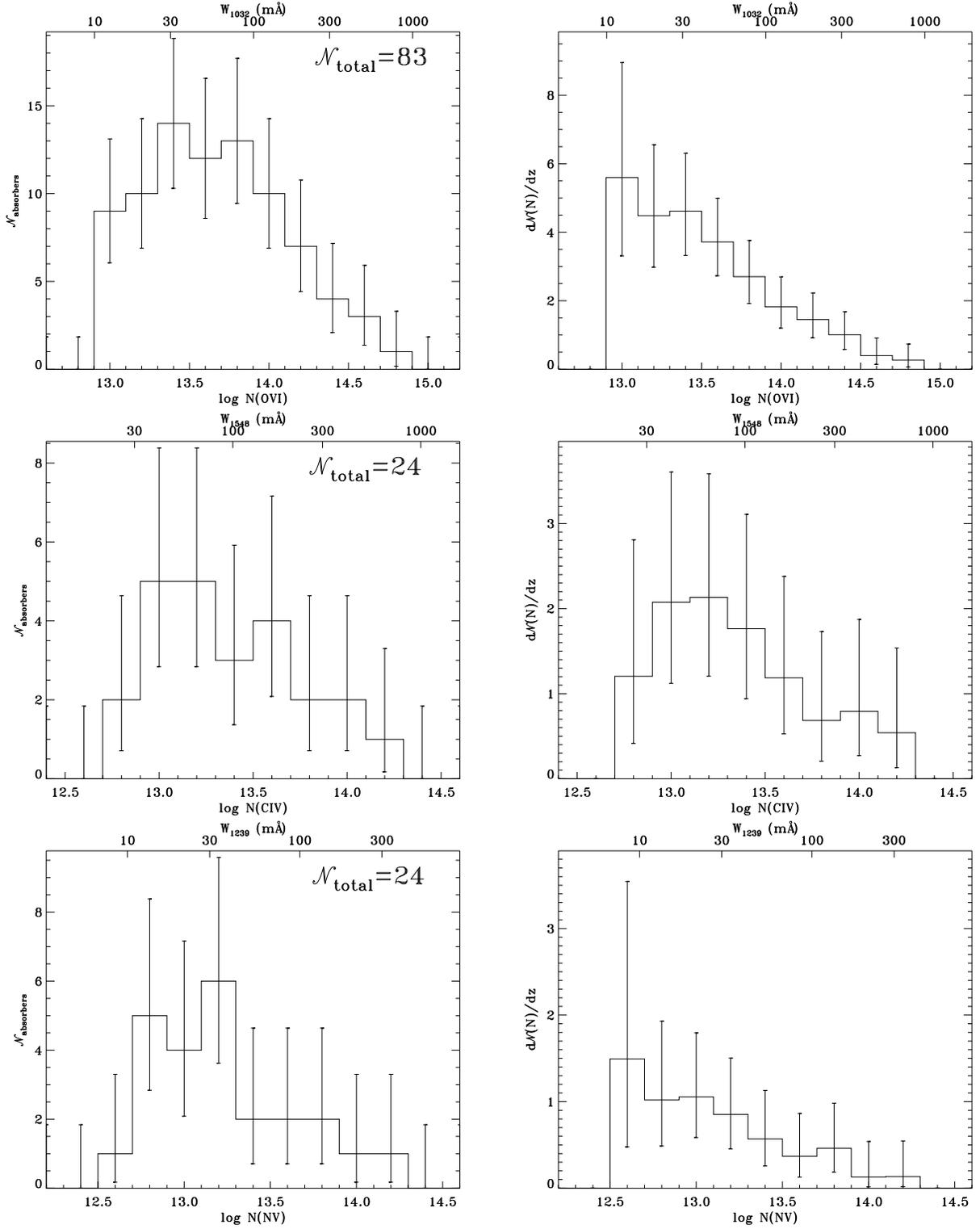} 
  \caption{Detection statistics and \dndz\ for O\,VI (top), C\,IV
  (middle), and N\,V (bottom).  All column densities are measured
  directly from the data (profile fits or AOD method) and not from
  curves of growth.}  \label{fig:highions}
\end{figure*}

Our \OVI\ results (Figure~\ref{fig:highions}a,b) are compatible with those in Paper~1.  In that work, we found a column density distribution with $\beta_{\rm OVI}=2.2\pm0.1$, slightly steeper\footnote{Results in Papers~1 and~2 did not use the distributed-power histogram technique reported here.  This technique is known to decrease $\beta$ values by 0.05--0.10 dex, however, we have not adjusted the literature values.} than seen in the current, larger sample ($\beta_{\rm OVI}=1.98\pm0.11$).  The multiphase slope parameter $\alpha_{14}$ differs between the two samples as well (0.7 vs.\ 0.9).  In Paper~1 we derived $\alpha_{14}=0.9\pm0.1$, consistent with no \NHI\ correlation; but we now find $\alpha_{14}=0.71\pm0.03$ in the larger sample, consistent with some degree of correlation.  However, the two studies probe different redshift ranges.  If the current sample is limited to $z_{\rm abs}<0.15$, as in Paper~1, we have 35 absorbers with $\beta=2.26\pm0.20$ and $\alpha_{14}=0.81\pm0.06$, in closer agreement with Paper~1.  This suggests some redshift evolution in the \OVI\ absorbers between $0<z_{\rm abs}<0.4$, as suggested in some simulations of the rate at which the metal-seeded WHIM increases at low redshift \citep{Dave99,CenOstriker99}.  

The new sample of \OVI\ absorbers shows a strong power-law distribution (Figure~\ref{fig:highions}b).  In Paper~1, we reported hints of a turnover in the \dndz\ column-density distribution at log\,$N_{\rm OVI}<13.4$ ($W_{1032}<30$~m\AA).  However, there were only five absorbers in these low-$N$ bins, and our conclusion was tentative.  Our new study features 26 absorbers with log\,$N_{\rm OVI}<13.4$, and the distribution continues as a smooth power law, with no apparent turnover down to log\,$N_{\rm OVI}\approx13.0$ ($W_{1032}>12.5$~m\AA).  Even the $z<0.15$ subsample has six absorbers with log\,$N_{\rm OVI}<13.4$ and also shows a similarly smooth power law down to log\,$N_{\rm OVI}\sim13.1$, with no suggestion of a turnover.  The continued rise in this distribution can be used to constrain the extent of metal transport away from dwarf galaxies \citep{Stocke06b,TumlinsonFang05}.

\begin{figure}[t]
  \epsscale{1.1}\plotone{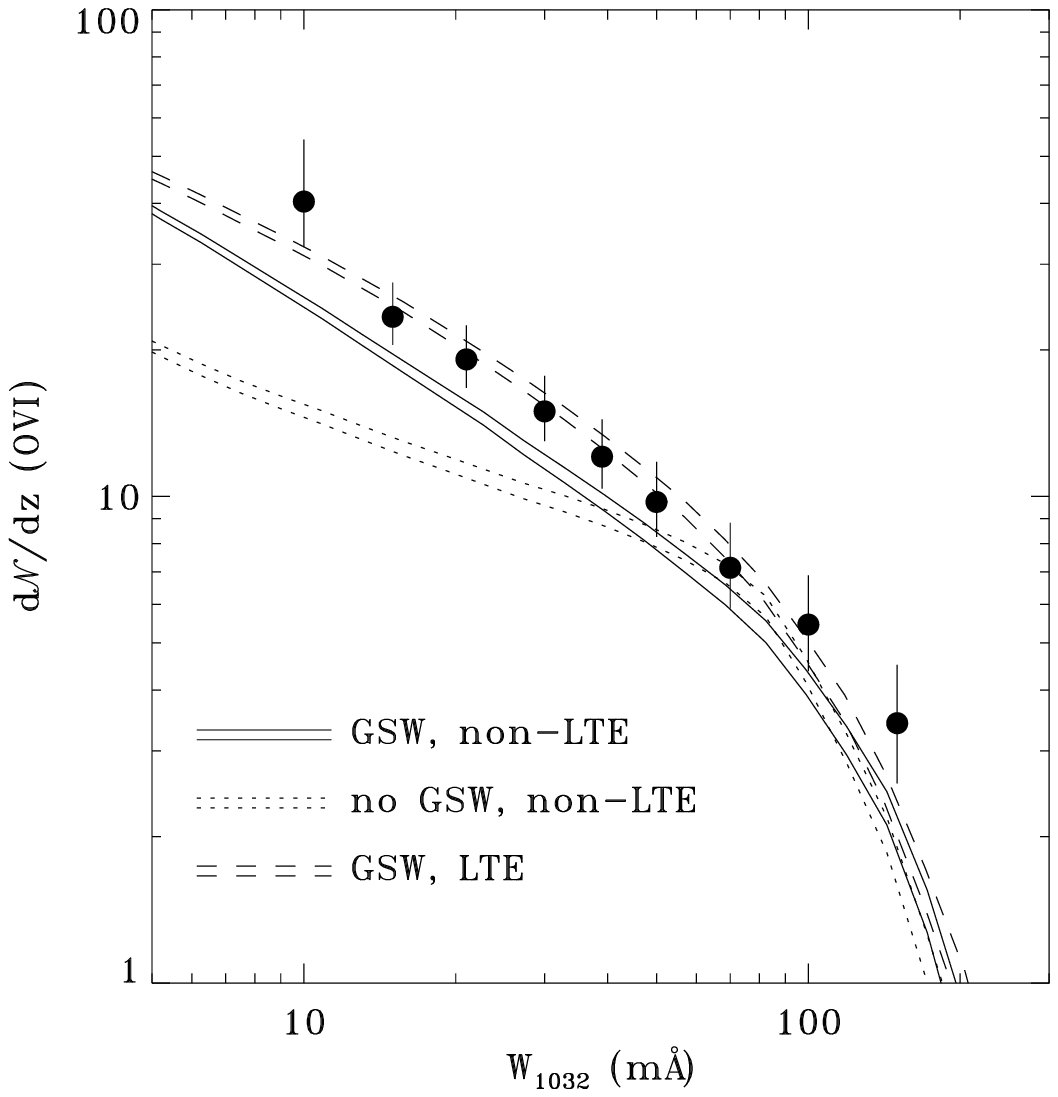} 
  \caption{Observed cumulative \dndz\ for \OVI\ at different
  equivalent widths, compared with several simulations by
  \citet{CenFang06}.  The simulation labeled LTE (actually collisional
  ionization equilibrium) and galactic stellar winds (GSW) matches the
  observations most closely.  However, note the slight excess in
  observed strong systems above $W>100$~m\AA.  This may be due to
  enhanced metallicity in the strongest systems, or it may be an
  artifact of the finite size of the simulation volume.}
  \label{fig:cenfangfig2}
\end{figure}

As in Paper~1, we compare the observed \OVI\ statistics to theoretical predictions.  \citet{CenFang06} compute several different sets of models of the evolution of the WHIM with (and without) realistic galactic superwind (GSW) feedback and with non-equilibrium evolution of major metal ions (including \OVI\ and other WHIM tracers).  They compare their models (their Figure~2) to the observed cumulative \dndz\ distribution at different \OVI\ equivalent widths (Paper~1).  They see reasonably good agreement with the data in the model with GSW and non-equilibrium ionization.  In Figure~\ref{fig:cenfangfig2}, we compare the \citet{CenFang06} models to the current \dndz\ data versus equivalent width.  Minimizing the \Lya\ bias has generally increased \dndz\ and now is best described by the GSW model with species in ionization equilibrium.  The non-equilibrium model underpredicts \OVI\ absorber density at all column densities.  While we do not believe that the models are sufficiently accurate to discriminate between CIE and non-equilibrium ionization, it is clear that GSW are required to produce the observed enhancement in number of weak, low-\NOVI\ absorbers.  Changes in metallicity tend to shift the models up and down, and a higher metal production rate might match the observations more closely.  There are certain additional systematics which must be mentioned.  For instance, the models were calculated at $z=0$, while our \OVI\ observations are at $\langle z\rangle\sim0.2$.  The WHIM fraction is expected to increase as a function of time \citep{Dave99}; if the models were integrated over the same redshift range as the observations, they would likely predict lower \dndz\ values.  Also, if we account for the full \OVI\ census including lines with no measurable \Lya\ counterpart, the \dndz\ values would increase as discussed below.  Both of these systematics would decrease the agreement between the data and models.

We see a distribution of \OVI\ $b$-values that covers a wide range.  This distribution is of physical importance, since explaining observed line width as thermal broadening provides an upper limit to the temperature of \OVI-bearing gas in the IGM.  \OVI\ has a thermal line width 
\begin{equation}
   b_{\rm OVI}=\sqrt{\frac{2kT}{m_O}}\approx(17.05~{\rm km~s^{-1}})\left(\frac{T}{T_{\rm max}}\right)^{1/2} \; ,
\end{equation} 
where $T_{\rm max}=10^{5.45}$ K and $m_O=16\,m_H$.  Turbulence and unresolved velocity components may artificially broaden the line profiles, but $b_{\rm OVI}$ does provide a useful upper limit to the gas temperature.  For the most part, \OVI\ lines were measured by a profile fitting routine that takes instrumental resolution into account.  The distribution looks much like those reported in Fig.~12 of \citet{Tripp08} and in Fig.~6 of Paper~2.  Of our 83 \OVI\ detections, only 11 have $b_{\rm OVI}\geq50$~\kms\ and only four systems have $b_{\rm OVI}\geq60$~\kms\ (each of these four systems shows multiple components).  The mean is $\langle b_{\rm OVI}\rangle=30\pm16$~\kms\ with a median value lower by 3~\kms, nearly identical to our results in Paper~2.  Nineteen of our absorbers (23\%) show $b_{\rm OVI}\leq17$~\kms\ (temperatures at or below the peak CIE temperature) while only five (6\%) show $b_{\rm OVI}<10$~\kms, corresponding to $T\leq10^5$ K, consistent with photoionization.  By comparison, \citet{Tripp08} find that 37\% of their intervening \OVI\ systems have $b<10$~\kms, implying low temperatures in a substantial fraction of their absorbers.  As shown by \citet{CenFang06}, one expects some fraction of the WHIM to lie at temperatures below $T_{\rm max}$, owing to time-dependent radiative cooling to $T\leq 10^5$~K.  However, Tripp \etal\ subdivided many of their absorption systems into individual components, and some of the fits may be artificially narrow.

As in Paper~2, we can use the distribution of \HI\ line $b$-values to provide an upper bound on the size of IGM clouds.  Non-virialized clouds of greater than $\sim1$~Mpc extent would show significant Hubble expansion.  Assuming that the observed line width is due to instrumental, thermal, and Hubble expansion effects, and that all \OVI\ absorbers exist at their peak CIE temperature, we calculate 
\begin{equation}
   \langle\ell \rangle \leq [\langle b_{\rm OVI}\rangle^2-17^2]^{1/2}\,[H(z)]^{-1}\sim350\,h_{70}^{-1}\rm~kpc,
\end{equation} 
consistent with previous results \citep{TumlinsonFang05,Paper2}.  Here, we have accounted for the small increase in Hubble expansion, $H(z)=H_0\,[\Omega_m\,(1+z)^3+\Omega_\Lambda]^{1/2}\approx75~{\rm km~s}^{-1}~{\rm Mpc}^{-1}$, at the mean \HI\ absorber redshift, $\langle z \rangle=0.14$.


Throughout this analysis, we assume that \OVI\ is predominantly thermally ionized.  For photoionization to dominate collisional ionization for \OVI, the gas density must be fairly low, $n_H \approx 10^{-6}$ cm$^{-3}$.  A typical \HI\ absorber has $N_{\rm HI}\sim10^{14}$ \cd\ and, at WHIM temperatures, a neutral fraction $f_{\rm HI}<10^{-5}$.  This gives a characteristic length scale
\begin{equation}
L_{\rm HI}=\frac{(N_{\rm HI}/f_{\rm HI})}{n_H}
~\sim~(3~{\rm Mpc})\,\frac{N_{\rm HI}}{10^{14}~{\rm cm^{-2}}}
\end{equation}
The high ionization parameter for photoionized \OVI\ requires extremely diffuse clouds ($\delta_{\rm H}\sim5$) and thus extremely long pathlengths ($L_H\sim3$~Mpc) for typical observed column densities.  The Hubble broadening over this length for an unvirialized cloud, $v_H=L_{\rm HI}\,H_0\approx200\,h_{70}$~\kms, is many times larger than the observed line widths of any observed absorber.  Photoionized \OVI\ may be feasible in high radiation environments near AGN at observed column densities, or in extremely low column density diffuse clouds, but is largely inconsistent with the observed IGM.

In most instances of metal line absorption, we observe an apparent kinematic association of \HI\ and \OVI\ (typical $\Delta v\la50$ \kms), suggesting that collisionally ionized gas and photoionized gas are spatially associated in a multiphase system, as simulations predict \citep{CenOstriker99,Dave01}.  We assume that the two phases have the same metallicity and differ only in ionization state.  Even at WHIM temperatures, there exists a small hydrogen neutral fraction ($f_{\rm HI}\sim10^{-6}$) and thus there should exist a hot gas signature in the \Lya\ profile as well as our standard WHIM metal ions:  
\[ N_{\rm HI}^{\rm(hot)}=\frac{N_{\rm OVI}}{Z_O\,(O/H)_\sun}\,\frac{f_{\rm HI}(T)}{f_{\rm OVI}(T)} \]
\begin{equation}=(1.5\times10^{13}~{\rm cm^{-2}})\,\biggl(\frac{N_{\rm OVI}}{10^{14}~{\rm cm^{-2}}}\biggr)\,\biggl(\frac{Z_O}{0.1\,Z_\sun}\biggr)^{-1}.
\end{equation}
Here, we adopt the CIE-peak ratio $f_{\rm HI}/f_{\rm OVI}=7.5\times10^{-6}$ at $T=10^{5.45}$~K \citep{SutherlandDopita93} and scale \NHI\ to $N_{\rm OVI}=10^{14}$~\cd\ and $Z_O=0.1\,Z_\sun$.  Typical \OVI\ detections are $N_{\rm OVI}\sim4\times10^{13}$ \cd, so the expected hot \HI\ line would have $N_{\rm HI}^{\rm(hot)}\sim6\times10^{12}$ \cd\ with doppler width $b_{\rm HI}\sim70$ \kms, equivalent width $W_{\rm Ly\alpha}\sim33$~m\AA, and a central line depth only $\sim7$\% below the continuum ($\tau_0\sim0.07$).  This is marginally detectable by itself in high S/N data, but it is very difficult to separate from the stronger, narrower \Lya\ profile in a multiphase absorber and might be mistaken for the wings of the spectrograph point spread function.  Even a strong \OVI\ absorber ($N_{\rm OVI}=10^{14}$~\cd) would show $W_{\rm Ly\alpha}=78$~m\AA\ and a fractional depth of $\sim15$\% ($\tau_0\sim0.16$).  

To investigate the existence of blended hot-plus-photoionized \Lya\ profiles, we compare the \bHI\ distribution of \HI\ systems with \OVI\ detections with those having clean \OVI\ nondetections ($N_{\rm OVI}<10^{13.2}$~\cd).  For the 83 systems with an \OVI\ detection, the median and standard deviation are $b_{\rm HI}=31\pm15$ \kms\ while the 273 \OVI\ nondetections show $b_{\rm HI}=26\pm13$ \kms.  The sample means have uncertainties of $\pm3$ \kms\ and $\pm2$ \kms, respectively, which suggests that shallow, weak \Lya\ lines {\it are} broadening the overall \HI\ absorption profile.  Hot \HI\ is an ideal method of tracking the WHIM and does not rely on metal enrichment.  However, it is unclear whether a difference between the two populations would be apparent without the \OVI\ diagnostic.   

The \OVI\ census in Paper~1 included only absorbers for which $W_{\rm Ly\alpha}\ge80$~m\AA, yet \NOVI\ is seen to have little correlation with \NHI\ and low-\NHI\ absorbers often have comparable values of \NOVI\ in the multiphase plot.  What effect does this ``\Lya\ bias'' have on the WHIM census?  Weak \Lya\ absorbers ($W_{\rm Ly\alpha}<80$~m\AA) make up 45\% of our current survey, while 19 out of 83 \OVI\ detections (23\%) were seen in weak \Lya\ lines, which we would not have considered in Paper~1.  In this survey, for the \FUSE\ sample of 35 absorbers at $z<0.15$, we find $d{\cal N}/dz=19^{+6}_{-4}$ down to 30~m\AA\ and $d{\cal N}/dz=60^{+38}_{-19}$ down to 10~m\AA.  These compare to our finding in Paper~1, $d{\cal N}/dz=17\pm3$ (down to 30~m\AA) where weak \HI\ systems were not taken into account.  Our estimate, $\Omega_{\rm OVI}$ (Table~5) is also $\sim25$\% higher in the low-$z$ sample than was seen in Paper~1.

Our current survey does not fully correct for \Lya\ bias, since we still require an \HI\ detection at some level to trigger a search for \OVI. \citet{Tripp08} performed an ambitious ``blind'' \OVI\ survey of 16 AGN sight lines including many of those in this work.  They find $d{\cal N}/dz=18.3^{+3.0}_{-2.6}$ down to 30~m\AA, compared to our $d{\cal N}/dz=15.0^{+2.7}_{-2.0}$ to the same limit ($z<0.4$ sample).  In the same sub-sample, we also find $d{\cal N}/dz=40^{+14}_{-8}$ down to 10~m\AA\ (\dndx$=54^{+12}_{-9}$).  Ignoring systematic differences between the two surveys, such as our respective criteria for absorber path length ($\Delta z$) and S/N-limited equivalent width limits, this implies that an additional $\sim20$\% of \OVI\ absorbers do not have any measurable corresponding \HI\ absorption.  This effect was already apparent in a qualitative sense in the low-\NHI\ end of the multiphase plot (Figure~1b of Paper~1), where we found many absorbers with \NOVI\ $\geq$ \NHI.  When combined with the $\sim20$\% effect implied by the differences between this work and Paper~1 noted above, we conclude that ``\Lya-bias'' can have a significant effect on a complete WHIM baryon survey.  


Following the example of \citet{TumlinsonFang05}, we can use the \OVI\ line frequencies, \dndz, to constrain the distribution of metals in the low-$z$ IGM.  Referring to their Figure~2, we find that our value for the $z<0.4$ sample, $d{\cal N}/dz=40^{+14}_{-8}$, integrated down to 10~m\AA, is consistent with a mean spatial extent of metals out to 250--300 kpc from galaxies in the Sloan Digital Sky Survey (SDSS).  In this picture, and that proposed by \citet{Stocke06b}, metal enrichment  occurs around dwarf galaxies of SDSS $r$-band absolute magnitudes down to $M_r\approx-17$ to $-18$ (0.03--0.05 $L^*$). 

\subsubsection{\CIV\ and \NV } 

The ions \NV\ and \CIV\ are both Li-like, with strong UV resonance ($2s-2p$) doublets.  Like \OVI, they may provide good WHIM tracers, at longer wavelengths accessible in the \HST\ spectral band.  These ions have been studied toward many Galactic high-velocity clouds \citep{SembachSavage94,IndebetouwShull04,Sembach03,Collins03,Collins07}.  However, to date, no large, low-redshift IGM surveys have been conducted in these ions.  \CIV\ is a strong absorber of an abundant element, but the STIS/E140M data are limited to $z_{\rm abs}\la0.116$, at which redshift \CIV\ $\lambda 1548$ shifts out of the passband.  The longer wavelength STIS/E230M grating covers $0.01<z_{\rm CIV}<1.0$, but this band was not often used in AGN observations. 

In our new survey, we detected 24 \CIV\ absorbers in the STIS/E140M data over a total redshift pathlength $\Delta z=2.42$ and inferred $(d{\cal N}/dz)_{\rm CIV}=10^{+4}_{-2}$ and $\Omega_{\rm CIV}=(7.67\pm1.45)\times10^{-8}$ down to $W\ge30$~m\AA.  At high redshift ($1.5<z<3.1$) \citet{Scannapieco06} inferred $\Omega_{\rm CIV}=7.54\times10^{-8}$ with no sign of evolution over that redshift range.  Similarly, \citet{Songaila97,Songaila01} inferred $\Omega_{\rm CIV}\approx(2-7)\times 10^{-8}$ at $z=3.0-3.5$.  \citet{Pettini03} reported $\Omega_{\rm CIV}=(4.3\pm2.5)\times10^{-8}$ at $z\sim5$ using similar column density limits as are used in this paper.  The consistency of our low-$z$ measurement to those in the early universe implies little or no evolution in \CIV\ since reionization.  This may imply a changing metal enrichment rate and/or changing ionizing flux with redshift.  However, since the ionization corrections depend on both photon flux and gas density, further speculation is beyond the scope of this paper.  For carbon metallicity $Z_C=0.1\,Z_{\odot}$, where (C/H)$_{\odot}=2.45\times 10^{-4}$, and $f_{\rm CIV}\approx 0.29$, the maximum ion fraction at log\,$T_{\rm max}=5.0$ in CIE, $\Omega_{\rm CIV}$ corresponds to a total baryon fraction associated with \CIV\ of $\Omega_{\rm IGM}^{\rm (CIV)}=(2.6\pm0.5)\times10^{-3}$ or 25-30\% that seen with \OVI.

As with \OVI\ and \HI, the \CIV\ absorbers follow a power-law distribution in column density with $\beta_{\rm CIV}=1.79\pm0.17$ (Figure~\ref{fig:highions}c,d), consistent with literature values at high redshift, $\beta_{\rm CIV}=1.5-1.8$ \citep{Songaila97,Songaila01,Ellison00}.  This slope is shallower than is seen in \OVI\ and more consistent with those of lower-ionization species.  However, the uncertainty on $\beta_{\rm CIV}$ precludes any definite conclusions as to whether \CIV\ exhibits a steep, WHIM-like distribution in column density, or a more moderate, \HI-like distribution.  The ionization energy required to produce \CIV\ is only 3.52 ryd (vs.\ 8.37~ryd to produce \OVI), thus \CIV\ may be produced partly by photoionization by AGN and hot stars at low $z$ \citep{GirouxShull97}.  On the other hand, we see little correlation between \NCIV\ and the column densities of lower ions, even \CIII\ (see below).  In contrast, there is some degree of \NCIV--\NOVI\ correlation, suggesting that \CIV\ may primarily trace shock-heated material.  The baryon fraction traced by \CIV\ is less than half that of \OVI\ for a given equivalent width limit, consistent with the solar abundance ratio, (C/O)$_{\sun}=0.50\pm0.07$ \citep{Allende02,Asplund05}, and the implied sub-solar (C/O) ratio seen in the IGM (Paper~2).  In the  pure thermal (collisional ionization) interpretation of \CIV, the cooler  portions of WHIM (at $T\sim10^5$~K) may contain a smaller portion of the baryons than at $T\sim10^{5.5}$~K, consistent with more efficient cooling at temperatures near or below the peak of the cooling curve.

The rest wavelengths of the \NV\ doublet (1238.821, 1242.804~\AA) are well placed in the STIS/E140M data for redshift coverage out to $z_{\rm abs}\approx0.4$.  However, the solar abundance of nitrogen is less than 20\% that of oxygen.  Moreover, nitrogen is a product of secondary nucleosynthesis, and its abundance relative to C and O is observed to be sub-solar in Galactic HVCs \citep{Gibson01,Collins03,Collins07}.  We detected 24 \NV\ absorbers in one or both lines of the doublet (Figure~\ref{fig:highions}e,f) over a total pathlength $\Delta z=5.30$. Because of the lower relative abundance of nitrogen compared to carbon or oxygen, the detections are never strong, and we derive $d{\cal N}/dz=2\pm1$ down to 30~m\AA.  Weaker \NV\ lines are more numerous, and our total sample has $d{\cal N}/dz=7^{+3}_{-2}$ down to 10~m\AA.  We detected \NV\ and \OVI\ together in 11 systems, and their column densities are well correlated.  The ionization potential to produce \NV\ is 5.69 ryd, and its peak CIE ion abundance occurs at log\,$T_{\rm max}=5.25$, near to that (log\,$T_{\rm max}=5.45$) for \OVI\ \citep{SutherlandDopita93}.  We also note that these high ions are not necessarily in equilibrium, and non-equilibrium ionization effects may alter $f_{\rm ion}$.  We infer a value of $\Omega^{\rm (NV)}_{\rm IGM}/\Omega_b$ intermediate between those of \OVI\ and \CIV\ at both 30 and 10~m\AA\ (Table~\ref{tab:results2}).  We conclude that \NV\ is a reliable tracer of WHIM gas based on its correlation with \OVI\ and steep power-law slope, although with sizable error bars.  Future studies of these ions with the {\it Cosmic Origins Spectrograph} on {\it Hubble} have great promise, not only to trace the WHIM but to derive nucleosynthetic signatures from a range of heavy elements (e.g., C, N, O, Si, Fe).

\subsubsection{\CIII} 

\begin{figure*}
  \epsscale{1.1}\plotone{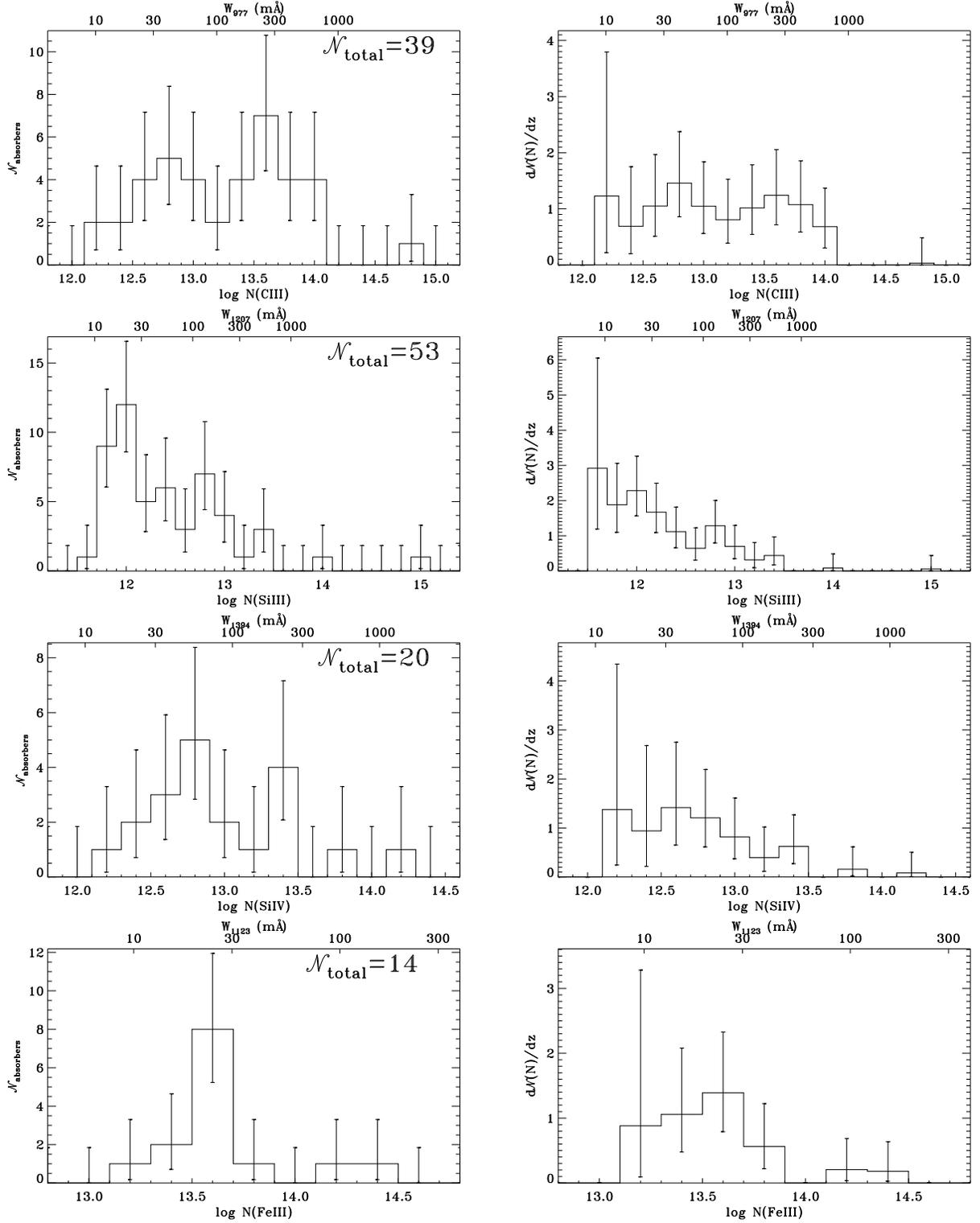} 
  \caption{Same as Figure~\ref{fig:highions} for C\,III (top), Si\,III
  (second row), Si\,IV (third row), and Fe\,III (bottom).}
  \label{fig:lowions}
\end{figure*}

We made 39 \CIII\ detections, yielding \dndz$=10^{+3}_{-2}$ ($d{\cal N}/dX=12^{+4}_{-2}$) down to 30~m\AA.  The column density distribution shows a power-law slope of $\beta_{\rm CIII}=1.79\pm0.10$ (Figures~\ref{fig:lowions}a,b).  Absorber frequency is compatible with that found in Paper~2, but spectral slope appears higher by $\sim0.2$ dex.  A subsample of absorbers at $z_{\rm abs}<0.21$, comparable to the sample used in Paper~2, shows 28 absorbers with a somewhat higher absorber frequency, \dndz$=14^{+5}_{-3}$ ($d{\cal N}/dX=14^{+7}_{-4}$), and a steeper column density distribution, $\beta_{\rm CIII}=2.06\pm0.15$.  The reason for this discrepancy may be related to our inclusion of weaker \Lya\ systems, which were not studied in our previous work.  Since much of the \CIII\ is expected to arise in the same photoionized gas traced by \HI, we would expect that the inclusion of weaker \HI\ systems in the survey would produce a larger number of \CIII\ systems as well.  However, this does not explain the steepening of the column-density distribution seen in the low-$z$ subsample.
			
In Paper 2, we saw a reasonably strong correlation between \NHI\ and \NCIII.  The multiphase ratio was fitted with a line of slope $\alpha_{14}=0.73\pm0.08$ and scale factor $C_{14}=1.06\pm0.04$.  From this, we incorrectly derived a metallicity for intergalactic carbon of $Z_C=0.12\,Z_\sun$, in good agreement with $Z_O=0.09\,Z_\sun$ derived in Paper~1.  The {\it correct} derivation was lower by a factor of ten and should read $Z_C=0.012\,Z_\sun$, for a peak CIE ionization fraction $f_{\rm CIII}=0.83$.  This inferred metallicity would increase if we chose a lower value of $f_{\rm CIII}$ more typical of photoionized gas.  While the corrected value is substantially different from the derived oxygen metallicity in the IGM, the implied abundance ratio, $(C/O)_{\rm IGM}\sim0.1\,(C/O)_\sun$, is compatible with the observed \CIII/\OVI\ ratio (see \S~4.4 and Figure~11 of Paper~2).  With more \CIV\ data, we could use both ions, \CIII\ and \CIV, as a joint constraint on ionization corrections and carbon abundance. 
  
The behavior of \CIII\ in our new study is consistent with the (corrected) previous results in Paper~2.  The correlation of \NHI\ and \NCIII\ is still strong, and metallicity implied by the multiphase fit ($Z_{\rm C}/Z_\sun\approx1.5$\%) is similar to previous results for both the full sample of 39 absorbers and the 28-absorber low-$z$ subsample.  When adjusted for the mean absorber redshift of each sample and corrected by a range of ionization fractions, $f_{\rm CIII}=$0.2--0.8, for either photoionization or collisional ionization equilibrium, these values imply $Z_{\rm C}=(0.015\pm0.002)\,Z_{\odot}\,(f_{\rm CIII}/0.83)^{-1}$ or $Z_{\rm C}/Z_\sun=1.5-6$\%.

\subsubsection{\SiIII, \SiIV, and \FeIII} 

Intergalactic \SiIII, \SiIV, and \FeIII\ at low redshift have received little or no attention.  Along with \CIII\ discussed above, all three ions are are presumed to trace primarily metal-enriched, photoionized material.  All have peak CIE temperatures below $10^5$~K and ionization potentials $>3$~ryd.  The cooling times at these temperatures are relatively short, and even at the low densities typical in the IGM, any collisionally heated material at $T<10^5$~K should quickly cool.  Such transient situations are likely places for non-equilibrium ionization effects.

Because \SiIII\ \lam1206.500 is a very strong transition, STIS observations are sensitive to extremely low column densities.  Despite the lower abundance of silicon relative to more common metals  ([Si/O]$_\sun\sim-1.4$), \SiIII\ is expected to be the dominant ionization stage in the photoionized or collisionally ionized IGM over a broad range of temperatures and photon energies.  We therefore expect it to be an effective tracer of IGM enrichment, complementary to \CIII\ or \OVI.  We detect \SiIII\ in 53 systems (Figures~\ref{fig:lowions}c,d) for \dndz$=6^{+2}_{-1}$ ($d{\cal N}/dX=7^{+2}_{-1}$) down to 30~m\AA\ or \dndz$=14^{+3}_{-2}$ ($d{\cal N}/dX=16^{+4}_{-2}$) down to 10~m\AA, about 40\% of the value for \OVI\ at either limit.  The column-density distribution, $(d^2{\cal N}/dz\,dN_{\rm SiIII})$, follows a clear power law with $\beta_{\rm SiIV}=1.80\pm0.09$, consistent with $\beta_{\rm CIII}$ and $\beta_{\rm CIV}$, slightly higher than $\beta_{\rm HI}$.

We made 20 \SiIV\ detections and derived \dndz$=4^{+2}_{-1}$ ($d{\cal N}/dX=5^{+2}_{-1}$) down to 30~m\AA\ (Figures~\ref{fig:lowions}e,f).  The column density \NSiIV\ does not show a particular correlation with \NHI, as do \NCIII\ and \NSiIII.  The column density distribution follows a rough power law and a fit to the cumulative distribution gives $\beta_{\rm SiIV}=1.92\pm0.17$, higher than other low-ionization species.  We measure $\Omega_{\rm SiIV}=(8.3\pm1.3)\times10^{-8}$ down to 30~m\AA.  For absorbers at at high redshift, \citet{Scannapieco06} measured $\Omega_{\rm SiIV}=6.0\times10^{-9}$ to log\,\NSiIV$>12$ at $\langle z\rangle=2.4$, a factor of 14 lower than our most sensitive limit at $\langle z\rangle\sim0.1$.  They see no evidence for evolution in \SiIV\ in the redshift range $1.5<z<3.1$.  Similarly, \citet{Songaila97,Songaila01} measured $\Omega_{\rm SiIV}=(7-30)\times10^{-9}$ at $1.5<z<5.5$.  This factor of $2.5-12$ increase in the \SiIV\ fraction between the early and modern universe is opposite to the behavior of \CIV\ above, which remains essentially unchanged.  This may imply some combination of changes in the metagalactic ionization field and/or IGM enrichment.

\FeIII\ is a relatively weak transition \citep[$f=0.0544$,][]{Morton03} of a relatively low-abundance element ([Fe/O]$_\sun\approx-1.4$) and thus it comes as no surprise that we make only 14 detections (Figure~\ref{fig:lowions}g,h) with $d{\cal N}/dz\sim1$ down to 30~m\AA.  The power-law slope is the steepest of any in our survey ($\beta_{\rm FeIII}=2.2\pm0.4$) but with such large uncertainty as to render any comparison speculative.  As with \NOVI, \NFeIII\ shows only cursory correlation with \NHI\ and covers a range of 1.5 dex in column density, while \NHI\ is seen over almost three decades.  This evidence is exactly what we used in Paper~1 to argue that \OVI\ and \HI\ were tracing different IGM phases.  However, given the low ionization energy required to produce \FeIII\ (16.18 eV), it is likely that \FeIII\ and \HI\ arise in similar photoionized gas.  We may be seeing preferential iron enrichment in material that has been shocked and cooled to ambient temperatures.  We will return to these issues and their redshift dependence in a future paper that deals with IGM absorber evolution.

\subsection{Ion Ratios and Comparison with Models}

With such a rich data set, many absorbers show detections in more than one ion, allowing us to compare observed line ratios with predictions from a set of simple CLOUDY models.  In Paper~2 we compared the observed \CIII/\OVI\ ratios for 13 absorbers with models and concluded that a single thermal-plus-photoionized phase could not realistically account for the observations.  We found that \CIII\ and \OVI\ absorbers probably included both photoionized and collisionally ionized gas, although \OVI\ was likely to be mostly collisionally ionized.   

In this work, we use the same set of models:  a four-dimensional grid calculated using CLOUDY v96.01 with parameters temperature $T$, ionization parameter $U$, metallicity $Z$ (relative to solar values), and hydrogen number density $n$ (cm$^{-3}$).  We simulate the effects of collisional ionization by setting the cloud to a constant temperature in each model.  Low temperatures, $T\approx(1-3)\times10^4$~K, approximate a pure photoionization model, while low photoionization parameters, log\,$U<-4$, are equivalent to a pure CIE model.  These models were calculated for a 400~kpc thick optically thin slab of gas illuminated on one side by an AGN radiation field with a power-law continuum $F_\nu\propto\nu^{-\alpha_s}$ in the EUV (1--4~ryd) and soft X-ray (10--22~ryd) with $\alpha_s=1.8$ \citep{HaardtMadau96,Fardal98,Telfer02}.  Since a more realistic isotropic radiation field illuminates the slab from all sides, the actual photon density is a factor of 4 higher ($+0.6$ dex) than that output by the models;  we have adjusted all citations to reflect this.  We also computed a second grid of CLOUDY models assuming $n=10^{-5}~\rm cm^{-3}$, $Z=0.1\,Z_\sun$, and pure photoionization (no thermal ionization component), but we allowed the ionization parameter $U$ and spectral index $\alpha_s$ to vary between $1<\alpha_s<3$ \citep{Telfer02,Shull04}.  The equilibrium temperatures of the pure photoionization models was log\,$T=4.1-4.3$.

\begin{figure*}
  \plottwo{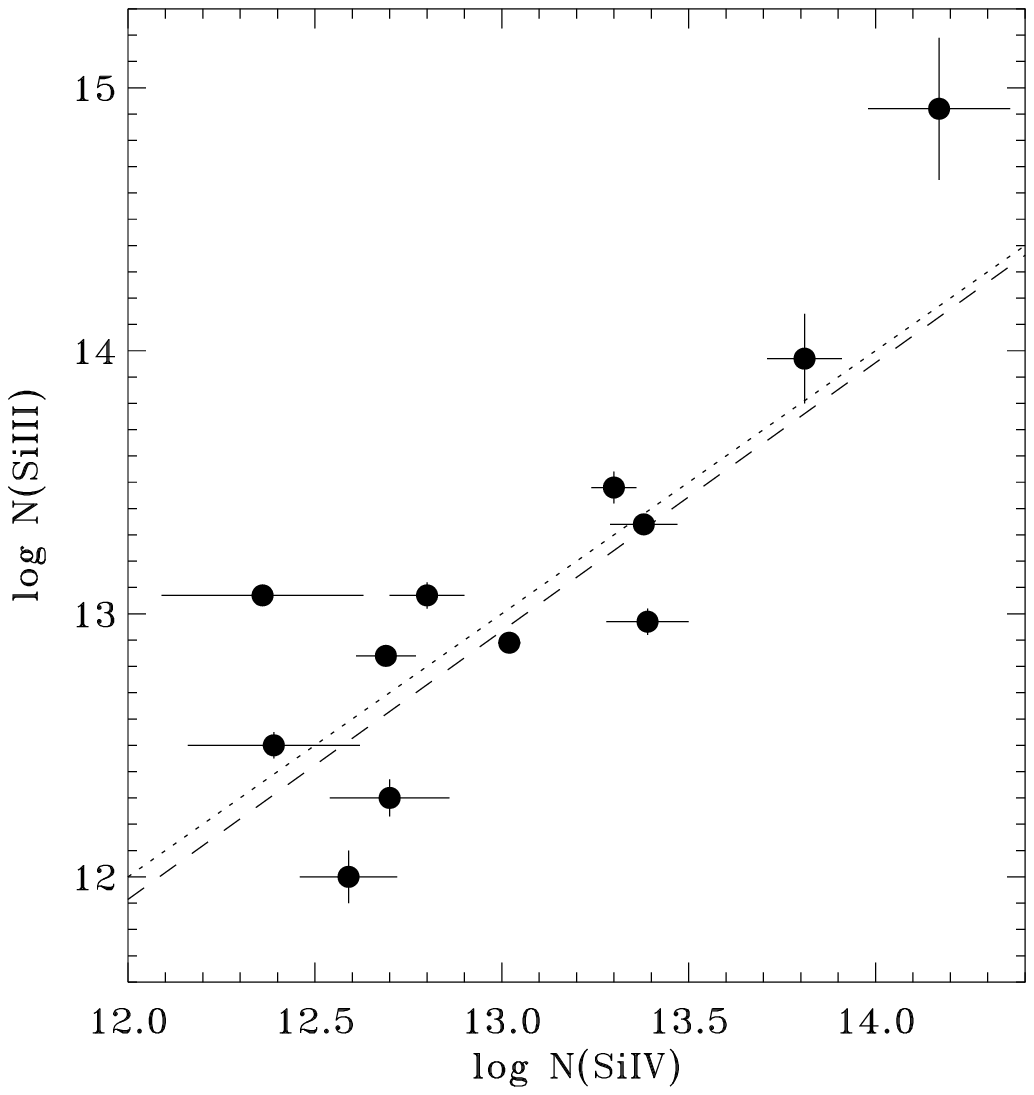}{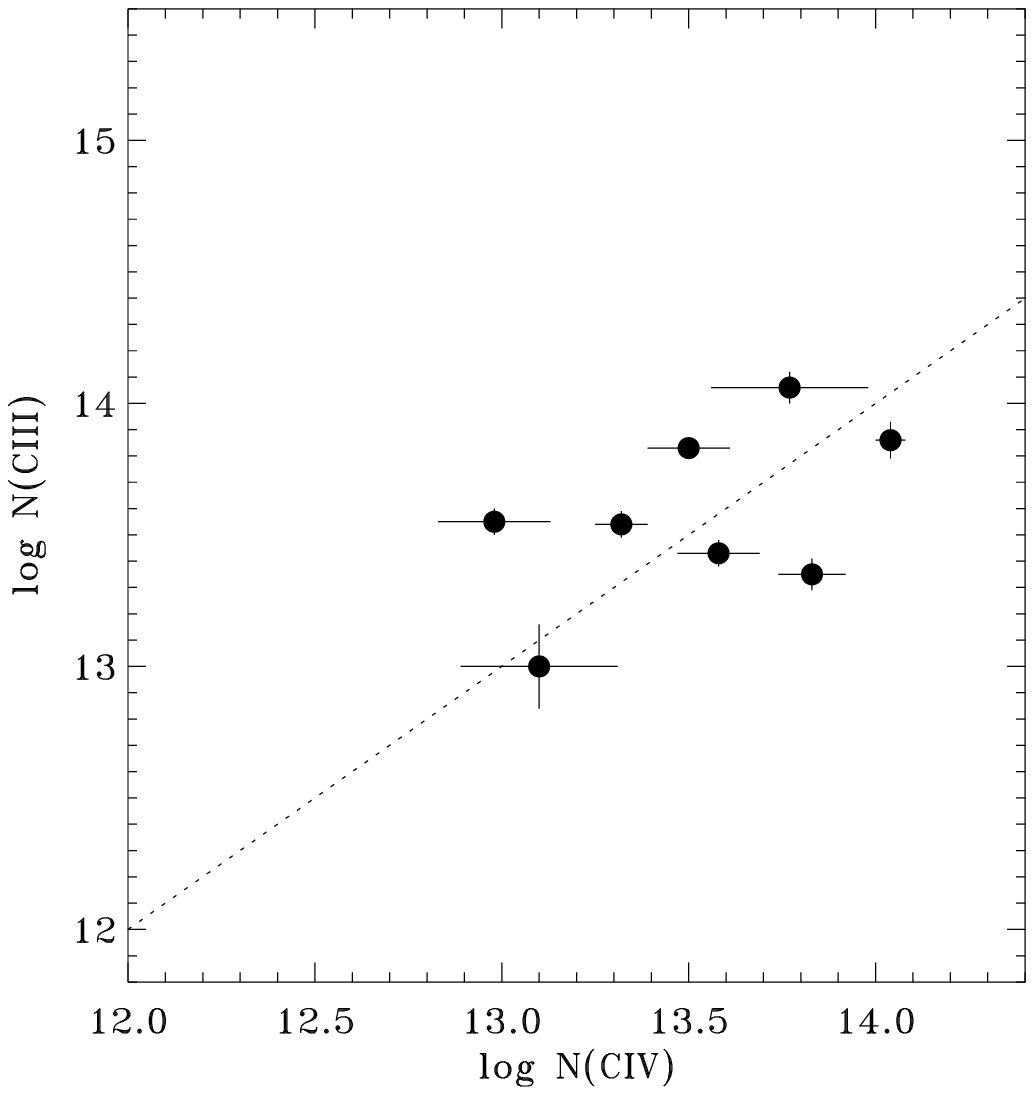} 
  \caption{Detections of both Si\,III and Si\,IV (left) and C\,III and
  C\,IV (right).  The column densities of the Si ions are
  well-correlated ($R=0.87$), and the linear fit (dashed) has a slope
  $1.1\pm0.1$, consistent with a 1:1 ratio (dotted).  The correlation
  is much less for the carbon ions ($R=0.52$), although the data are
  also consistent with a 1:1 ratio (dotted).  The absolute scatter for
  each pair of ions is similar.} \label{fig:ratios1}
\end{figure*}

Observed \SiIII\ and \SiIV\ show a strong correlation in column density ($R=0.85$, \NSiIII\ $\propto$ \NSiIV$^{1.0\pm0.1}$, Fig.~\ref{fig:ratios1}a), suggesting that the two ions trace the same gas.  Since \SiIII/\SiIV\ are adjacent ionization states of the same element, we can avoid complications arising from abundance and metallicity and use \NSiIII/\NSiIV\ as a tracer of the ionization state of the cloud.  The observed ratios show some scatter, $\langle \log\,(N_{\rm SiIII}/N_{\rm SiIV})\rangle=0.1\pm0.4$, but all 12 systems in which both ions appear can be explained as a single-phase absorber, either collisionally ionized at $T=10^{4.8\pm0.2}$~K or photoionized with log\,$U=-1.6\pm0.7$.  Variations in model metallicity and density do not produce significant shifts in allowed parameter space, nor does changing the photoionizing spectral index $\alpha_s$ significantly alter the ($U,T$) solution.  Unfortunately, the doppler $b$-values for the lines are of little help in constraining the temperature solution; the median line width for these Si ions is $b\sim30$~\kms, corresponding to $T>10^6$~K for a case of pure thermal broadening.  We conclude from this that the low-ionization material is consistent with photoionized clouds with log\,$U\sim-1.6$.  Similar CLOUDY models and silicon ion ratios in Galactic high velocity clouds (HVCs) show log\,$U\sim-3$, presumably due to much higher gas densities \citep[see][]{Collins03,Collins07,Westbrook08}.

\CIII\ and \CIV\ are the other pair of adjacent ions in this work.  Detections in both species occur in only eight absorbers (Figure~\ref{fig:ratios1}b), and the column densities are not well correlated ($R\sim0.5$).  Still, the detections are consistent with a 1:1 column-density ratio, and the absolute scatter in log\,(\NCIII/\NCIV) is actually smaller than for log\,(\NSiIII/\NSiIV).  Compared to the models, the range of observed \NCIII/\NCIV\ is consistent with either log\,$U=-1.4\pm0.5$ or log\,$T=5.0\pm0.1$, near the peak CIE fractional abundance of \CIV.  However, since \NCIV\ and \NCIII\ do not show the same degree of correlation present in the Si ions and since we believe \CIV\ to be collisionally ionized in at least some cases, we hesitate to assign too much significance to the \CIII/\CIV\ model comparisons.

\begin{figure*}
  \plottwo{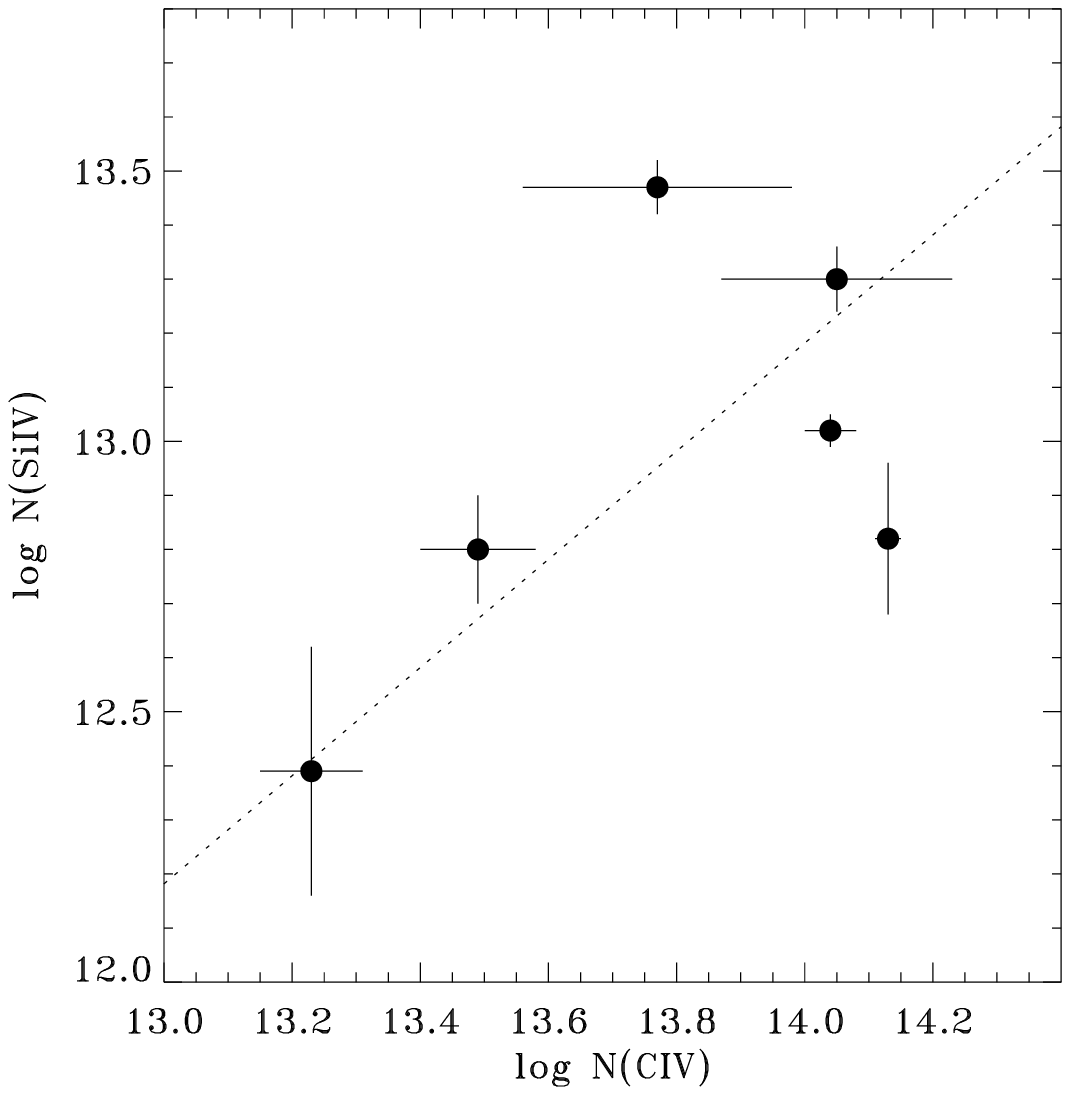}{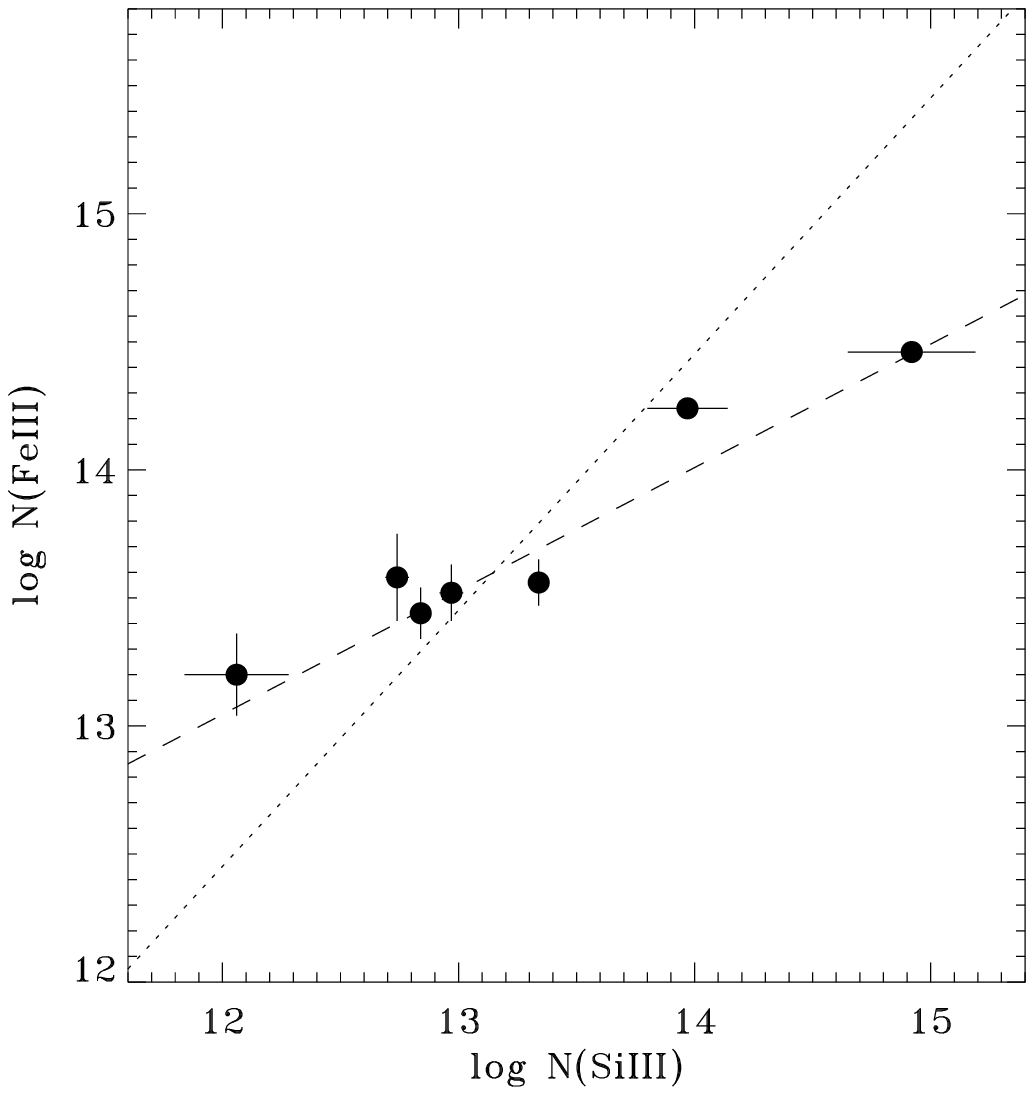} 
  \caption{Line ratio plots for the same ionization stages of
  different elements.  Six absorbers with both C\,IV and Si\,IV
  detections show some correlation (left), but the general trend is
  consistent with $N_{\rm CIV}\propto N_{\rm SiIV}$ (dotted).  Si\,III
  and Fe\,III (right) appear in seven absorbers and are extremely
  well-correlated over several decades, but with $N_{\rm FeIII}\propto
  N_{\rm SiIII}^{0.5\pm0.1}$ (dashed), not 1:1 proportionality
  (dotted).}
\label{fig:ratio2}
\end{figure*}

Because of the expected redshift evolution of the thermal state, ionization state, and metallicity of the IGM, it is useful to compare our low-$z$ results with those seen in the early universe.  High redshift ($z>2$) \SiIV\ is often compared to \CIV\ as a probe of the metagalactic ionizing background \citep{Songaila98,Kim02,Scannapieco06}, or as a measure of the $\alpha$-process versus RGB stellar nucleosynthesis \citep{McWilliam97}, however both analyses assume that both ions are photoionized.  For young nucleosynthetic systems, one expects to observe abundance enhancements in $\alpha$-process elements (Si, O) relative to (C, Fe).  However, deriving accurate (Si/O) and (Si/C) abundances is difficult, owing to the dependence of ionization corrections on the shape of the ionizing spectra \citep{GirouxShull97}.  We see six absorbers with both \SiIV\ and \CIV\ detections with $-1.3\le\log\,\bigl( N_{\rm SiIV}/N_{\rm CIV}\bigr)\le-0.3$, similar to the range seen at higher redshift.  The correlation is moderate ($R=0.6$), but the general trend is consistent with a linear relationship between the two species (Figure~\ref{fig:ratio2}a).

The ion pairs \FeIII/\SiIII\ and \CIII/\SiIV\ have similar ionization potentials and thus their column density ratios should be relatively insensitive to the ionizing spectrum.  These two pairs provide probes of relative C/O/Si/Fe abundance ratios in the IGM.  \CIII\ and \SiIV\ are seen together in six absorbers and are poorly correlated.  However, the observed $\langle\log(N_{\rm CIII}/N_{\rm SiIV})\rangle=0.7\pm0.6$ is slightly lower than model predictions at log\,$U\approx-1.5$, suggesting, if the single-phase assumption is used, an overabundance of silicon/carbon by $\sim0.3$ dex with respect to solar values.

\FeIII\ and \SiIII\ are seen together in 7 systems and their column densities are well-correlated ($R>0.9$) though with $N_{\rm FeIII}\propto N_{\rm SiIII}^{0.5\pm0.1}$ (Fig.~\ref{fig:ratio2}b).  The modeled \NFeIII/\NSiIII\ values depend on $U$, but for log\,$U\approx-1.5$ and solar [Fe/Si], the observed values, $\langle\log(N_{\rm FeIII}/N_{\rm SiIII})\rangle=0.5\pm0.5$, are too high by a factor of 2-3 dex.  If \FeIII\ and \SiIII\ truly arise in the same gas, normal photoionization corrections require an enormous overabundance of iron in the IGM, $\rm[Fe/Si]_{IGM}=2.6\pm1.2$.  A similar iron overabundance is found by comparing \FeIII/\CIII.  

The basic issue here arises because of the low ionization threshold (30.63 eV) and high cross section for photoionizing \FeIII\ to \FeIV.  If Fe/Si and Fe/C are in solar abundance ratios, it is difficult to obtain $N_{\rm FeIII}>N_{\rm SiIII}$.  In many of the IGM absorbers detected in both ions (Fig.~\ref{fig:ratio2}b), the observed column density ratio is [\FeIII/\SiIII]$=0.5\pm0.3$, whereas CLOUDY models predict much lower ratios.  The solution to the \FeIII\ problem may lie in details of the assumed AGN spectrum; we use a continuous power-law spectrum from 1-22 ryd, which leads to \FeIII\ being photoionized to \FeIV.  A power law with a slope break at 3-5 ryd, as proposed by \citet{HaardtMadau01}, results in relatively less flux at higher energies and boosts the predicted \FeIII\ columns.  However, this would make the production of photoionized \CIV\ and other high ions even more difficult.  With more data and a larger survey, we might be able to distinguish among several competing effects:  metallicity enhancements of [Fe/Si] due to nucleosynthesis, changes in the ionizing radiation field changes around 3-5 ryd, and the relative portions of \FeIII, \SiIII, and \CIII\ in various ionization phases. 

\begin{figure*}[t]
  \plottwo{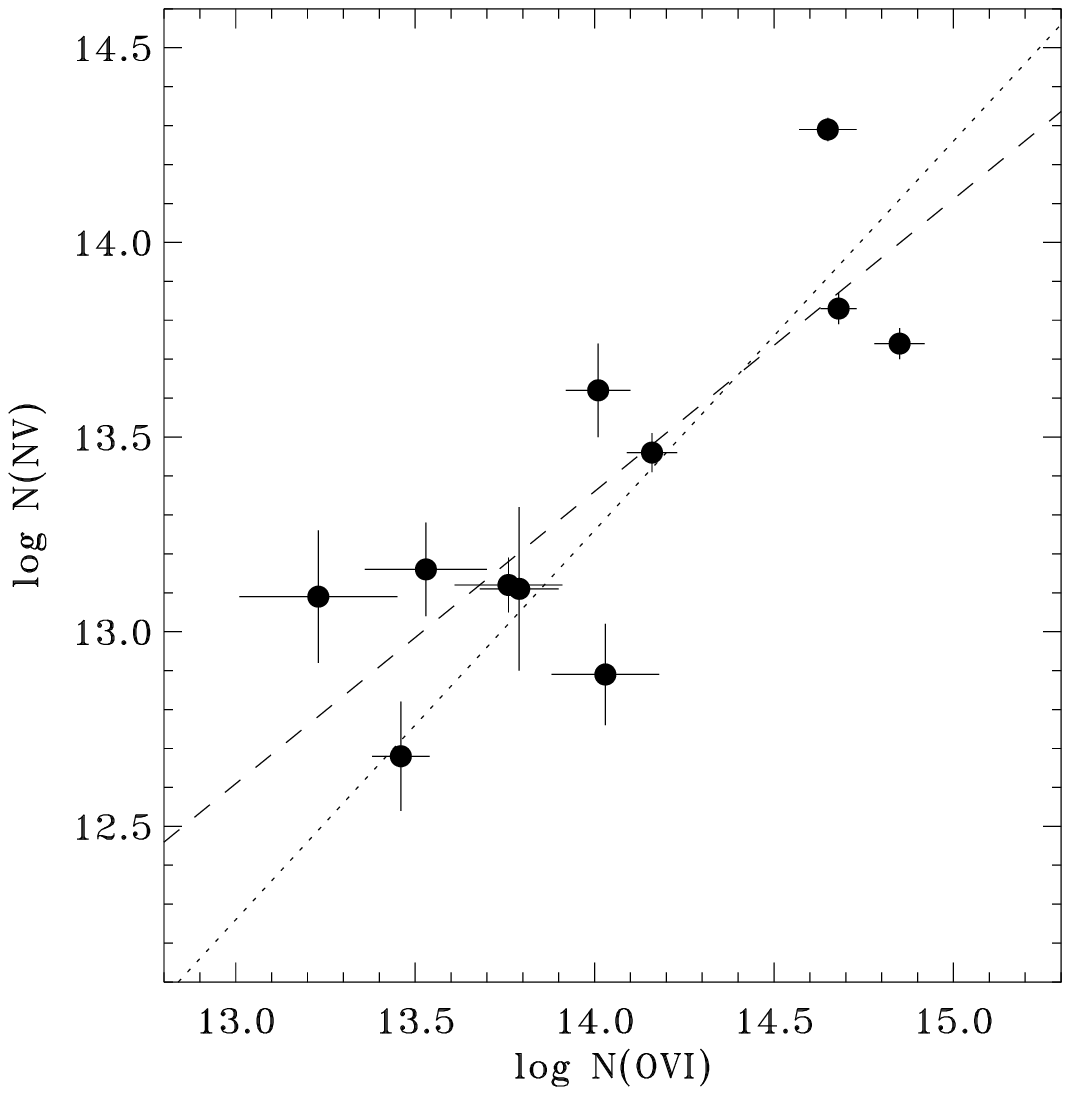}{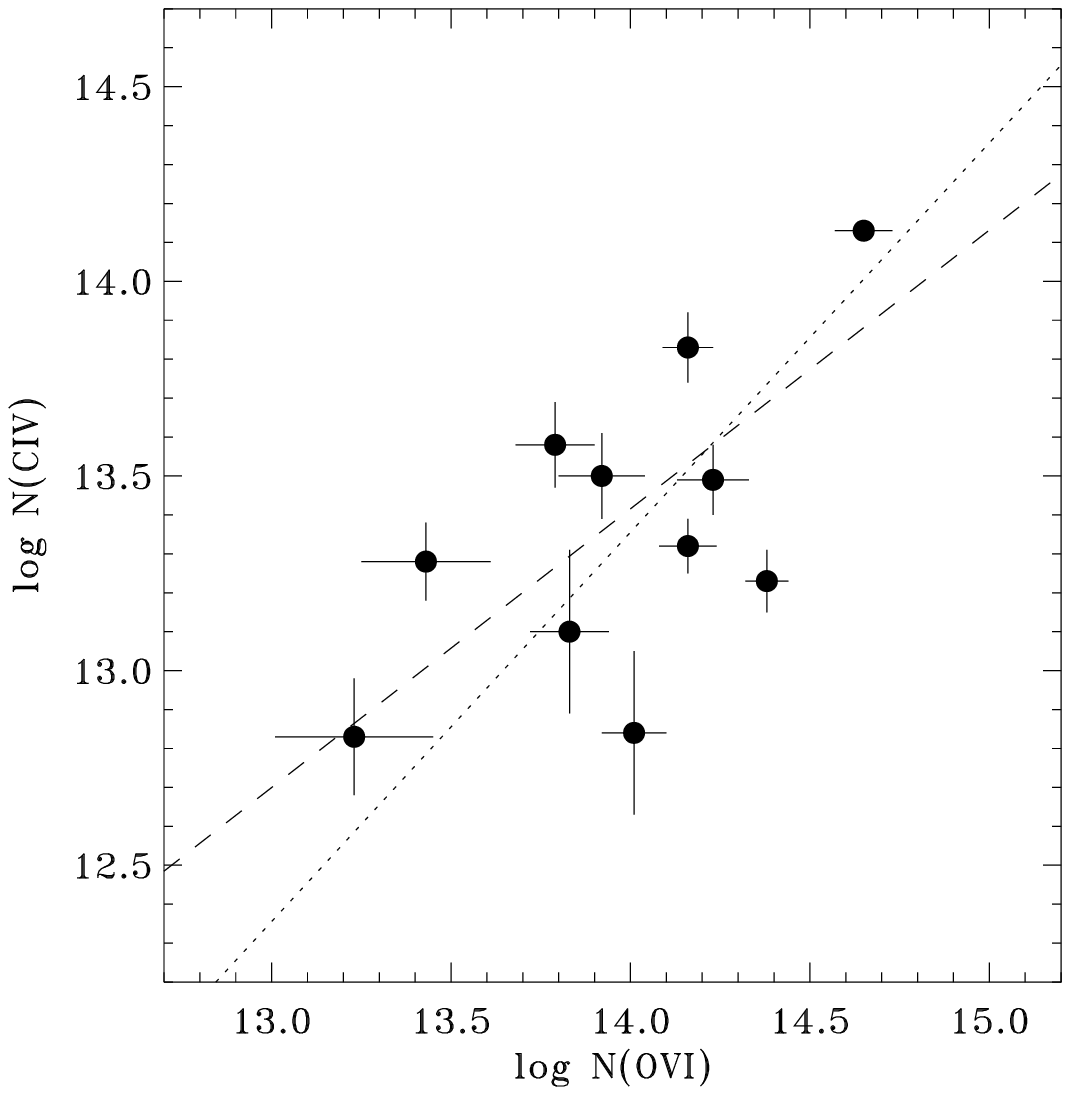} 
  \caption{Correlations between highly ionized species.  The column
  densities of N\,V and O\,VI (left) are well correlated in eleven
  absorbers ($R=0.81$), and the best fit (dashed) has a slope of
  $0.75\pm0.08$.  However, a 1:1 fit is plausible (dotted), with an
  offset of $-0.74\pm0.03$ dex, consistent with log\,(N/O)$_\sun
  \approx-0.7$.  C\,IV and O\,VI appear together in eleven absorbers
  (right), but the correlation is only moderate ($R=0.62$).  Much like
  N\,V vs.\ O\,VI, the best fit slope is $\sim0.7$ (dashed) but the
  data are are consistent with a 1:1 relationship with an offset of
  $\sim-0.6$ (dotted).}  \label{fig:ratio3}
\end{figure*}

The CLOUDY models can also be applied to the high ions, believed to reside in the WHIM.  The correlation is much better for the eleven systems with both \NV\ and \OVI\ absorption ($R=0.8$, Fig.~\ref{fig:ratio3}a).  The observed best fit has a slope of $0.75\pm0.08$, but a fit with slope of 1 is quite plausible with an offset of $-0.74\pm0.03$ dex, comparable to the solar abundance, log\,$(N/O)_{\odot}\sim-0.7$.  Unfortunately, these three ions are neither of the same element nor do they have overlapping ionization potentials, so there is an abundance/ionization degeneracy in the $(U,T)$ solutions.  Assuming a solar N/O abundance, we find that the observations are consistent with collisionally ionized systems at $T=10^{5.35\pm0.10}$~K, midway between the peak CIE abundance temperatures of the two ions.  Changing the N/O abundance to the sub-solar values typically seen in HVC gas \citep{Gibson00,Gibson01, Collins03,Collins07} does not change the temperature solution appreciably.

Observed \NV/\OVI\ ratios are also consistent with pure photoionization models with log\,$U=-0.4\pm0.5$.  However, this is a factor of $\sim10$ higher than that derived from the low ion ratios.  This pattern is familiar from the \OVI/\CIII\ results in Paper~2, where \CIII\ suggested log\,$U\sim-1.5$, while photoionized \OVI\ required log\,$U\sim-0.5$.  Altering the relative N/O abundance in the IGM varies the ionization parameter solution.  Matching the observed ratios to the derived log\,$U=-1.5$ requires a relative N/O overabundance of 0.3--1.3 dex, opposite to the trend observed in HVCs.  Varying the slope of the ionizing spectrum will also alter the predicted ratio, but an extremely hard incident spectrum ($\alpha_s<1$) is required for a log\,$U=-1.5$ solution.  Based on observations of nearby AGN, a spectral slope this steep is deemed unrealistic.

\CIV\ and \OVI\ appear together in eleven systems though the correlation is not as clear as with \NV\ vs.\ \OVI\ ($R=0.6$, Fig.~\ref{fig:ratio3}b).  The best fit slope is $0.7\pm0.1$ though a 1:1 fit with an offset of $-0.64\pm0.04$ is plausible.  This offset may be a manifestation of the sub-solar C/O ratio implied in Paper~2, or it may reflect the faster cooling times experienced by lower temperature shocked gases.  

The model ($U,T$) solution is similar to that for \NV/\OVI\ above; log\,$U\sim-0.5$ or $T\sim10^{5.4\pm0.1}$~K for a solar C/O ratio.  Again, matching observed C/O to a log\,$U\sim-1.5$ solution requires C to be overabundant with respect to O by $\sim0.6$ dex, opposite the subsolar C/O ratio implied in Paper~2.

\subsection{WHIM Cosmology and Metallicity of the IGM}

Integrated down to the lowest level of detectable absorbers ($W\sim10$~m\AA), our survey of three high metal ions estimates that they trace baryon contents of approximately  $\Omega^{\rm (ion)}_{\rm IGM}\approx9\%$ (\OVI), 5\% (\NV), and 3\% (\CIV) of the total.  These estimates scale inversely with an assumed IGM metallicity of 10\% solar and the adopted ionization corrections characteristic of maximum abundance in CIE:  $f_{\rm ion}=0.22$ (\OVI), 0.24 (\NV), and 0.29 (\CIV).  We see that the WHIM baryon fraction, $\Omega^{\rm(ion)}_{\rm IGM}$, implied by \OVI\ absorption is greater than that from \NV, which in turn is greater than that from \CIV.  However, these differences should not be taken too seriously, since there are many assumptions that go into these estimates.  First, the ionization fractions may differ from CIE, particularly if photoionization plays some role in the \CIV\ absorbers.  Second, these ions have different sensitivity limits in column density owing to line strengths and their location at different wavelengths.  Alternatively, the differences may represent a true gradient in $\Omega_{\rm WHIM}$ as a function of temperature.  The latter interpretation is plausible since \OVI, \NV, and \CIV\ have peak CIE temperatures of 3, 2, and $1\times10^5$~K, respectively, and radiative cooling rates have a steep dependence on temperature in that regime \citep{SutherlandDopita93}.  Given its poor correlation with \OVI\ and \NV, many \CIV\ absorbers may be photoionized gas rather than thermal in nature.  Thus, $\Omega\rm_{WHIM}^{(CIV)}$ should probably be taken as an upper limit until we can refine the multiphase nature of the absorbers with additional data from larger \CIII\ and \CIV\ surveys. 

At our most sensitive \OVI\ limit ($W>10$ m\AA), we find $\Omega_{\rm WHIM}=8.6\pm0.8$\%.  This is higher than the value quoted in Paper~1 ($\sim5$\%), probably because we now include weaker \Lya\ systems (the \Lya-bias) and have better statistics in the weak \OVI\ absorbers themselves.  However, the power-law slope of the WHIM ions is steep, with a similar amount of mass in the numerous weak absorbers as in the infrequent strong absorbers.  Future WHIM surveys sensitive to lower \NOVI\ limits may well increase the baryon mass fraction in the WHIM closer to the 30-50\% predicted by simulations \citep{Dave99,CenFang06}.  

X-ray probes of the WHIM at $T>10^6$~K, where a considerable fraction of WHIM is predicted to exist, are not yet available with the same fidelity as FUV WHIM tracers (\OVI, etc).  For example, \OVII\ $\lambda21.60$ is 9.1 times weaker ($f\lambda$) and \OVIII\ $\lambda18.97$ is 17.4 times weaker than \OVI\ $\lambda1032$.  Partially offsetting these atomic ratios are the CIE peak ionization fractions of \OVII\ and \OVIII, which are 3--4 times that of \OVI.   However, the primary contribution to the low X-ray spectral sensitivity comes from the low resolution of the grating spectrographs aboard \Chandra\ (700~\kms\ at 20~\AA) and {\it XMM-Newton} ($\sim$1000~\kms), compared to those on {\it Hubble} (7--20~\kms) or \FUSE\ (20~\kms).  Thus, the \OVI\ surveys probe hot gas down to $N_{\rm OVI}\approx 10^{13}$~\cd, whereas the recent claimed \OVII\ and \OVIII\ detections are at the level $N_{\rm OVIII}\approx5\times10^{15}$~\cd\ \citep[toward PKS\,2155$-$304,][]{Fang02,Fang07} and $N_{\rm OVII}\approx1\times10^{15}$~\cd\ \citep[toward Mrk\,421,][]{Nicastro05} with significant uncertainties.  However, these redshifted IGM absorbers have not been seen in {\it XMM/Newton} observations \citep{Rasmussen07,Kaastra06} and their verification is therefore in doubt \citep{Bregman07}.  Thus, for the moment, the X-ray detection statistics of remain highly uncertain while FUV spectroscopy provides our best (only?) measurements of the WHIM, limited to the regime with $T<10^6$~K.

\begin{figure}[b]
  \epsscale{1}\plotone{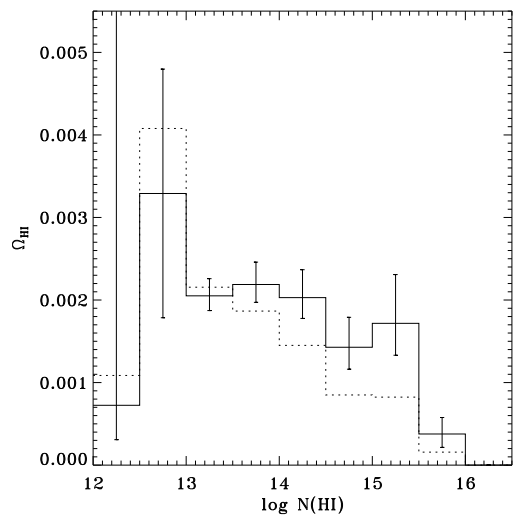} 
  \caption{Differential $\Omega_{\rm HI}$ as a function of log\,\NHI\
  for 650 \Lya\ absorbers calculated by the methods of Penton et
  al. (2000, solid) and Schaye (2001, dotted).  The log\,$N<12.5$ bin
  contains only 17 absorbers and is not included in our calculations
  of cumulative $\Omega_{\rm HI}$ in Table~\ref{tab:omega_lya}.}
  \label{fig:omega_lya}
\end{figure}

%
%
\begin{deluxetable*}{lccccc}
\tabletypesize{\footnotesize}
\tablecolumns{6} 
\tablewidth{0pt} 
\tablecaption{Baryon Content of the Local \Lya\ Forest}
\tablehead{\colhead{$\log\,N_{\rm HI}$ Range}    &
           \colhead{$\cal N$}     &
           \colhead{$\Omega_{\rm Ly\alpha}$\,\tnma}     &
           \colhead{$\Omega_{\rm Ly\alpha}/\Omega_b~(\%)$\,\tnma}   &
           \colhead{$\Omega_{\rm Ly\alpha}$\,\tnmb}     &
           \colhead{$\Omega_{\rm Ly\alpha}/\Omega_b~(\%)$\,\tnmb}   }
\startdata
$ 12.5-13.5 $&$ 373 $&$ 0.0053\pm0.0015 $&$ 11.7\pm 3.3 $&$ 0.0062\pm0.0019 $&$ 13.7\pm 4.1 $\\
$ 13.5-14.5 $&$ 206 $&$ 0.0042\pm0.0004 $&$  9.3\pm 0.8 $&$ 0.0033\pm0.0003 $&$  7.3\pm 0.7 $\\
$ 14.5-15.5 $&$  50 $&$ 0.0031\pm0.0006 $&$  6.9\pm 1.3 $&$ 0.0017\pm0.0003 $&$  3.7\pm 0.7 $\\
$ 15.5-16.5 $&$   4 $&$ 0.0004\pm0.0002 $&$  0.8\pm 0.4 $&$ 0.0002\pm0.0001 $&$  0.3\pm 0.2 $\\
&&&&\\
$ 12.5-14.5 $&$ 579 $&$ 0.0096\pm0.0016 $&$ 21.0\pm 3.4 $&$ 0.0096\pm0.0019 $&$ 21.0\pm 4.2 $\\
$ 14.5-16.5 $&$  54 $&$ 0.0035\pm0.0006 $&$  7.7\pm 1.4 $&$ 0.0018\pm0.0003 $&$  4.0\pm 0.7 $\\
&&&&\\
$ 12.5-16.5 $&$ 633 $&$ 0.0131\pm0.0017 $&$ 28.7\pm 3.7 $&$ 0.0114\pm0.0019 $&$ 25.0\pm 4.2 $\\
\enddata
\tablenotetext{a}{Method of Penton, Shull, \& Stocke (2000)}
\tablenotetext{b}{Method of Schaye (2001)}
  \label{tab:omega_lya}
\end{deluxetable*}

In the photoionized \HI\ gas in the \Lya\ forest, \citet{Penton4} found that $29\pm4$\% of the local baryon mass could be accounted for down to $\log\,N_{\rm HI}=12.5$ ($W_{\rm Ly\alpha}=17$~m\AA).  We determine values of $\Omega$ for \HI\ via equations~\ref{eq:omega_penton} and \ref{eq:omega_schaye} and present the results in Table~\ref{tab:omega_lya} and Figure~\ref{fig:omega_lya} for different log\,\NHI\ values.  Our survey of 650 \Lya\ absorbers gives $\Omega_{\rm IGM}^{(\rm HI)}/\Omega_b=29\pm4$\% integrated down to the same column density limit as Penton \etal\ but with $\sim3$ times higher pathlength $\Delta z$.  Both results are consistent with cosmological simulations, which find that $\Omega_{\rm IGM}^{(\rm HI)}/\Omega_b=20-40$\% \citep{Dave99,Dave01,CenOstriker99,CenFang06}.

Low ions such as \CIII\ and \SiIII\ likely trace the subset of the \Lya\ forest that is enriched by star formation and galaxy feedback.  Using these ions as probes of metallicity will allow us to look for evolution of metallicity with $z$ and \NHI.  We can also look for patterns indicative of nucleosynthesis near galaxies and effects on ion state arising from multiphase photoionized and collisionally ionized gas.  In the doubly-ionized species, we see $\Omega_{\rm IGM}^{(\rm ion)}/\Omega_b=1-4$\% (Table~\ref{tab:results2}), well below the $\sim30$\% value for the \Lya\ forest as a whole if we take the fiducial CIE values for $f_{\rm ion}$ listed in Table~3.  The assumed ion fractions for \CIII, \SiIII, and \FeIII\ are all close to unity.  Lower values of $f_{\rm ion}$, as might be the case in non-equilibrium ionization, would produce smaller baryon fractions and we conclude that $\Omega_{\rm IGM}^{(\rm ion)}$ is an upper limit in the case of the doubly-ionized species.  Likewise, the baryon fraction probed by \SiIV\ at low $z$ is surprisingly high, $\Omega_{\rm IGM}^{\rm (SiIV)}/\Omega_{\rm b}=8\pm1$\%, comparable to that of \OVI.  However, \SiIV\ is most likely photoionized, and thus the CIE ion fraction $f_{\rm SiIV}\sim0.2$ assumed in this calculation may be too low.  These two limits suggest $1\%\la(\Omega_{\rm IGM}^{\rm (ion)}/\Omega_b)\la8\%$, or that roughly 3-25\% of the local \Lya\ forest by mass is enriched at a currently detectable level.  More sophisticated models may shed light on this problem.

As in Papers 1 and 2, we estimate the metallicity of IGM material based on multiphase ratios (Table~\ref{tab:results1}).  The values generated cover a wide range from $Z_{\rm Fe}=0.7\,Z_\sun$ to $Z_{\rm CIII}=0.015\,Z_\sun$.  With $Z_{\rm OVI}$ and $Z_{\rm CIII}$ we reproduce the result, $\rm [C/O]\approx0.1\,[C/O]_\sun$, reported in Paper~2, however, \CIV\ reports $Z_C=0.055$, a factor of 3 higher than estimated by \CIII.  Similarly, adjacent Si ions disagree on IGM Si abundance by a factor of eight.  As with $\Omega_{\rm IGM}^{(\rm ion)}$ above, each metallicity estimate scales as $[f_{\rm ion}]^{-1}$, so that the inferred metallicities for the low ions may be upper limits.  However, this puts the adjacent-ion metallicity estimates even further from agreement.  With larger data bases, it may be better to perform a joint analysis of ion pairs such as \CIII/\CIV\ and \SiIII/\SiIV, to arrive at consistent values for $\log U$ and $Z$.

If we restrict ourselves to species with the most detections (\OVI, \CIII, \SiIII) we see that the WHIM tracer (\OVI) shows $Z\approx0.15\,Z_\sun$ while the \Lya\ forest tracers (\CIII, \SiIII) have a lower metallicity ($Z\sim0.02\,Z_\sun$).  The pattern can be forced a bit by including \CIV\ ($Z\sim0.05\,Z_\sun$).  This would imply that shocked material traced by high ions tends to have somewhat higher metallicity than the ambient IGM, as one would expect if both shocks and enrichment were products of galactic winds and star formation feedback.  However, \NV, \SiIV, and \FeIII\ break the metallicity-ionization potential pattern with anomalously high values.  Future nearest-neighbor-galaxy studies \citep[e.g.,][]{Prochaska06,Stocke06b} may help make sense of these confusing multiphase ratio metallicity estimates.  In particular, nucleosynthetic effects may show up in galactic outflows which may be partially responsible for IGM shock heating traced by \OVI.

\section{Conclusions and Summary}

We present the results of the largest low-redshift IGM survey to date. In total, we analyzed 650 \HI\ absorbers at $z<0.4$ along 28 AGN sight lines.  For each IGM system, we measured detections or upper limits in 13 transitions of 7 metal ions as well as \HI\ (\Lya, \Lyb, \OVI, \NV, \CIV, \CIII, \SiIV, \SiIII, and \FeIII).  

Our \OVI\ results reinforce Danforth \& Shull (2005) results but over a broader redshift range, and to a slightly deeper sensitivity limit.  We found 40 \OVI\ absorbers in Paper~1 at $z<0.15$ and expand the sample to 83 \OVI\ absorbers at $z<0.4$ in this work.  The \dndz\ turnover below log\,$N_{\rm OVI}=13.4$ seen in Paper~1 is not reproduced in the larger dataset, at least down to log\,$N_{\rm OVI}=13.0$.  Detection statistics from the low-$z$ bin of this survey are consistent with equivalent values from Paper~1.  Comparison of \dndz, $\alpha_{14}$, and $\Omega$ between high and low redshift bins hits at $z$-evolution in \dndz, $\alpha_{14}$, and derivative quantities.  

The \NV\ detections are not as common as \OVI\ (24 vs.\ 83) owing to the lower cosmic abundance of nitrogen.  However, \NV\ shows excellent correlation with \OVI\ in column density and displays many of the same characteristics (steep power-law slope, lack of correlation with \HI) and we conclude that \NV\ is a reliable tracer of WHIM material.  \CIV\ is a strong transition of an abundant metal, but our detection statistics are hampered by the limited redshift range in which we can observe ($z<0.112$).  Nevertheless, \NCIV\ does not appear to be well-correlated with either WHIM tracers (\OVI, \NV) nor lower ions (\CIII, \HI) and thus may arise from both photoionization and shocks.  We conclude that \CIV\ is not a conclusive WHIM tracer, though it is a good probe of metal enrichment.  Longer wavelength observations of AGN sight lines would improve the statistics of this ion.

We made numerous detections of both \CIII\ and \SiIII\ in IGM absorbers (39 and 53, respectively).  Column densities of both ions show reasonable correlation with \NHI, and the \dndz\ distributions are similar, leading us to classify both ions as predominantly photoionized.

Detections of both \SiIV\ and \FeIII\ are not as plentiful as the other low ions, and conclusions are harder to reach.  Still, the majority of the \SiIV\ detections (12/20) also show \SiIII\ absorption, and \NSiIII\ and \NSiIV\ are very well correlated.  We compare our observed line ratios with a set of CLOUDY models and find them consistent with a photoionized IGM with log\,$U=-1.6\pm0.7$.  Our other adjacent ion pairs (C\,III/IV) appear together in only eight absorbers and their column densities are not well-correlated.  However, they are consistent with models with log\,$U=-1.4\pm0.5$.  Based on these diagnostics, we conclude that typical IGM clouds have an ionization parameter of log\,$U\approx -1.5\pm0.5$.  Photoionization models with $U$ ten times higher would be required to explain the observed \NNV/\NOVI\ and \NCIV/\NOVI\ ratios, and this bolsters our interpretation of those species as collisionally ionized.

\NFeIII\ appears stronger than predicted by CLOUDY models by a factor of $\sim100$ or more when compared to either \NSiIII\ or \NCIII.  This may be indicative of a softer ionizing spectrum above $\sim3$ ryd than was assumed in the models, but that explanation causes a set of other problems.  More sophisticated modeling is required to explain these observations.


\subsection{Status of the Baryon Census}

With an unprecedented number of \HI\ absorbers, we conclude that the \Lya\ forest makes up $29\pm4$\% of the local baryonic material by mass.  This is compatible with previous surveys \citep{Penton4} as well as simulations \citep{Dave99,Dave01}.  \CIII\ and \SiIII\ detection statistics suggest that 3-25\% of the \Lya\ forest material is enriched with metals at a detectable level at low $z$.

At our most sensitive limit, log\,$N_{\rm OVI}\geq13.0$, we can further account for $8.6\pm0.7$\% of the baryons residing in the WHIM at temperatures around $T=10^{5.5}$~K via \OVI\ observations.  Our \NV\ detections suggest a smaller WHIM fraction, consistent with nitrogen's lower abundance and lower range of temperatures over which it is is abundant.  However, the \dndz\ distributions of both WHIM tracers are fitted with a power law of index $\beta\approx2$, implying that weak absorbers contribute as much to the mass fraction as strong absorbers.  More sensitive FUV surveys will likely increase the WHIM fraction.  

Additionally, there is the issue of ``\Lya\ bias''.  In Paper~1, we examined only strong \Lya\ systems ($W_{\rm Ly\alpha}>80$~m\AA), yet \NOVI\ was shown to have little or no correlation with \NHI.  In this work, we set out to minimize this bias by including all detected \HI\ systems, no matter how weak.  Inclusion of weaker \Lya\ lines in the survey boosts \OVI\ detection statistics and all related quantities (\dndz, $\Omega_{\rm IGM}^{\rm (ion)}$) by $\sim20$\% compared with the previous survey.  However, even this survey results in some \Lya\ bias since \HI\ detections are required at some level, and the finite S/N of the data limits this to $W_{\rm Ly\alpha}\ga15$ m\AA\ in even the best data.  The results of \citet{Tripp08} suggest an additional $\sim20$\% increase if \OVI\ systems are measured with {\it no} corresponding \HI\ detection.  Combining these two effects, we suggest that the \Lya\ bias on WHIM surveys is a real effect at the level of 20-40\%.  Thus we suspect the true value of $\Omega_{\rm WHIM}$ as detected via \OVI\ absorption is probably closer to 10\% of $\Omega_b$.

Even reliable WHIM tracers such as \OVI\ and \NV\ can only detect material that has been significantly enriched above primordial metallicity.  Thus, there is a metallicity bias in this survey.  If we take the very rough 3\%-25\% value from the \SiIII\ and other low ion estimates discussed above as a typical fraction of the IGM enriched to detectable levels, that may boost the true fraction of material at $T=10^{5-6}$~K by 3-25 times ($\sim30\%-250\%$ of the baryons)!  Clearly the latter value is unphysical, but $\Omega_{\rm WHIM}\sim30$\% is plausible and consistent with some simulations.

But what about the balance of the baryon census?  If condensed objects (stars, galaxies, etc) account for $\sim7$\% of cosmological baryons \citep{SalucciPersic99}, then $\sim50\%$ of the local baryons remain unaccounted for.  Cooling of diffuse gas is fastest at lower temperatures, so it is likely that there is a vast hot reservoir of gas in the IGM \citep{CenOstriker99,Shull03}.  This hot WHIM ($T>10^6$~K) is inaccessible through FUV absorption lines, and the primary metal ions (\OVII, \OVIII) are only detected in the X-ray.  While \OVII\ and \OVIII\ absorption has been seen associated with the Galaxy and certain AGN \citep{Nicastro05,Fang07}, there are no confirmed measurements of these lines in the IGM.  This remains a challenge for sensitive X-ray spectrographs of the future.

The scheduled August 2008 installation of the {\it Cosmic Origins Spectrograph} on HST will provide a great opportunity to study the low-$z$ IGM.  With greater than ten times the sensitivity of STIS and a resolution comparable to that of \FUSE, high resolution, high signal-to-noise observations of faint quasars will become feasible in very reasonable observing times.  High throughput will enable us to observe many fainter AGN targets, many with very large pathlengths ($z<0.45$).  This will increase the available total IGM path length by perhaps an order of magnitude or more.  With this increase will come a proportional boost in detection statistics of WHIM tracers such as \OVI\ and \NV\ as well as \Lya, \SiIII, \CIV, and a host of other astrophysical interesting lines.

\acknowledgements

This work was supported by \FUSE\ grant NNG06GI91G, and NASA Theory grant NNX07AG77G.  The authors wish to acknowledge several fruitful conversations with Mark Giroux on the subject of line ratios and metagalactic ionizing radiation and John Stocke regarding cosmic variance and line identifications.  We thank Steve Penton for his expertise in reducing the STIS Echelle data, Renyue Cen and Taotao Fang for supplying the data for the simulations used in Figure~\ref{fig:cenfangfig2}, and Gary Ferland for feedback on our CLOUDY models.  


\end{document}